# Super-tetragonal $Sr_4Al_2O_7$: a versatile sacrificial layer for high-integrity freestanding oxide membranes

Jinfeng Zhang[1,9], Ting Lin[2,9], Ao Wang[1,9], Xiaochao Wang[3,9], Qingyu He[4], Huan Ye[1], Jingdi Lu[1], Qing Wang[1], Zhengguo Liang[1], Feng Jin[1], Shengru Chen[2], Minghui Fan[1], Er-Jia Guo[2], Qinghua Zhang[2], Lin Gu[5], Zhenlin Luo[4], Liang Si[3,6]*, Wenbin Wu[1,7,8]*, and Lingfei Wang[1]*

[1]Hefei National Research Center for Physical Sciences at Microscale, University of Science and Technology of China, Hefei 230026, China

[2]Beijing National Laboratory for Condensed Matter Physics, Institute of Physics, Chinese Academy of Sciences, Beijing 100190, China

[3]School of Physics, Northwest University, Xi'an 710127, China

[4]National Synchrotron Radiation Laboratory, University of Science and Technology of China, Hefei 230026, China.

[5]Beijing National Center for Electron Microscopy and Laboratory of Advanced Materials, Department of Materials Science and Engineering, Tsinghua University, Beijing 100084, China.

[6]Institut für Festkörperphysik, TU Wien, 1040 Vienna, Austria.

[7]Institutes of Physical Science and Information Technology, Anhui University, Hefei 230601, China.

[8]Collaborative Innovation Center of Advanced Microstructures, Nanjing University, Nanjing 210093, China

[9]These authors contributed equally: Jinfeng Zhang, Ting Lin, Ao Wang, and Xiaochao Wang.

*Email: liang.si@ifp.tuwien.ac.at; wuwb@ustc.edu.cn; wanglf@ustc.edu.cn

**Abstract**

Releasing the epitaxial oxide heterostructures from substrate constraints leads to the emergence of various correlated electronic phases and paves the way for integrations with advanced semiconductor technologies. Identifying a suitable water-soluble sacrificial layer, compatible with the high-quality epitaxial growth of oxide heterostructures, is currently the key to the development of large-scale freestanding oxide membranes. In this study, we unveil the "super-tetragonal" $Sr_4Al_2O_7$ ($SAO_T$) as a promising water-soluble sacrificial layer. The distinct low-symmetric crystal structure of $SAO_T$ enables a superior capability to sustain epitaxial strain, thus allowing for broad tunability in lattice constants. The resultant structural coherency and defect-free interface in perovskite $ABO_3$/$SAO_T$ heterostructures effectively restrain crack formations during the water-assisted release of freestanding oxide membranes. For a variety of non-ferroelectric oxide membranes, the crack-free areas can span up to a few millimeters in length scale. These compelling features, combined with the inherent high-water solubility, make $SAO_T$ a versatile and feasible sacrificial layer for producing high-quality freestanding oxide membranes, thereby boosting their potential for innovative oxide electronics and flexible device designs.

Transition-metal oxide-based heterostructures are characterized by a vast of emergent interfacial phenomena, stimulated by the coupling of spin, charge, orbital, and lattice degrees of freedom at the heterointerfaces[1,2]. Examples include 2-dimensional electron/hole gas[3], interfacial superconductivity[4], improper ferroelectricity[5], magnetic/polar skyrmions[6,7], etc. Although these interfacial phenomena hold rich physics and functionalities[8–10], the strong covalent bonds at film/substrate interfaces largely limit their integrations with other low-dimensional material systems, thus hindering the potential device applications[11,12]. In recent years, freestanding oxide membrane exfoliating and transferring technologies developed rapidly[13–17]. Among these advancements, the water-assisted exfoliation of freestanding oxide membranes using cubic $Sr_3Al_2O_6$ ($SAO_C$) epitaxial sacrificial layers has emerged as one of the most prominent and feasible approaches[14]. Since its discovery in 2016, $SAO_C$ has significantly boosted research on integrating $ABO_3$ perovskite oxide heterostructures with Van de Waals materials and advanced semiconductor technologies, signifying a great potential for next-generation electronic/spintronic devices[18–20]. Moreover, $SAO_C$ provides a step forward in exploiting fascinating functionalities that exclusively exist in the freestanding membrane form, including ferroelastic domain transformation-induced superelasticity[21,22], ferroelectricity in the monolayer limit[23], correlated electronic phase under extreme tensile strain[24,25], novel lateral twisting/boundary states[26,27], and high-density switchable polar skyrmions[28].

Despite these exciting advancements, the crystallinity and integrity of the freestanding oxide membranes remain unsatisfactory compared to the typical Van de Waals materials like graphene and transition-metal dichalcogenides[29,30]. Particularly for the non-ferroelectric (non-FE) oxides, the water-assisted release processes are often accompanied by degraded crystalline coherent length and high-density crack formation[31–34]. Millimeter-sized crack-free membranes were rarely achieved[24,25]. Such brittle fractures in released oxide membranes can be attributed to two main factors: 1) the intrinsic structural characteristics, including strong ionic/covalence bonds and lack of slip system; 2) extrinsic defect formation due to unavoidable interfacial lattice mismatch and strain relaxation[34–37]. Because of the strong electron correlation nature, such unwilling structural changes also lead to considerate degradations of physical properties in freestanding oxide membranes, limiting their potential in next-generation electronic device applications. Facing this challenge, several new sacrificial layer materials have been developed recently, aiming to reduce the interfacial lattice mismatch and crack density. But the improvement is still limited by the discrete lattice constants, poor solubility, or non-generic etchant[13,37–39].

In this work, we systematically explored the growth phase diagram of $SAO_C$ films and discovered a new $Sr_4Al_2O_7$ phase (denoted as $SAO_T$). The biaxial-strained $SAO_T$ film has a tetragonal structural symmetry and Sr-rich stoichiometry, distinct from the well-recognized cubic $SAO_C$ phase. Such a low-symmetric crystal structure of $SAO_T$ enables superior flexibility under epitaxial strain and thus a wide-range tunability of in-plane lattice constants. The resultant coherent growth of high-quality $ABO_3$/$SAO_T$ epitaxial heterostructures considerably improves the crystallinity and integrity of water-released freestanding oxide membranes. For the widely-investigated non-FE nickelates, manganites, titanates, ruthenates, and stannates with a broad lattice constant range (3.85~4.04 Å), the crack-free areas of the membranes released from $SAO_T$ can span up to a few millimeters in scale. The corresponding functionalities can be comparable to the epitaxial counterparts. Moreover, the unique atomic structure of $SAO_T$ lead to an inherent high water-solubility, thus ensuring an effective water-assisted exfoliation process. These compelling advantages make $SAO_T$ film a versatile and viable water-soluble sacrificial layer for fabricating a broad array of high-quality freestanding oxide membranes, which offer fertile grounds for the development of innovative electronic devices.

**Growth window of $SAO_T$ films**

The strontium aluminate (SAO, including $SAO_C$ and $SAO_T$) thin films and $ABO_3$/SAO multilayer heterostructures were epitaxially grown by pulsed laser deposition (PLD). We first grow the SAO films by laser ablation of a polycrystalline stoichiometric $SAO_C$ target (the molar ratio Sr:Al:O = 3:2:6) on (001)-oriented $(LaAlO_3)_{0.3}$–$(SrAl_{0.5}Ta_{0.5}O_3)_{0.7}$ [LSAT(001)] single crystalline substrates (see Methods Section for details). The epitaxial quality and stoichiometry of the SAO films predominantly depend on two parameters: oxygen partial pressure ($P_{O2}$) and laser fluence ($F_L$). During the growth of a series of SAO films, we altered the $P_{O2}$ from $10^{-4}$ to 20 Pa and adjusted the $F_L$ from 2.5 to 1.0 J/cm$^2$. A full set of x-ray diffraction (XRD) $2\theta$-$\omega$ scans of SAO/LSAT(001) films near the LSAT(002) diffraction are shown in Supplementary Fig. 1, and two representative curves are shown in Fig. 1a. Most curves exhibit a clear film diffraction peak near 45.8°, in line with previously reported $SAO_C(008)$ diffractions[14]. The out-of-plane $d$-spacing of $SAO_C(008)$ and peak intensity are summarized in Supplementary Fig. 2 and Fig. 3. Based on these parameters and the sharpness of Laue fringes, we can evaluate the epitaxial quality of SAO films and construct the growth phase diagram (Fig. 1b). Using the stoichiometric $SAO_C$ target, $SAO_C$ films can grow well within a dome-like $F_L$-$P_{O2}$ range, consistent with the large variety of $SAO_C$ growth conditions reported in the literature[13]. Film deposition beyond the $F_L$-$P_{O2}$ boundaries mostly leads to off-stoichiometry and poor crystallinity. However, within a narrow window near $F_L = 1.0$ J/cm$^2$ and $P_{O2} = 5$ Pa, sharp Laue fringes reemerge and the SAO diffraction unexpectedly shifts to a much lower Bragg angle of ~41.8° (Fig. 1a and Supplementary Fig. 4), signifying the emergence of a new structural phase, denoted as $SAO_T$. Subsequent structure characterizations further reveal that the epitaxial quality of the $SAO_T$ film is comparable to the optimized $SAO_C$ films (Supplementary Fig. 5). Using energy-dispersive spectroscopy measurements, we further confirmed that the Sr:Al molar ratio of $SAO_T$ films is ~2.05, close to the nominal value of a $Sr_4Al_2O_7$ compound (Supplementary Table 1 and Supplementary Fig. 6). The large deviation in chemical stoichiometry between the $SAO_T$ film and $SAO_C$ (Sr:Al ~ 1.51) target should relate to the kinetic processes during the laser ablation and deposition in such a narrow $L_F$-$P_{O2}$ window (see Supplementary Section 1 for detailed discussions). Accordingly, we refined the film deposition using a $Sr_4Al_2O_7$ target and substantially expanded the growth window of high-quality $SAO_T$ film (Fig. 1c and Supplementary Fig. 7). The wide $P_{O2}$ range ($10^{-4}$ ~ 20 Pa) and $F_L$ range (1.0 ~ 3.0 J/cm$^2$) should align with the capabilities of most research groups using PLD systems.

We further characterized the epitaxial strain states of the $SAO_C$ and $SAO_T$ films through the reciprocal space mappings (RSMs). For the $SAO_C$/LSAT(001) film, (Fig. 1d), the $SAO_C$(4012) and LSAT(103) diffractions have unequal in-plane and out-of-plane reciprocal space vectors ($Q_x$ and $Q_z$). Both the in-plane and out-of-plane lattice constants in pseudocubic perovskite notation derived from the RSMs are 3.96 Å ($a^*_{SAO-C}$). The $a^*_{SAO-C}$ value is the same as that of bulk $SAO_C$, signifying a fully-relaxed epitaxial strain. For the $SAO_T$/LSAT(001) film (Fig. 1e), however, the elongated $SAO_T$(2218) diffraction shares the same $Q_x$ with that of the LSAT(103) diffraction, demonstrating a coherent strain state. The in-plane and out-of-plane lattice constants in pseudocubic perovskite notation ($a^*_{SAO-T}$ and $c^*_{SAO-T}$) are 3.87 and 4.32 Å, respectively. Such a biaxially- and compressively-strained $SAO_T$/LSAT(001) film is expected to show a tetragonal symmetry (Supplementary Figs. 8-10). The tetragonality ($c/a$ ratio) reaches a significant value up to ~1.12. Drawing an analogy to the compressive-strain-induced isosymmetric phase transition in $BiFeO_3$ films[40], we suggest that the $SAO_T$ phase could be a "super-tetragonal" structural polymorph, featuring dramatic atomic arrangement changes from the parent $SAO_C$.

The ability to sustain epitaxial strain seems to be the most prominent feature of the newly discovered $SAO_T$ phase. For the $SAO_T$/LSAT(001) films, the coherent strain state can be maintained up to 100 nm thick (Supplementary Fig. 11). The resultant structural coherency further ensures a sharp and defect-free $SAO_T$/LSAT(001) interface (Supplementary Fig. 12). In addition to the LSAT(001), high-quality $SAO_T$ film can be epitaxially grown on and coherently strained to a variety of (001)-oriented $ABO_3$ perovskite substrates (Supplementary Fig. 13). As summarized in Fig. 1f, the in-plane lattice constant $a^*_{SAO-T}$ can be continuously adjusted in a wide range from 3.84 Å [grown on $SrLaGaO_4$(001)] to 3.95 Å [grown on $DyScO_3$(001)]. Further, by introducing Ba and Ca doping to the $SAO_T$, we successfully grow coherently-strained $Ba_2Sr_2Al_2O_7$/$KTaO_3$(001) and $Ca_2Sr_2Al_2O_7$/$LaAlO_3$(001) films (Supplementary Fig. 14). Consequently, the strain-tuning range of $a^*_{SAO-T}$ is expanded to 3.79~3.99 Å. More interestingly, the structure coherency can be maintained even in the (110)-oriented $SAO_T$ films grown on $SrLaGaO_4$(100) substrate, in which an elongated $c$-axis lying in-plane (Supplementary Fig. 15). Such superior strain adaptability, rarely observed in the $SAO_C$ counterparts, should be an inherent property of $SAO_T$, stemming from its distinctive crystal structure.

## Crystal structure of $SAO_T$

After identifying the compelling structural flexibility of $SAO_T$, we turn to probe its structural differences with $SAO_C$ at the atomic scale using cross-sectional scanning transmission electron microscopy (STEM). As shown in Fig. 2a,b, the STEM image of the epitaxial $SAO_C$/LSAT(001) film measured in high-angle annular dark-field (HAADF) mode displays a rhombus-like contrast along the LSAT[100] zone axis, stemming from alternating "B-site" cations (Sr and Al) and regularly ordered oxygen/cation vacancies[14]. As schematically depicted in Fig. 2c, such a cation ordering quadruples the lattice constant of $SAO_C$ ($a_{SAO-C}$) compared to that of cubic $ABO_3$ perovskite. In sharp contrast, the HAADF-STEM images of the $SAO_T$ film along LSAT[100] zone-axis (Fig. 2d,e) display a perovskite-like atomic contrast and an intensity modulation along the out-of-plane [001] axis. The "A-site" atomic columns exhibit higher intensity alternatively for every three perovskite-like unit-cells (see Supplementary Fig. 16 for details). More interestingly, the HAADF-STEM image captured along the LSAT[110] zone axis (Fig. 2f) displays a complex cation ordering: "A-sites" are fully occupied and display a similar 3 unit-cell intensity modulations, while the "B-sites" are alternatively occupied and show a blurry in-plane intensity modulation.

We now determine the atomic structure of the $SAO_T$ phase based on the STEM images and density-functional-theory (DFT) calculations. To the best of our knowledge, none of the reported strontium aluminates can match the STEM and XRD results. We searched numerous possible structure candidates and eventually found that the $SAO_T$ could share a similar atomic structure and orthorhombic symmetry as the $Ba_4Al_2O_7$ compound[41]. DFT-level structure relaxations further confirm that this orthorhombic $Sr_4Al_2O_7$ unit-cell is thermodynamically stable, with simulated lattice constants $a_{SAO-T} = 10.798$ Å, $b_{SAO-T} = 11.238$ Å, and $c_{SAO-T} = 25.732$ Å (See Methods and Supplementary Section 2 for details). The atomic arrangements of simulated $SAO_T$ ($Sr_4Al_2O_7$) unit-cell viewed along both $SAO_T$[100] and [010] axes match perfectly with the HAADF-STEM image along LSAT[110] zone-axis (Fig. 2f and Supplementary Fig. 17). Hence, the epitaxial relationship of coherently-grown $SAO_T$/LSAT(001) films should be $SAO_T$[100]//LSAT[110] and $SAO_T$[001]//LSAT[001] (Supplementary Fig. 18). Assuming that the $SAO_T$ unit-cell is $2\sqrt{2} \times 2\sqrt{2} \times 6$ times as large as the cubic perovskite unit-cell (Fig. 2g), we can obtain the reduced lattice constants as $a^*_{SAO-T} = b^*_{SAO-T} = 3.896$ Å, and $c^*_{SAO-T} = 4.288$ Å. These simulated values are very close to those derived from the XRD results (Supplementary Table 2), further supporting the validity of our proposed structure. For the $SAO_T$ films grown on cubic substrates [e.g. LSAT(001) or STO(001)], the biaxial strain could convert the lattice symmetry from orthorhombic to tetragonal. For the films grown on orthorhombic substrates [e.g. $NdGaO_3$(001) or $DyScO_3$(001)], the original orthorhombic symmetry preserves (Supplementary Fig. 10).

The unique atomic structure of $SAO_T$ also plays a deterministic role in enabling the coherent epitaxial strain at $ABO_3$/$SAO_T$ interfaces. To address this point, we first performed DFT calculations on the relative energy changes ($\Delta E$) of $SAO_C$ and $SAO_T$ unit-cells under manually-imposed bi-axial and anisotropic strain (See Methods Section for details). As shown in Fig. 3a,b and Supplementary Fig. 19, for both biaxial and anisotropic strain configurations, the $\Delta E$ calculated from $SAO_T$ is consistently smaller than that of $SAO_C$. Namely, the low-symmetry $SAO_T$ unit-cell is more flexible to accommodate the misfit strain imposed by either cubic or orthorhombic substrates, while the cubic $SAO_C$ unit-cell could be more rigid against strain-induced lattice distortion. Moreover, we also constructed $SAO_C$/STO(001) and $SAO_T$/STO(001) heterostructures as model systems (Supplementary Fig. 20) and evaluated their interfacial bonding strength through DFT calculation. As shown in Fig. 3c, the DFT-calculated bonding energy at $SAO_T$/STO(001) interface (1.97 eV) is more than twice higher than that at $SAO_C$/STO(001) interfaces (0.82 eV), and even comparable with that at $LaAlO_3$/STO interface (2.34 eV). Such a strong interfacial bonding strength and the inherent structural flexibility make $SAO_T$ a versatile structure template for the coherent growth of various oxide films. From the perspective of water-soluble sacrificial layers, it could be the key to minimizing the interfacial misfit strain and improving the quality of exfoliated freestanding oxide membranes.

**Freestanding oxide membranes released from $SAO_T$**

We now examine the potential of $SAO_T$ as a water-soluble sacrificial layer. An "optimal" water-soluble sacrificial layer must satisfy three key requirements. Firstly, it must enable the successful growth of target oxide films. The high crystallinity and integrity should be maintained in the freestanding membranes after the lift-off and transfer processes. Secondly, the representative functionalities of the exfoliated freestanding oxide membranes should be comparable to those grown epitaxially on single-crystalline substrates. Lastly, it should dissolve easily in water, allowing efficient membrane exfoliation within reasonable durations. To evaluate these criteria for the $SAO_T$ film, we correspondingly grow several typical perovskite oxide films on both $SAO_T$ and $SAO_C$ epitaxial films, and then conduct comparative studies on the integrity, crystallinity, functionalities, and exfoliation speed of the freestanding membranes.

We first performed comparative characterizations on the integrity and crystallinity of various perovskite oxide membranes released from both $SAO_C$ and $SAO_T$. As summarized in Supplementary Table 3, these oxides include $NdNiO_3$ (NNO), $LaNiO_3$ (LNO), $La_{0.7}Ca_{0.3}MnO_3$ (LCMO), $SrTiO_3$ (STO), $SrRuO_3$ (SRO), $SrSnO_3$ (SSO), and $BaTiO_3$ (BTO), with a broad range of bulk lattice constants ($a_p$, in pseudocubic notation) from 3.81 to 4.12 Å. To minimize the lattice mismatch between SAO and $ABO_3$ perovskite, we choose to grow the $ABO_3$/SAO bilayers on either LSAT(001) or STO(001) substrates. As depicted in Fig. 4a, we use standard PDMS-assisted release and transfer procedures of freestanding oxide membranes from both $SAO_C$ and $SAO_T$ (see Methods section for procedural details). For the FE BTO membranes, the inherent superelasticity accommodates stress and deformations generated during the lift-off process. Hence, the choice of water-soluble sacrificial layers ($SAO_C$ or $SAO_T$) becomes less critical for the crystallinity and integrity (Supplementary Fig. 21)[21,22,42]. Nevertheless, for the other non-FE oxides, the membranes released from $SAO_C$ and $SAO_T$ show dramatic differences. According to the optical microscopic images (Fig. 4b), the freestanding oxide membranes released from 30 nm $SAO_C$ exhibit high-density and periodic cracks, which are qualitatively similar to the morphologies shown in the literature[24,37]. In contrast, the oxide membranes released from 30 nm $SAO_T$ (Fig. 4c) show crack-free regions spanning up to several hundred micrometers in scale. For the membranes released from 10 nm $SAO_T$, the crack-free region can be expanded further to a few millimeters in scale (Supplementary Figs. 22-28), which could be attributed to the improved structure coherency and slower release speed (see Supplementary Section 4). Intriguingly, the optical microscopic images reveal micrometer-scale and periodic wrinkling morphologies in the crack-free regions, which were commonly observed in the FE oxide membranes with superelasticity. Accordingly, the freestanding oxide membranes released from $SAO_T$ should possess superior integrity and flexibility, even withstanding the large lattice deformation in the wrinkled microstructures.

To quantify the $SAO_T$-induced integrity improvements in oxide membranes, we calculated the average equivalent diameters ($D_E$) of the uncracked areas in the freestanding membranes. The $D_E$ versus $a_p$ curves are summarized in Fig. 4d. Within a broad $a_p$ range from 3.81~4.04 Å, the $D_E$ values of membranes released 10 nm (30 nm) $SAO_T$ reach few millimeters (hundreds of micrometers) in scale, which are orders of magnitude higher than the corresponding $D_E$ values in the $SAO_C$ case. According to the aforementioned structural differences between $SAO_C$ and $SAO_T$, we speculate that the strain coherency at $ABO_3$/SAO interfaces may play a crucial role in determining the crystallinity and integrity of these freestanding membranes. Taking LCMO/SAO/LSAT(001) as model systems, we verified this hypothesis by detailed strain analyses. For LCMO film grown on $SAO_C$/LSAT(001), RSM near LSAT(103) diffraction (Fig. 4e) shows a partial strain relaxation at the LCMO/$SAO_C$ interfaces. The weak and broad LCMO(116) diffractions of the exfoliated freestanding LCMO membrane (Supplementary Fig. 22) further signify an unsatisfactory crystallinity. In contrast, RSM characterizations (Fig. 4f) from the LCMO/$SAO_T$/LSAT(001) bilayer demonstrate a coherent strain state and high epitaxial quality. The strong and sharp LCMO(116) Bragg diffraction of the freestanding LCMO membranes confirms a persistent high crystallinity even after water-assisted exfoliation (Supplementary Fig. 22).

The strong correlation between the high integrity/crystallinity of oxide membranes and the coherent strain state at the $ABO_3$/SAO interface can be understood by a simple scenario as depicted in Fig. 4g. For the $ABO_3$/$SAO_C$ epitaxial heterostructures, the robust cubic lattice of $SAO_C$ inhibits the epitaxial strain propagation from the substrate to $ABO_3$ epitaxial films. The unavoidable lattice mismatch between $ABO_3$ and $SAO_C$ must be accommodated by the formation of periodic dislocations[34,36]. During the water-assisted exfoliation process, these defects inevitably rupture the film lattice, leading to the formation of periodic cracks. The correlation between lattice mismatch and membrane integrity is further implied by the steep slope-like line shape of the $D_E$ - $a_p$ curve. For the $ABO_3$/$SAO_T$ heterostructures, on the contrary, the inherent structural flexibility of $SAO_T$ and strong interfacial bonding enable a coherent strain state. The resultant sharp and dislocation-free $ABO_3$/$SAO_T$ interface should effectively hinder crack formation during exfoliation. And the release of compressive strain could be the main driving force of the wrinkling microstructure.

Secondly, we characterized the evolutions of physical properties in various oxide membranes released from $SAO_C$ and $SAO_T$. For the ferromagnetic metal LCMO/$SAO_T$/LSAT(001) film and corresponding membrane, the temperature-dependent magnetization ($M$-$T$) and resistivity ($\rho$-$T$) curves reveal a sharp paramagnetic-insulator (PMI) to ferromagnetic-metal (FMM) transition (Fig. 5a,b). The Curie temperature ($T_C$), saturated magnetization, and residual resistivity are comparable to the LCMO/LSAT(001) epitaxial films, consistent with the observed high crystallinity and integrity. In contrast, for both the LCMO/$SAO_C$/LSAT(001) epitaxial film and the corresponding LCMO membrane, the PMI-to-FMM transition becomes more slanted. These degradations in ferromagnetism and metallicity can be attributed to the residual tensile strain at the LCMO/$SAO_C$ interface and the high-density cracks formed during exfoliation (Fig. 4)[24,43]. The itinerant ferromagnet SRO films also show a similar trend. The $M$-$T$ and $\rho$-$T$ curves of the membranes released from

$SAO_T$ reveal a sharp FMM transition at $T_C$ ~150 K. The residual resistance ratio (RRR) value is up to 4.83, comparable with PLD-grown SRO/STO(001) epitaxial films (Fig. 5c-h)[6,44,45]. The SRO membrane also exhibits a strong perpendicular magnetic anisotropy (MA) dominated by the intrinsic magnetocrystalline anisotropy[46], which also signifies a high crystallinity. In contrast, the SRO membrane exfoliated from $SAO_C$ shows a demagnetization-dominated in-plane MA and a reduced RRR down to 1.93, which are consistent with the degradation of crystallinity and crack formation. Following the same scenario, we also characterized the electrical transport of LNO and NNO membranes (Supplementary Figs. 29,30). Consistently, their representative transport properties in membranes released from $SAO_T$ are well maintained or even improved. Namely, $SAO_T$ can universally ensure both the high integrity and epitaxial-film-like functionalities of oxide membranes.

Lastly, we examined the exfoliation efficiency of oxide membranes from $SAO_T$. To perform an equitable comparison between $SAO_C$ and $SAO_T$, we opted to exfoliate the 50 nm thick BTO membranes. Due to intrinsic superelasticity, the exfoliation speed of BTO membranes should not be influenced by crack formation and associated extrinsic water penetration. According to the *in-situ* monitoring by optical microscope (Supplementary Movies 1,2), the water-assisted exfoliation speed of BTO membranes released from $SAO_T$ is approximately one order of magnitude faster than that released from $SAO_C$. The trend of faster exfoliation speed is universally applicable for all the other oxide membranes we grew (Supplementary Fig. 31), signifying a higher water-solubility of $SAO_T$ than $SAO_C$. The dissolution speed of $SAO_T$ film also highly depends on the film thickness and Ca/Ba doping (Supplementary Fig. 32), which provides independent parameters for simultaneously optimizing the exfoliation efficiency and quality. The high water-solubility of $SAO_T$ also has a structural origin. As depicted in Supplementary Fig. 33, the Al-O networks in $SAO_T$ comprise discrete $AlO_4^{5-}$ and $Al_3O_{10}^{11-}$ groups, which hydrolyze in water more readily[41,47] than the $Al_6O_{18}^{18-}$ rings in $SAO_C$[14]. Despite such a high water-solubility, the $SAO_T$ film exhibits remarkable stability against ambient moisture over 40 days when incorporating an ultrathin STO protective layer (Supplementary Fig. 34). Consistently, this long-term stability of $SAO_T$ could be attributed to the high epitaxial quality of $ABO_3/SAO_T$ heterostructure[48].

**Summary and Outlook**

By exploring the phase diagram of SAO film growth, we identified the super-tetragonal $SAO_T$ as a promising water-soluble sacrificial layer with several compelling advantages. Firstly, $SAO_T$ showcases remarkable structural flexibility to adapt the epitaxial strain imposed by various perovskite substrates, providing wide-range tunability of in-plane lattice constant. Such inherent structural flexibility further ensures high-quality epitaxy of a broad spectrum of $ABO_3/SAO_T$ heterostructures with coherent strain states and dislocation-free interfaces, which restrain crack formation during water-assisted exfoliation. For various non-FE oxide membranes with lattice constants ranging from 3.81 to 4.04 Å, the crack-free areas can spans up to few millimeters in scale. Moreover, the strain-tunability of $SAO_T$ persists with Ba/Ca doping and (110)-oriented epitaxial growth, broadening its potential in developing novel freestanding oxides beyond traditional perovskites. Next, the $SAO_T$ has a wide and stable growth window, accessible for standard PLD techniques and compatible with the growth of most perovskite oxides. Lastly, its high water-solubility streamlines the membrane exfoliation process. With these attributes, $SAO_T$ sacrificial layer offers a versatile and feasible experimental approach to producing large-scale, crack-free freestanding oxide membranes, of which crystallinity and functionalities are comparable to the epitaxial films. The discovery of $SAO_T$ introduces a pivotal complement to the widely-used $SAO_C$ sacrificial layer, which may substantially promote the potential of freestanding oxide membranes for innovative, low-dimensional, and flexible device applications[18–20].

## Reference

1. Hwang, H. Y. et al. Emergent phenomena at oxide interfaces. *Nat. Mater.* **11**, 103–113 (2012).
2. Yu, P., Chu, Y. -H. & Ramesh, R. Oxide interfaces: pathways to novel phenomena. *Mater. Today* **15**, 320–327 (2012).
3. Ohtomo, A. & Hwang, H. Y. A high-mobility electron gas at the $LaAlO_3/SrTiO_3$ heterointerface. *Nature.* **427**, 423–427 (2004).
4. Caviglia, A. D. et al. Electric field control of the $LaAlO_3/SrTiO_3$ interface ground state. *Nature.* **456**, 624–627 (2008).
5. Bousquet, E. et al. Improper ferroelectricity in perovskite oxide artificial superlattices. *Nature.* **452**, 732–736 (2008).
6. Wang, L. et al. Ferroelectrically tunable magnetic skyrmions in ultrathin oxide heterostructures. *Nat. Mater.* **17**, 1087–1094 (2018).
7. Das, S. et al. Observation of room-temperature polar skyrmions. *Nature.* **568**, 368–372 (2019).
8. J. Mannhart & D. G. Schlom, Oxide interfaces—An opportunity for electronics. *Science.* **327**, 1607–1611 (2010).
9. Manipatruni, S. et al. Scalable energy-efficient magnetoelectric spin-orbit logic. *Nature.* **565**, 35–42 (2019).
10. Garcia, V. & Bibes, M. Ferroelectric tunnel junctions for information storage and processing. *Nat. Commun.* **5**, 5289 (2014).
11. Bakaul, S. R. et al. Single crystal functional oxides on silicon. *Nat. Commun.* **7**, 10547 (2016).
12. Lu, H. et al. Ferroelectric tunnel junctions with graphene electrodes. *Nat. Commun.* **5**, 6518 (2014).
13. Chiabrera, F. M. et al. Freestanding Perovskite Oxide Films: Synthesis, Challenges, and Properties. *Ann. Phys. (Berlin)* **534**, 2200084 (2022).
14. Lu, D. et al. Synthesis of freestanding single-crystal perovskite films and heterostructures by etching of sacrificial water-soluble layers. *Nat. Mater.* **15**, 1255–1260 (2016).
15. Chu, Y. H. Van der Waals oxide heteroepitaxy. *npj Quantum Mater.* **2**, 67 (2017).
16. Kum, H. S. et al. Heterogeneous integration of single-crystalline complex-oxide membranes. *Nature.* **578**, 75–81 (2020).
17. Sambri, A. et al. Self-Formed, Conducting $LaAlO_3/SrTiO_3$ Micro-Membranes. *Adv. Funct. Mater.* **30**, 1909964 (2020).
18. Yang, A. J. et al. Van der Waals integration of high-κ perovskite oxides and two-dimensional semiconductors. *Nat. Electron.* **5**, 233–240 (2022).
19. Kim, B. S. Y., Hikita, Y., Yajima, T. & Hwang, H. Y. Heteroepitaxial vertical perovskite hot-electron transistors down to the monolayer limit. *Nat. Commun.* **10**, 5312 (2019).

20. Lu, D., Crossley, S., Xu, R., Hikita, Y. & Hwang, H. Y. Freestanding Oxide Ferroelectric Tunnel Junction Memories Transferred onto Silicon. *Nano Lett.* **19**, 3999–4003 (2019).
21. Dong, G. et al. Super-elastic ferroelectric single-crystal membrane with continuous electric dipole rotation. *Science.* **366**, 475–479 (2019).
22. Peng, B. et al. Phase transition enhanced superior elasticity in freestanding single-crystalline multiferroic $BiFeO_3$ membranes. *Sci. Adv.* **6**, eaba5847 (2020).
23. Ji, D. et al. Freestanding crystalline oxide perovskites down to the monolayer limit. *Nature.* **570**, 87–90 (2019).
24. Hong, S. S. et al. Extreme tensile strain states in $La_{0.7}Ca_{0.3}MnO_3$ membranes. *Science.* **368**, 71–76 (2020).
25. Xu, R. et al. Strain-induced room-temperature ferroelectricity in $SrTiO_3$ membranes. Nat. Commun. **11**, 3141 (2020).
26. Wu, P. C. et al. Twisted oxide lateral homostructures with conjunction tunability. *Nat. Commun.* **13**, 2565 (2022).
27. Chen, S. et al. Braiding Lateral Morphotropic Grain Boundaries in Homogenetic Oxides. *Adv. Mater.* **35**, 2206961 (2023).
28. Han, L. et al. High-density switchable skyrmion-like polar nanodomains integrated on silicon. *Nature.* **603**, 63–67 (2022).
29. Liu, Y., Huang, Y. & Duan, X. Van der Waals integration before and beyond two-dimensional materials. *Nature.* **567**, 323–333 (2019).
30. Kim, G. et al. New Approaches to Produce Large-Area Single Crystal Thin Films. *Adv. Mater.* **35**, 2203373 (2023).
31. Li, Y. et al. Electrostatically Driven Polarization Flop and Strain-Induced Curvature in Free-Standing Ferroelectric Superlattices. *Adv. Mater.* **34**, 2106826 (2022).
32. Zhang, B., Yun, C. & MacManus-Driscoll, J. L. High Yield Transfer of Clean Large-Area Epitaxial Oxide Thin Films. *Nano-Micro Lett.* **13**, 363–376 (2021).
33. Wang, Q. et al. Towards a large-area freestanding single-crystal ferroelectric $BaTiO_3$ membrane. *Crystals.* **10**, 733 (2020).
34. Baek, D. J., Lu, D., Hikita, Y., Hwang, H. Y. & Kourkoutis, L. F. Mapping cation diffusion through lattice defects in epitaxial oxide thin films on the water-soluble buffer layer $Sr_3Al_2O_6$ using atomic resolution electron microscopy. *APL Mater.* **5,** 096108 (2017).
35. Hong, S. S. et al. Two-dimensional limit of crystalline order in perovskite membrane films. *Sci. Adv.* **3**, eaao5173 (2017).
36. Lu, D. et al. Strain Tuning in Complex Oxide Epitaxial Films Using an Ultrathin Strontium Aluminate Buffer Layer. *Phys. Status Solidi - Rapid Res. Lett.* **12**, 1700339 (2018).
37. Singh, P. et al. Large-Area Crystalline $BaSnO_3$ Membranes with High Electron Mobilities. *ACS Appl. Electron. Mater.* **1**, 1269–1274 (2019).
38. Bourlier, Y. et al. Transfer of Epitaxial $SrTiO_3$ Nanothick Layers Using Water-Soluble Sacrificial Perovskite Oxides. *ACS Appl. Mater. Interfaces.* **12**, 8466–8474 (2020).
39. Peng, H. et al. A Generic Sacrificial Layer for Wide-Range Freestanding Oxides with Modulated Magnetic Anisotropy. *Adv. Funct. Mater.* **32**, 2111907 (2022).
40. Hatt, A. J., Spaldin, N. A. & Ederer, C. Strain-induced isosymmetric phase transition in $BiFeO_3$. *Phys. Rev. B - Condens. Matter Mater. Phys.* **81**, 054109 (2010).
41. Kahlenberg, V. Crystal structure of $Ba_8[Al_3O_{10}][AlO_4]$, a novel mixed-anion Ba aluminate related to kilchoanite. *Mineral. Mag.* **65**, 533–541 (2001).
42. Jin, C. et al. Super-Flexible Freestanding $BiMnO_3$ Membranes with Stable Ferroelectricity and Ferromagnetism. *Adv. Sci.* **8**, 2102178 (2021).

43. Gao, G., Jin, S. & Wu, W. Lattice-mismatch-strain induced inhomogeneities in epitaxial La$_{0.7}$Ca$_{0.3}$MnO$_3$ films. *Appl. Phys. Lett.* **90**, 012509 (2007).
44. Lee, H. G. et al. Atomic-Scale Metal-Insulator Transition in SrRuO$_3$ Ultrathin Films Triggered by Surface Termination Conversion. *Adv. Mater.* **32**, 1905815 (2020).
45. Lu, J. et al. Defect-Engineered Dzyaloshinskii-Moriya Interaction and Electric-Field-Switchable Topological Spin Texture in SrRuO$_3$. *Adv. Mater.* **33**, 2102525 (2021).
46. Koster, G. et al. Structure, physical properties, and applications of SrRuO$_3$ thin films. *Rev. Mod. Phys.* **84**, 253–298 (2012).
47. Kahlenberg, V., Lazić, B. & Krivovichev, S. V. Tetrastrontium-digalliumoxide (Sr$_4$Ga$_2$O$_7$)-Synthesis and crystal structure of a mixed anion strontium gallate related to perovskite. *J. Solid State Chem.* **178**, 1429–1439 (2005).
48. Li, D. et al. Stabilization of Sr$_3$Al$_2$O$_6$ Growth Templates for Ex Situ Synthesis of Freestanding Crystalline Oxide Membranes. *Nano Lett.* **21,** 4454–4460 (2021).

**Methods**

**Epitaxial film growth.** We prepared the $Sr_3Al_2O_6$ ($SAO_C$) and $Sr_4Al_2O_7$ ($SAO_T$) polycrystalline targets by first sintering a mixture of stoichiometric amounts of $SrCO_3$ and $Al_2O_3$ at 1350 °C for 24 h, and then immediately grinding and pressing into pellets. We also prepare the $NdNiO_3$ (NNO), $LaNiO_3$ (LNO), $La_{0.7}Ca_{0.3}MnO_3$ (LCMO), $SrTiO_3$ (STO), $SrRuO_3$ (SRO), $SrSnO_3$ (SSO), and $BaTiO_3$ (BTO) targets using similar procedures. Using these oxide targets, we grow the epitaxial thin films on single-crystalline substrates through a pulsed laser deposition (PLD) system with a KrF excimer laser (248 nm). During the growth of strontium aluminate (SAO, including $SAO_C$ and $SAO_T$) films, we maintained the laser spot size, repetition rate, substrate temperature, and target-substrate distance at 4 $mm^2$, 3 Hz, 750 °C, and 55 mm, respectively. And we varied the laser fluence ($F_L$) from 1.0 to 3.0 $J/cm^2$ by inserting attenuators and altered the oxygen partial pressure ($P_{O2}$) from $10^{-4}$ to 30 Pa. We controlled the film thickness ($t_{SAO-C}$ or $t_{SAO-T}$) by the laser ablation duration. We selected (001)-oriented $(LaAlO_3)_{0.3}$-$(SrAl_{0.5}Ta_{0.5}O_3)_{0.7}$ [LSAT(001)] substrate to deposit SAO films and explore the growth phase diagram. The thermal stability of LSAT(001) substrates in a reduced environment (e.g. high vacuum) at high temperature (> 650 °C) surpasses that of the widely used STO(001) substrate. The other oxides (NNO, LNO, LCMO, STO, SRO, SSO, and BTO) were subsequently deposited at their optimal conditions according to previous publications. The thickness of all the non-ferroelectric films is kept at 35 nm, and the thickness of the BTO films is 50 nm.

**Release of freestanding membranes.** Before the exfoliation, we attached the surfaces of $ABO_3$/SAO heterostructures onto 0.5 mm-thick polydimethylsiloxane (PDMS) films. We then immersed the entire sample in deionized water at room temperature until the sacrificial layer was completely dissolved, thus lifting off the oxide membranes from the substrate. To conduct the characterizations on structure and functionalities, the freestanding oxide membranes were transferred and attached to atomically-flat sapphire glass slides. It is worth noting that PDMS is an elastomer. Compared to other widely used adhesive materials (e.g. epoxy), it can minimize the support and constraint to the freestanding oxide films to the maximum extent. Nevertheless, its elastic nature also easily causes the formation of cracks.

**Structure characterizations.** The crystallographic analysis of the epitaxial films and freestanding membranes were examined by a high-resolution X-ray diffractometer (PANalytical Empyrean X-ray diffractometer, Cu K$\alpha$1 radiation) with both the $2\theta$-$\omega$ linear scan and off-specular reciprocal space mapping (RSM) mode. The thicknesses of SAO and other oxide layers were obtained by fitting Laue fringes. The growth rate of each layer was calculated and controlled precisely by counting the number of laser pulses. The atomic structure of as-grown $SAO_C$ and $SAO_T$ films was characterized at room temperature using an ARM-200CF transmission electron microscope, operated at 200 keV and equipped with double spherical aberration (Cs) correctors. The HAADF images were measured in the scanning mode. Cross-sectional TEM specimens of STO/$SAO_T$/LSAT(001) and STO/$SAO_T$/LSAT(001) heterostructures were prepared using $Ga^+$ ion milling after the mechanical thinning down to ~20 μm.

**Magnetism and electrical transport measurements.** Temperature-dependent magnetization (M-T) curves were measured via a vibrating sample magnetometer (VSM, Quantum Design MPMS3). The external H was applied perpendicular or parallel to the film plane. The T-dependent resistivity ($\rho$-T) curves were characterized by a Physical Properties Measurement System (PPMS, Quantum Design). Before measurements, the stripe-like Pt electrodes were patterned onto the membrane/thin film surface by photolithography.

**Density-functional theory calculations.** DFT-level structural relaxations and electronic structure computations for both $SAO_C$ and $SAO_T$ phases are performed using WIEN2K and VASP codes with the Perdew-Burke-Ernzerhof version for solids of the generalized gradient approximation (GGA-PBESol) on a dense momentum grid with 200 k-points[49–54]. The initial structural parameters of $SAO_C$ are from Ref. 14 and

55, while the parameters of $SAO_T$ are adopted from the previously reported $Ba_4Al_2O_7$ compound[41]. The structural relaxation calculations for both structures are convergent, providing two stable crystal structures. The DFT-relaxed lattice constants (listed in Supplementary Table 3) are close to the experimental values.

To evaluate the structural flexibility of $SAO_C$ and $SAO_T$ unit-cells, we calculated the energy changes of these unit-cells under both biaxial and anisotropic strains. We first calculate the total energy ($E_{total}$) of $SAO_C$ and $SAO_T$ unit-cells with fully relaxed lattices. Then we manually altered the in-plane lattice constants $a_{SAO}$ and $b_{SAO}$ to impose biaxial/anisotropic strain, relaxed the lattice along the $c_{SAO}$ axis, and calculated the $E_{total}$ of the biaxially or anisotropically strained $SAO_C$ and $SAO_T$ unit-cells. The differences in $E_{total}$ between the strained and strain-free cases are defined as $\Delta E$. To directly compare the $SAO_C$ and $SAO_T$ unit-cells, the energy changes ($\Delta E$) are normalized by the $E_{total}$ in the strain-free case. To impose biaxial strain, we impose a constraint of $a_{SAO} = b_{SAO}$, and alter the $a_{SAO}$ by $\pm 1\sim 3\%$. To impose biaxial strain, we impose a constraint of $a_{SAO} \times b_{SAO}$ = constant and alter the orthorhombicity $b_{SAO}/a_{SAO}$ from 0.9 to 1.1.

To investigate the bonding strength at $SAO/ABO_3$ interfaces, we construct three (001)-oriented heterostructure supercells: 1) $SAO_C$(1 unit-cell)/STO(3 unit-cells), 2) $SAO_T$(1 unit-cell)/STO(3 unit-cells), and 3) $LaAlO_3$(3 unit-cells)/STO(3 unit-cells) (LAO/STO). The LAO/STO supercell is a control group for evaluating the interfacial bonding strength at SAO/STO interfaces. The interfacial bonding energies are calculated by the following formula:

$$E_{bond} = - [E_{SAO-STO} - (E_{slabs-SAO} + E_{slabs-STO})]/2n_{STO}$$

The $E_{SAO-STO}$ is the total energy of the constructed SAO/STO heterostructure supercell; $E_{slabs-SAO}$ and $E_{slabs-STO}$ are the total energy values of SAO ($SAO_C$ or $SAO_T$) and 3-unit-cell-thick STO slabs; The obtained value of $E_{bond}$ reflects the bonding strength between the surface of SAO and STO slabs. The $E_{bond}$ value should be divided by 2 as there are two interfaces in the computational models; To directly compare the bonding strengths at $SAO_T$/STO and $SAO_C$/STO interfaces, the $E_{bond}$ should be further normalized by the $n_{STO}$, which is the number of STO unit-cells bonded to the interface. Note that we intentionally add a negative sign in front of formula (1). Hence, if $E_{bond}$ is positive (negative), the SAO and STO tend to form stable (unstable) chemical bonds at the interfaces. For comparison, we also calculated the $E_{bond}$ at the LAO/STO interface. The corresponding results are included in Supplementary Section 2 and discussed in the main text.

**References**

49. Kresse, G. & Hafner, J. Ab initio molecular dynamics for liquid metals. *Phys. Rev. B*. **47**, 558–561 (1993).
50. Kresse, G. & Furthmüller, J. Efficiency of ab-initio total energy calculations for metals and semiconductors using a plane-wave basis set. *Comput. Mater. Sci.* **6**, 15–50 (1996).
51. Kresse, G. & Furthmüller, J. Efficient iterative schemes for ab initio total-energy calculations using a plane-wave basis set. *Phys. Rev. B*. **54**, 11169–11186 (1996).
52. Perdew, J. P. et al. Restoring the density-gradient expansion for exchange in solids and surfaces. *Phys. Rev. Lett.* **100**, 136406 (2008).
53. Blaha, P. et al. WIEN2K: An augmented plane wave plus local orbitals program for calculating crystal properties. *Techn. Univ. Wien* (2001).
54. Schwarz, K., Blaha, P. & Madsen, G. K. H. Electronic structure calculations of solids using the WIEN2k package for material sciences. *Comput. Phys. Commun.* **147**, 71–76 (2002).
55. Alonso, J. A., Rasines, I. & Soubeyroux, J. L. Tristrontium dialuminum hexaoxide: An intricate superstructure of perovskite. *Inorg. Chem.* **29**, 4768–4771 (1990).

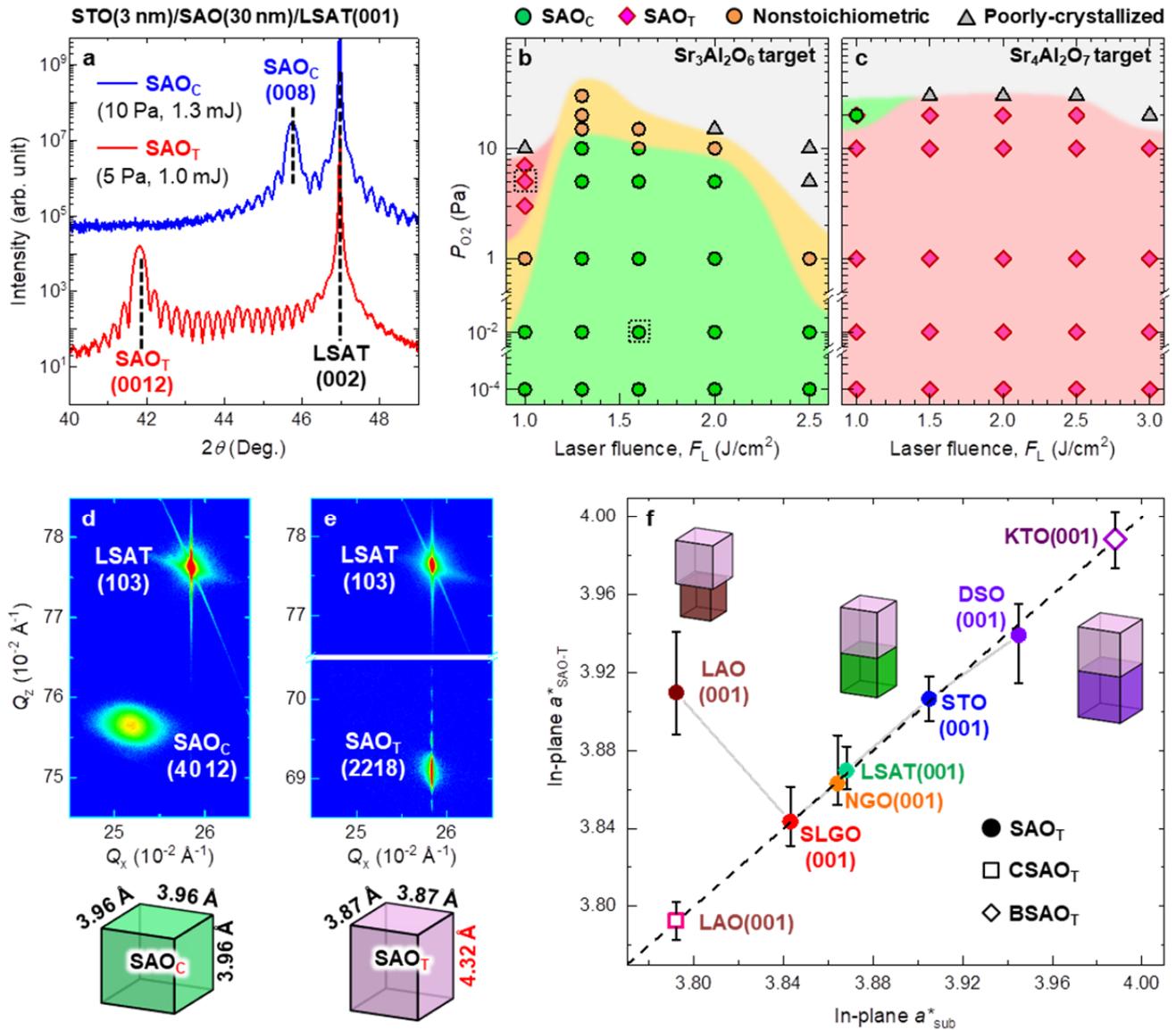

**Fig. 1. Growth of SAO epitaxial films.** **a**, X-ray diffraction (XRD) $2\theta$-$\omega$ linear scans measured from 30 nm thick SAO films grown on (001)-oriented $(LaAlO_3)_{0.3}$-$(SrAl_{0.5}Ta_{0.5}O_3)_{0.7}$ [LSAT (001)] substrates. The cubic and super-tetragonal-like SAO phases are denoted as $SAO_C$ and $SAO_T$, respectively. **b, c**, Laser fluence-oxygen partial pressure ($F_L$-$P_{O2}$) phase diagram of SAO film growth using (**b**) $Sr_3Al_2O_6$ and (**c**) $Sr_4Al_2O_7$ targets. Using a $Sr_3Al_2O_6$ target, $SAO_T$ film can be grown in a narrow $F_L$-$P_{O2}$ window, while film deposition using a $Sr_4Al_2O_7$ target enables a much broader $F_L$-$P_{O2}$ window for growing high-quality $SAO_T$ film. The growth conditions for the samples in (**a**) are marked by dashed boxes. **d, e**, Reciprocal-space mappings (RSMs) from (**d**) $SAO_C$/LSAT(001) and (**e**) $SAO_T$/LSAT(001) film. The lower panels of (**d**) and (**e**) depict schematics of pseudocubic $SAO_C$ and $SAO_T$ unit-cells, with lattice constants labeled. According to the crystal structure analyses shown in **Fig. 2**, the diffractions of $SAO_T$ in (**a**) and (**e**) should be indexed as (0012) and (2218), respectively. **f**, Pseudocubic in-plane lattice constants ($a^*_{SAO-T}$) from the RSMs of 30 nm $SAO_T$ films grown various substrates, including $LaAlO_3$(001) [LAO(001)], $SrLaGaO_4$(001) [SLGO(001)], $NdGaO_3$(001) [NGO(001)], LSAT(001), $SrTiO_3$(001) [STO(001)], $DyScO_3$(001) [DSO(001)]. The $a^*_{SAO-T}$ values are plotted as a function of the in-plane lattice constants ($a^*_{sub}$). All the lattice constants are converted into pseudocubic notation. The $a^*_{SAO-T}$ values for Ba and Ca-doped $SAO_T$ films ($BSAO_T$/$KTaO_3$(001) and $CSAO_T$/LAO(001) films) are also included. The $a^*_{SAO-T}$ values for most of the $SAO_T$ films [except for the $SAO_T$/LAO(001) film] align with the dashed line $a^*_{SAO-T} = a^*_{sub}$, suggesting a fully strain state and broad strain-tuning range of $a^*_{SAO-T}$.

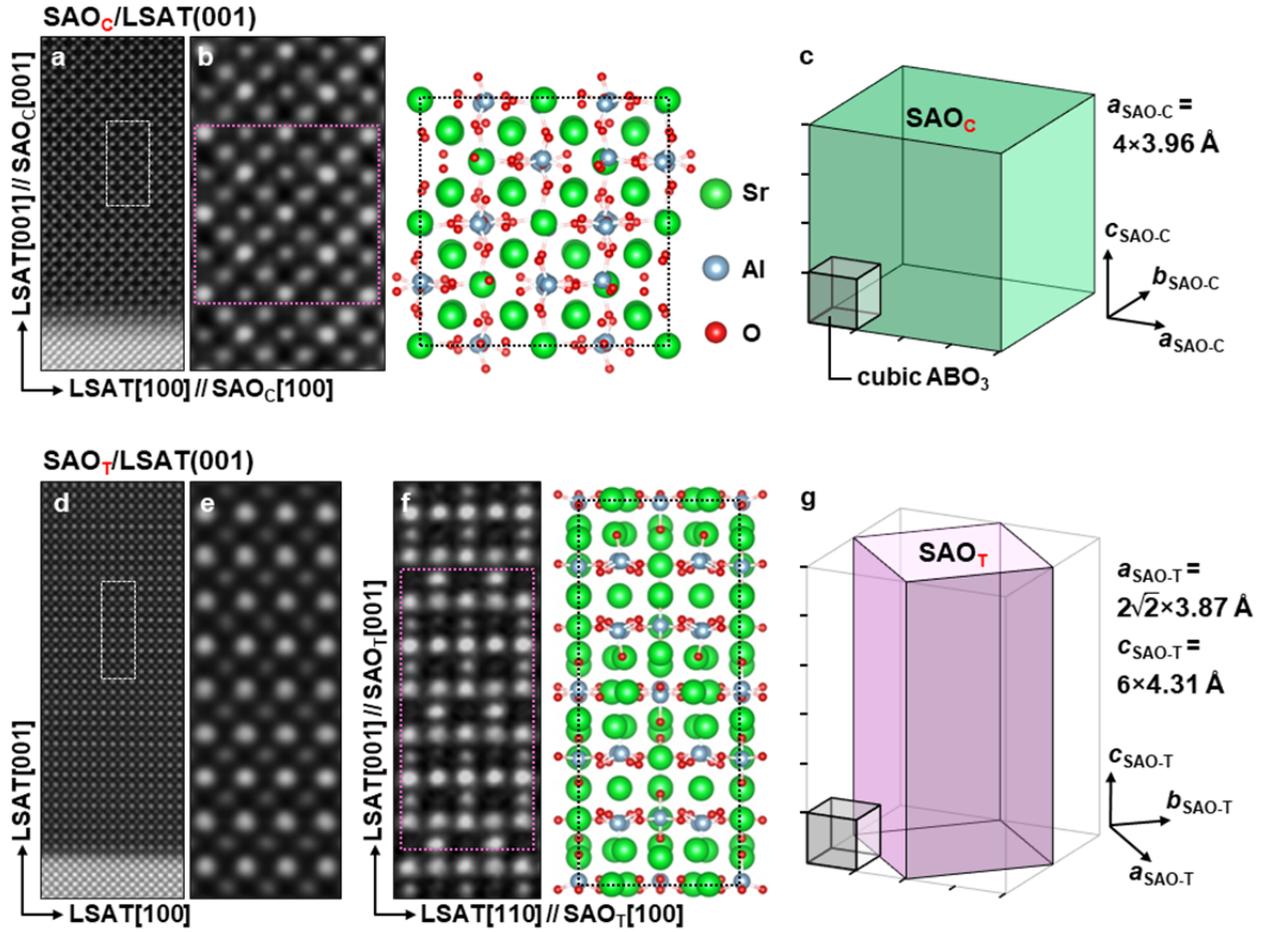

**Fig 2. Crystal structures of SAO$_C$ and SAO$_T$ films. a**, High-angle annular dark field scanning transmission electron microscopy (HAADF-STEM) images captured near the SAO$_C$/LSAT(001) interface, viewed along the LSAT[010] axis. **b**, Zoom-in HAADF-STEM image from the area marked by a white dashed box in (**a**) and DFT-relaxed atomic structure of SAO$_C$ unit-cell. **c**, Schematic illustration of the relative dimensions of the SAO$_C$ unit-cell and a cubic ABO$_3$ perovskite unit-cell. **d**, HAADF-STEM image captured near the SAO$_T$/LSAT(001) interface. **e**, Zoom-in HAADF-STEM image from the area marked by a white dashed box in (d). Both (d) and (e) are viewed along the LSAT[010] zone-axis. **f**, Zoom-in HAADF-STEM image measured from SAO$_T$ film and DFT-relaxed crystal structure, both viewed along SAO$_T$ [100] axis (parallel to LSAT[110]). **g**, Schematic illustration of the relative dimensions of the SAO$_T$ unit-cell and a cubic ABO$_3$ perovskite unit-cell. The lattice constants of SAO$_C$ and SAO$_T$ are labeled in (**c**) and (**g**). For the cubic SAO$_C$ unit-cell, $a_{SAO-C} = b_{SAO-C} = c_{SAO-C}$. And for the tetragonal SAO$_T$ unit-cell, $a_{SAO-T} = b_{SAO-T} \neq c_{SAO-T}$.

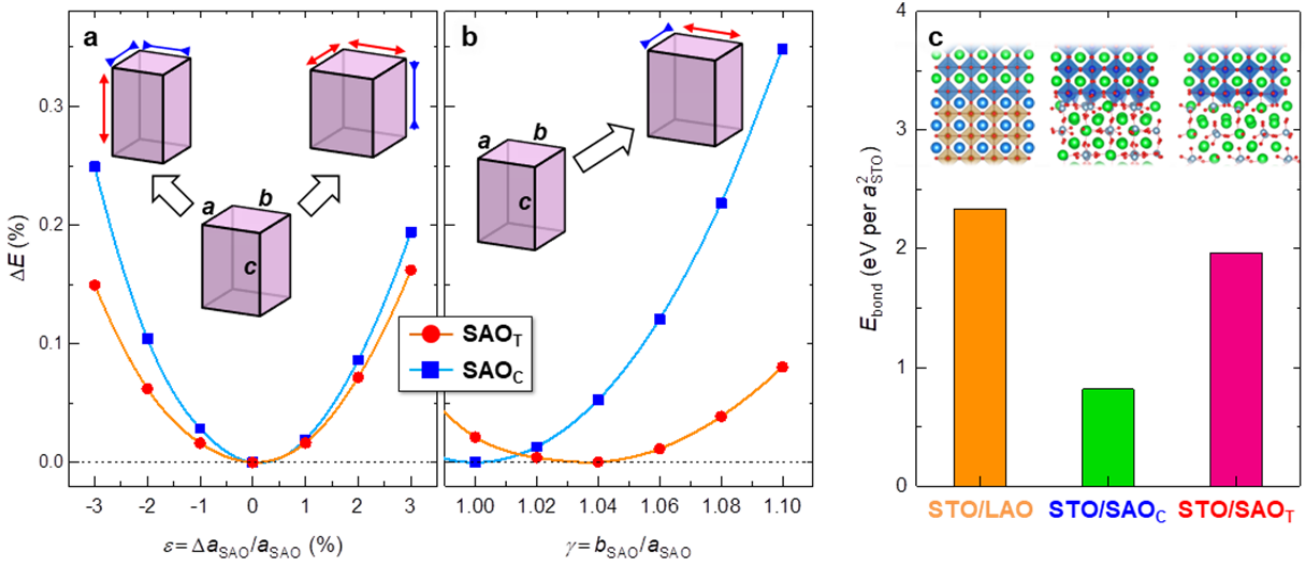

**Fig. 3. Strain-related density-functional-theory (DFT) calculations on SAO$_C$ and SAO$_T$ unit-cells. a**, DFT-calculated energy changes ($\Delta E$) of SAO$_C$ and SAO$_T$ unit-cells under bi-axial strain $\varepsilon$, determined from the relative in-plane lattice constant change ($\Delta a_{SAO}/a_{SAO}$). For direct comparison between SAO$_C$ and SAO$_T$ unit-cells, $\Delta E$ values are normalized by the total energy of the relaxed unit-cells. Both $\Delta E - \varepsilon$ curves display parabolic trends and local minima at $\varepsilon = 0$. Notably, the $\Delta E$ of SAO$_T$ unit-cell is consistently smaller than that of the SAO$_C$ unit-cell, particularly under compressive strain ($\varepsilon < 0$). **b**, DFT-calculated $\Delta E$ of SAO$_C$ and SAO$_T$ unit-cells under anisotropic strain $\gamma$, determined by the orthorhombicity ($b_{SAO}/a_{SAO}$). The $\Delta E - \gamma$ curve of SAO$_C$ unit-cell shows a local minimum at $\gamma = 1$, whereas the curve of SAO$_T$ unit-cell shows a local minimum at $\gamma = 1.04$, consistent with its inherent orthorhombic symmetry. Consistently, the $\Delta E$ of the SAO$_T$ unit-cell shows a much weaker dependence on the anisotropic strain $\gamma$. **c**, DFT-calculated interfacial bonding energy ($E_{bond}$) of the LAO/STO(001), SAO$_C$/STO(001), and SAO$_T$/STO(001) heterostructures. The $E_{bond}$ values are normalized by the number of STO unit-cells bonded at the heterointerface. The insets of (**c**) are schematics of the three interface structures used for DFT calculation.

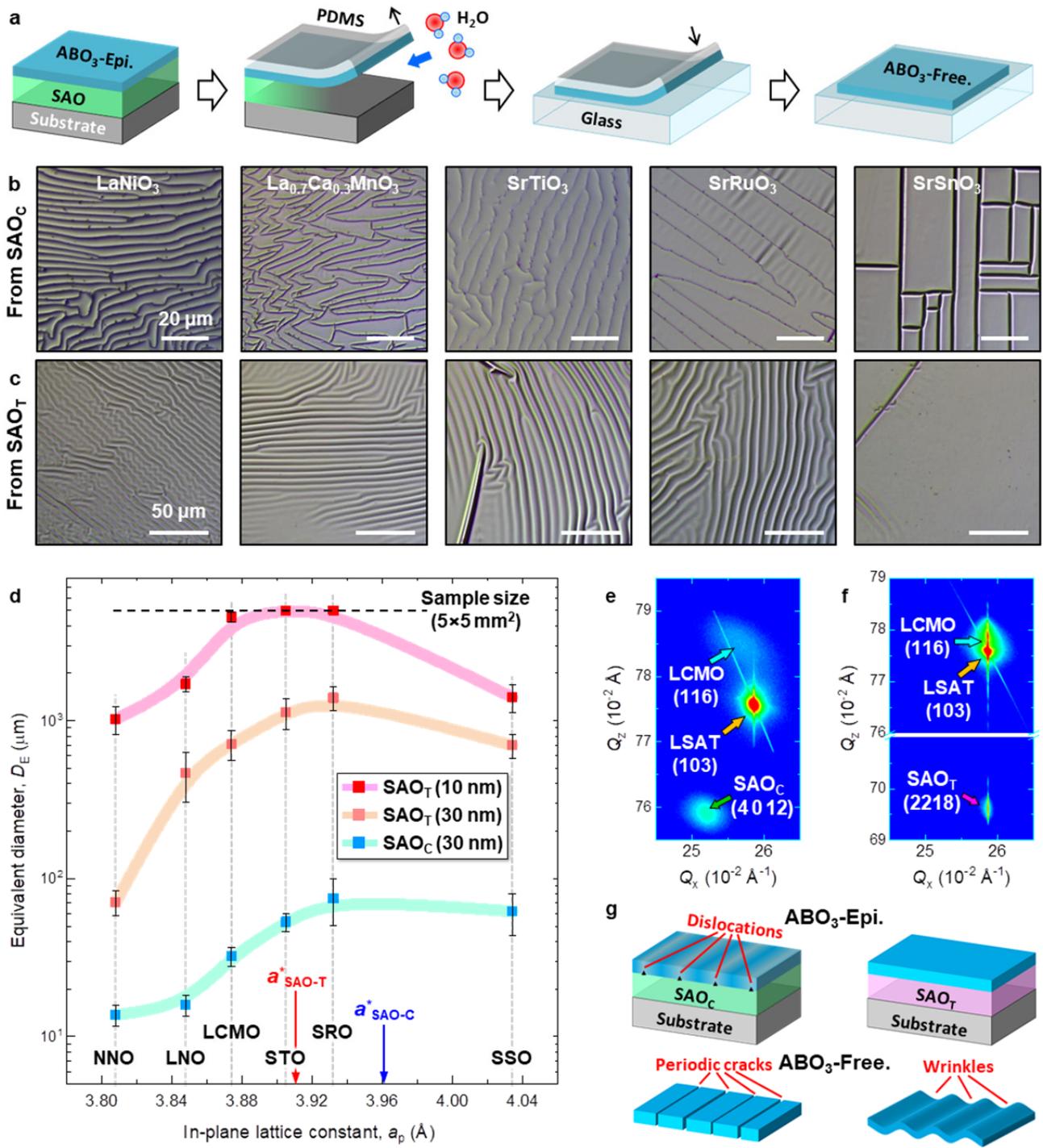

**Fig. 4. Freestanding oxide membranes released from SAO$_C$ and SAO$_T$. a**, Schematic illustration of PDMS and water-assisted exfoliation of freestanding oxide membranes from SAO$_C$ and SAO$_T$ sacrificial layers. **b, c**, Optical microscopic images of 35 nm thick LaNiO$_3$ (LNO), La$_{0.7}$Ca$_{0.3}$MnO$_3$ (LCMO), SrTiO$_3$ (STO), SrRuO$_3$ (SRO), and SrSnO$_3$ (SSO) films peeled from (**b**) SAO$_C$ and (**c**) SAO$_T$ layers. All of the freestanding oxide membranes exfoliated from SAO$_C$ show periodic and high-density cracks. By contrast, the membranes exfoliated from SAO$_T$ show large-scale crack-free but wrinkled morphology. **d**, Summary of averaged equivalent diameter $D_E = A_{free}^{1/2}$, where the $A_{free}$ is the uncracked area of the freestanding membranes. The averaged $D_E$ is plotted as a function of the in-plane lattice constant in pseudocubic perovskite notation ($a_p$), and the error bars represent the standard deviations of $D_E$. The $D_E$ values for NdNiO$_3$ (NNO) membranes are also included in (**d**). The in-plane lattice constants of bulk SAO$_T$ and SAO$_C$ in pseudocubic perovskite notations ($a^*_{SAO-T}$ and $a^*_{SAO-C}$) are marked in (**d**). The LCMO, STO, and SRO membranes released from 10 nm SAO$_T$ are almost crack-free. Thus the $D_E$ value approaches the sample size (5 mm, marked by a horizontal dashed line). **e, f**, RSMs around LCMO(116) diffractions measured from LCMO epitaxial film grown on (**e**) $SAO_C$/LSAT(001) and (**f**) $SAO_T$/LSAT(001). In (**e**), the in-plane reciprocal space vector ($Q_x$) value of LCMO(116) spans between the corresponding ones of LSAT and $SAO_C$, indicating that the LCMO/$SAO_C$ bilayer undergoes a partial strain-relaxation. In (**f**), the $Q_x$ values of LCMO(116) and $SAO_T$(4018) align with the one of LSAT(103) substrate, indicating that the LCMO/$SAO_T$ bilayer is coherently strained to the LSAT(001) substrate. **g**, Schematics elucidating the crack and wrinkle formations in freestanding oxide membranes. The perovskite oxide epitaxial films and freestanding membranes are marked as $ABO_3$-Epi. and $ABO_3$-Free., respectively.

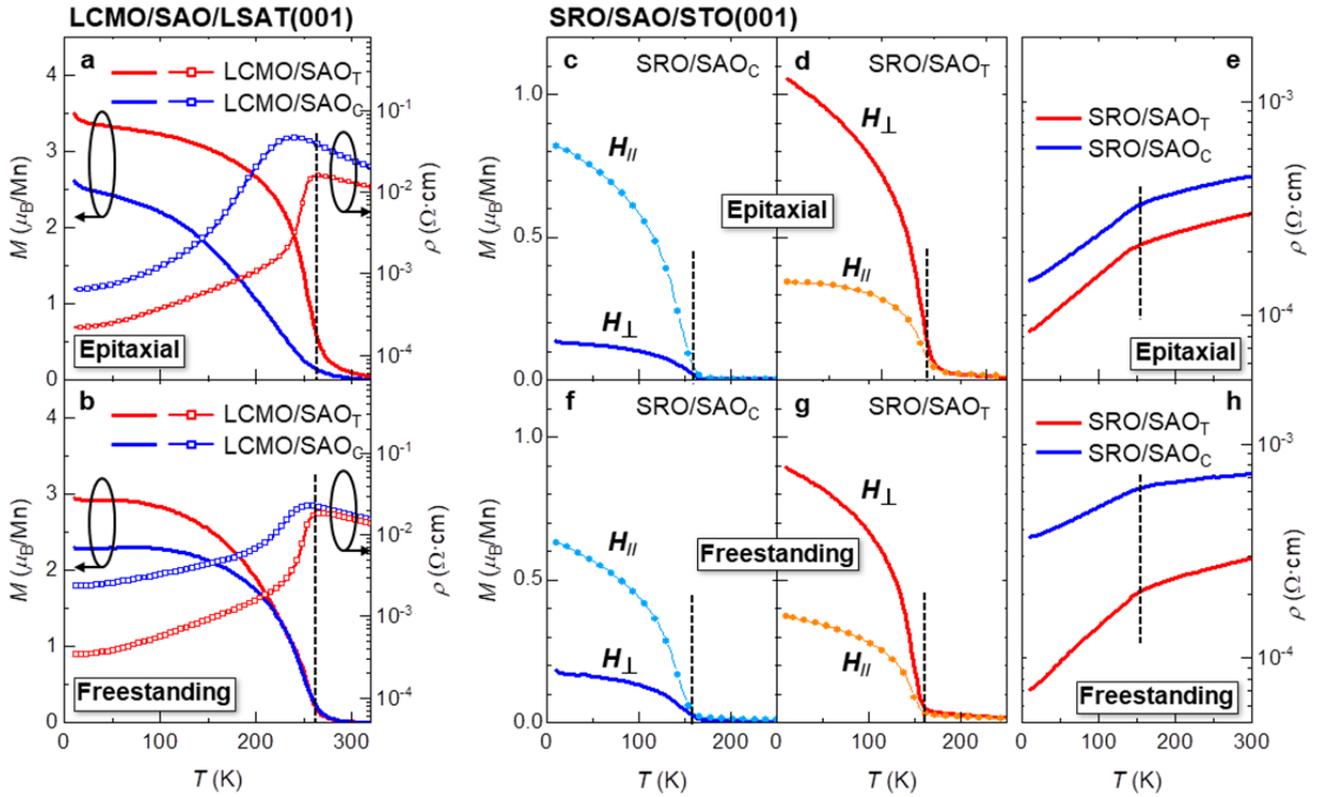

**Fig 5. Physical properties of LCMO and SRO epitaxial films/freestanding membranes. a**, Temperature-dependent resistivity ($\rho$-$T$) and magnetization ($M$-$T$) curves measured from the LCMO(35 nm)/SAO$_C$/LSAT(001) and LCMO(35 nm)/SAO$_T$/LSAT(001) epitaxial films. **b**, $\rho$-$T$ and $M$-$T$ curves measured from the freestanding LCMO membranes exfoliated from SAO$_C$ and SAO$_T$. The $M$-$T$ curves were measured with an in-plane magnetic field $\mu_0 H$ = 1000 Oe. **c-e**, $M$-$T$ (**c** and **d**) and $\rho$-$T$ (**e**) curves measured from the SRO(35 nm)/SAO$_C$/STO(001) and SRO(35 nm)/SAO$_T$/STO(001) epitaxial films. **f-h**, $M$-$T$ (**f** and **g**) and $\rho$-$T$ (**h**) curves measured from the SRO freestanding membranes released from SAO$_C$ and SAO$_T$. All the $M$-$T$ curves were measured with a magnetic field $\mu_0 H$ = 1000 Oe, applied both parallel ($H_{//}$) or perpendicular ($H_\perp$) to the film plane. The dashed lines are guidelines for the Curie temperature of LCMO and SRO.

Supplementary Information for

# Super-tetragonal $Sr_4Al_2O_7$: a versatile sacrificial layer for high-integrity freestanding oxide membranes

Jinfeng Zhang[1,9], Ting Lin[2,9], Ao Wang[1,9], Xiaochao Wang[3,9], Qingyu He[4], Huan Ye[1], Jingdi Lu[1], Qing Wang[1], Zhengguo Liang[1], Feng Jin[1], Shengru Chen[2], Minghui Fan[1], Er-Jia Guo[2], Qinghua Zhang[2], Lin Gu[5], Zhenlin Luo[4], Liang Si[3,6]*, Wenbin Wu[1,7,8]*, and Lingfei Wang[1]*

[1]Hefei National Research Center for Physical Sciences at Microscale, University of Science and Technology of China, Hefei 230026, China

[2]Beijing National Laboratory for Condensed Matter Physics, Institute of Physics, Chinese Academy of Sciences, Beijing 100190, China

[3]School of Physics, Northwest University, Xi'an 710127, China

[4]National Synchrotron Radiation Laboratory, University of Science and Technology of China, Hefei 230026, China.

[5]Beijing National Center for Electron Microscopy and Laboratory of Advanced Materials, Department of Materials Science and Engineering, Tsinghua University, Beijing 100084, China.

[6]Institut für Festkörperphysik, TU Wien, 1040 Vienna, Austria.

[7]Institutes of Physical Science and Information Technology, Anhui University, Hefei 230601, China.

[8]Collaborative Innovation Center of Advanced Microstructures, Nanjing University, Nanjing 210093, China

[9]These authors contributed equally: Jinfeng Zhang, Ting Lin, Ao Wang, and Xiaochao Wang.

*Email: liang.si@ifp.tuwien.ac.at; wuwb@ustc.edu.cn; wanglf@ustc.edu.cn

# Contents

# Section 1: Growth windows of SAO$_C$ and SAO$_T$ epitaxial films

In this Section, we first focus on determining the PLD growth window of SAO$_C$ and SAO$_T$ films. We then discuss the underlying mechanisms of the wide (narrow) growth window for the SAO$_C$ (SAO$_T$) films. In the last part, we will also evaluate the ability of SAO$_T$ films to sustain epitaxial strain.

## 1.1 Growth window of SAO films

We first investigate the PLD growth window of SAO films using Sr$_3$Al$_2$O$_6$ target. As mentioned in the Materials and Methods section, we grew a series of SAO films under various $P_{O2}$ ranging from 10$^{-4}$ to 30 Pa and various $F_L$ ranging from 1.0 to 2.5 J·cm$^{-2}$. Here, the STO capping layer of 3 nm is utilized to protect SAO film from moisture in the ambient for subsequent characterization. The full set of XRD 2$\theta$-$\omega$ linear scans is included in **Supplementary Fig. 1**. The *d*-spacing values of SAO$_C$(008) or SAO$_T$(0012) peaks in the 2$\theta$-$\omega$ linear scans are summarized in **Supplementary Fig. 2**. Deviations of the *d*-spacing from the bulk value commonly imply an off-stoichiometry in the films. The intensities of SAO$_C$(008) or SAO$_T$(0012) peaks in the 2$\theta$-$\omega$ linear scans are summarized in **Supplementary Fig. 3**. The peak intensity can reflect the crystallinity of the SAO films. Given a fixed $F_L \geq 2.0$ J/cm$^2$, the 2$\theta$-$\omega$ scan of the SAO$_C$/LSAT(001) film grown at $P_{O2}$ = 10$^{-4}$ Pa shows a strong SAO$_C$(008) peak at ~45.8° and sharp Laue fringes, implying a high epitaxial quality. As $P_{O2}$ increases to 1 Pa, the SAO$_C$(008) peak position shifts slightly towards a higher Bragg angle, which may originate from off-stoichiometry during film growth. Further increasing $P_{O2}$ up to 5 and 10 Pa, a considerate degradation of the film crystallinity occurs. By reducing the $F_L$ to 1.6 and 1.3 J/cm$^2$, the 2$\theta$-$\omega$ scans show similar but slower degradations of SAO film quality as $P_{O2}$ increases. At $F_L$ = 1.3 J/cm$^2$, high epitaxial quality persists even at $P_{O2}$ = 10 Pa. In a word, lowering the laser energy density could effectively broaden the $P_{O2}$ range for SAO$_C$ growth. Further lowering $F_L$ to 1.0 J/cm$^2$, at $P_{O2}$ = 5 Pa, the SAO$_C$ peak at ~45.8° disappears, and another strong peak appears in the 2$\theta$-$\omega$ scan at ~41.8°. According to the results shown in **Supplementary Section 2**, this peak can be attributed to the SAO$_T$(0012) peak. To precisely determine the growth window, we fixed the $F_L$ = 1.0 J/cm$^2$ and further scanned the $P_{O2}$ from 1 to 10 Pa. As shown in **Supplementary Fig. 4**, the high-quality SAO$_T$ epitaxial film can be stabilized in a rather narrow $P_{O2}$ window from 3 to 7 Pa.

According to the constructed SAO growth phase diagram, we selected three representative growth conditions to grow SAO epitaxial film samples (Sample **A**: 1.3 J·cm$^{-2}$, 10 Pa; Sample **B**: 2.5 J·cm$^{-2}$, 10$^{-4}$ Pa; and Sample **C**: 1.0 J·cm$^{-2}$, 5 Pa.) and further characterized their surface topography and crystallinity. As shown in **Supplementary Fig. 5**, the root-mean-square roughnesses derived from the AFM images of all three films are below 0.2 nm, and the full-width-of-half-maximum (FWHM) of the $\omega$-scan rocking curves around SAO$_C$(008) or SAO$_T$(0012) diffractions are below 0.02°, even comparable with the high-quality perovskite oxide films grown by PLD. The $\omega$-scan rocking curves of the SAO$_C$(008) diffraction (Samples **A** and **B**, shown in **Supplementary Fig. 5e** and **5f**) display broad shoulders underneath the sharp SAO$_C$(008) peak, which should correspond to strain relaxation-induced diffused scattering. In contrast, the $\omega$-scan rocking curve of the SAO$_T$(0012) diffraction (Samples **C** and **D**, shown in **Supplementary Fig. 5g** and **5h**) displays a sharp peak only, which suggests that the strain relaxation in SAO$_T$/LSAT(001) film is negligible. Based on these results, it is safe to say that the SAO$_T$ film exhibits a comparable or even better epitaxial quality than the SAO$_C$ film.

To further clarify the differences in PLD growth windows between SAO$_C$ and SAO$_T$ phases, we experimentally determined the chemical stoichiometry of the SAO$_C$ and SAO$_T$ phases by energy-dispersive X-ray spectroscopy (EDS) with an Oxford Instruments EDS (Ultim Max65) spectrometer, equipped with a field-emission scanning electron microscope (Zeiss Gemini 450). We also performed EDS measurements on SAO$_C$ as a reference. The typical EDS spectra are shown in **Supplementary Fig. 6**, and the measured molar ratios Sr:Al at different regions are summarized in **Supplementary Table 1**. In the SAO$_C$ (Sr$_3$Al$_2$O$_6$) polycrystalline target and epitaxial films, the molar ratio Sr:Al = 1.51±0.02, close to the nominal value of Sr$_3$Al$_2$O$_6$. However, in SAO$_T$ films, the molar ratio of Sr:Al = 2.05±0.03, close to the nominal value of Sr$_4$Al$_2$O$_7$ rather than Sr$_3$Al$_2$O$_6$. The large deviation in chemical stoichiometry between the SAO$_T$ film (Sr:Al ~ 2) and SAO$_C$ target (Sr:Al ~ 1.5) should be related to the narrow growth window (low $L_F$ and moderate $P_{O2}$). The detailed analyses and discussion are included below.

Since the SAO$_T$ phase turns out to be a Sr$_4$Al$_2$O$_7$ compound, replacing the Sr$_3$Al$_2$O$_6$ target with a Sr$_4$Al$_2$O$_7$ target may broaden the PLD growth window of high-quality SAO$_T$ films. Based on this consideration, we grew a series of SAO films by laser ablating a Sr$_4$Al$_2$O$_7$ target. During the growth, we also varied the $P_{O2}$ ranging from $10^{-4}$ to 30 Pa and $F_L$ from 1.0 to 3.0 J·cm$^{-2}$. Here, the STO capping layer of 3 nm is utilized to protect SAO film from moisture in the ambient for subsequent characterization. The full set of XRD $2\theta$-$\omega$ linear scans is included in **Supplementary Fig. 7**. Most of the $2\theta$-$\omega$ scans show a strong SAO$_T$(0012) peak at ~42.0° and sharp Laue fringes, implying a broad growth window of high-quality SAO$_T$ films. Such a wide growth window further makes SAO$_T$ an experimentally feasible water-soluble sacrificial layer.

**1.2 Phenomenological picture for the growth window of SAO$_C$ and SAO$_T$ films**

According to **Fig. 1** in the main text and **Supplementary Figs. 1-7**, we can conclude that the growth windows for high-quality SAO$_T$ and SAO$_C$ films are strongly dependent on the $F_L$ and $P_{O2}$. Furthermore, the energy-dispersive spectroscopy results shown in **Supplementary Fig. 6** demonstrate that the SAO$_T$ phase turns out to be a Sr$_4$Al$_2$O$_7$ compound. In other words, the molar ratio of Sr/Al increases from ~1.5 in the SAO$_C$ phase to ~2.0 in the SAO$_T$ phase.

Intuitively, PLD can effectively transfer the cation stoichiometry from the target to the film. Thus, laser ablating the $Sr_3Al_2O_6$ and $Sr_4Al_2O_7$ targets should result in the $SAO_C$ and $SAO_T$ films, respectively. Following this scenario, it is elusive why the $SAO_T$ film can be grown using the $Sr_3Al_2O_6$ target in a narrow growth window. In the following part, we will provide a phenomenological picture to explain the large deviation in chemical stoichiometry between the $SAO_T$ film and $SAO_C$ target (Sr:Al ~ 1.5).

During the PLD process, the photon energy from the excimer laser is absorbed by the target within a shallow surface layer, resulting in rapid heating up to several thousand Kelvins and subsequent material removal from the target. The kinetic energy of the ablated atoms/ions is in the range of tens of eV, and the velocity is in the range of $10^5$~$10^6$ cm/s. As these ablated specimens travel from the target to the substrate surface, they are scattered by the $O_2$ molecules in the PLD chamber. Upon reaching the substrate surface, a supersaturation-dominated vapor-solid phase transition occurs, leading to film growth[1]. According to earlier studies[2–8], the key parameters, $F_L$ and $P_{O2}$, can influence film quality and chemical stoichiometry at each stage. Based on these works, we attempt to elucidate the growth phase diagrams of $SAO_C$ and $SAO_T$ films using a phenomenological model.

We can categorize the SAO growth phase diagram into two regions based on $F_L$. In the first region, where $F_L \geq 1.3$ J/cm$^2$, the phase diagram displays the $SAO_C$ phase only. In this region, the $SAO_C$ film can grow well for $P_{O2} \leq 10^{-2}$ Pa, while the $SAO_C$(008) peak position shifts (signature of off-stoichiometry) and peak intensity decreases (signature of film quality degradation) occur as $P_{O2}$ increases further. The film quality is sensitive to the aforementioned vapor-solid transition at the film surface[2]. When the substrate temperature is constant, supersaturation increases with $P_{O2}$ and $F_L$. Namely, higher $P_{O2}$ and $F_L$ values result in a larger chemical potential difference of adatoms transitioning from their quasi-vapor phase (mobile adatoms on the surface and background oxygen in the gas phase) to their solid phase on the substrate[2]. Such a large difference in chemical potential is responsible for the degradation of the $SAO_C$ film's crystallinity. The chemical stoichiometry of $SAO_C$ film is governed by two physical processes. The first one is the scattering process of laser-ablated specimens by $O_2$ molecules[3,4]. Given a low $P_{O2} \leq 10^{-2}$ Pa, the mean free paths of both Sr and Al atoms are comparable or even larger than the substrate-target distance[5]. As a result, the scattering probability is rather low, allowing for the efficient transfer of chemical composition from the target to the film surface. When $P_{O2}$ exceeds 1 Pa, the mean free path of both Sr and Al atoms decreases significantly to the millimeter scale, leading to substantial scattering by $O_2$ molecules. In this case, the relative amount of atoms arriving at the film surface is dominated by the atom mass ($m_{Sr}$ or $m_{Al}$). Since $m_{Sr} > m_{Al}$, the film is expected to become Sr rich. The second process is resputtering at the target surface, which also contributes to the formation of a Sr-rich film[6,7]. For a given pulse duration, lower $F_L$ corresponds to a lower heating temperature at the target surface, reducing the kinetic energy of ablated atoms/ions and hindering the resputtering process. The combination of these two scenarios can well explain the experimentally observed trend: the $P_{O2}$ boundary for non-stoichiometric $SAO_C$ film shifts to higher values as $F_L$ decreases.

In the second region of the phase diagram, where $F_L \leq 1.0$ J/cm, the $SAO_T$ phase appears in the $P_{O2}$ range from 3 to 7 Pa. In this low $F_L$ region, the scattering-mediated Sr-rich atomic composition persists. Additionally, according to Dam et al.[8], when $F_L$ decreases below a certain threshold value, volume-diffusion-assisted preferential evaporation of Sr occurs at the target surface. In the SAO case, we speculate that the threshold value of $F_L$ could be close to 1.0 J/cm$^2$. Therefore, reducing $F_L$ down to 1.0 J/cm$^2$ will induce a significant Sr-rich composition, thereby stabilizing the $SAO_T$ phase. Moreover, the reduced $F_L$ also decreases the kinetic energy of adatoms and the film growth rate[7,8]. The resultant higher epitaxial quality may also help stabilize the $SAO_T$ film, which has a smaller lattice mismatch with LSAT(001) substrate. It is important to note that increasing $P_{O2}$ beyond 7 Pa can also enhance the supersaturation and degrade the film quality.

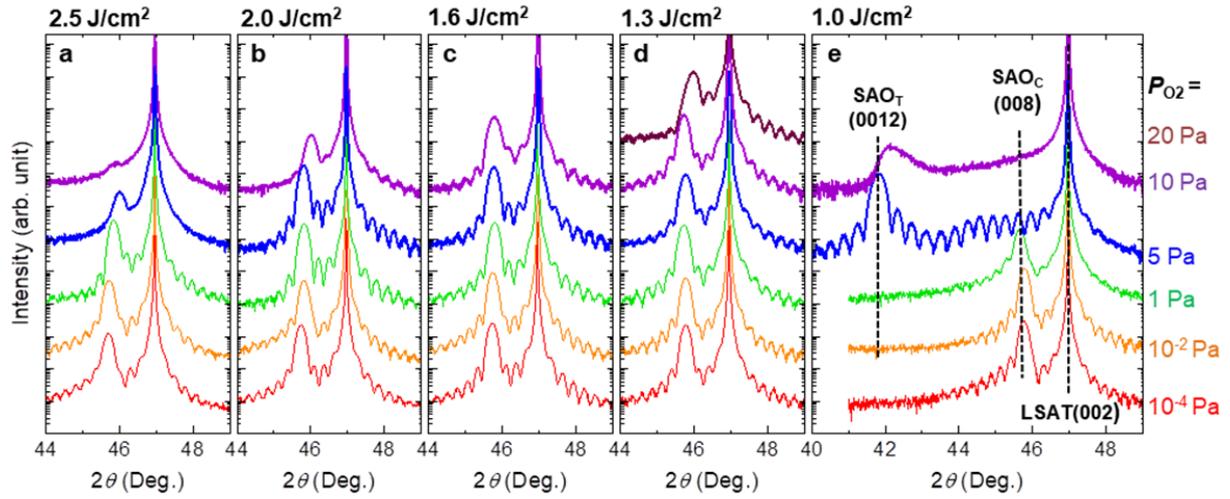

**Supplementary Fig. 1. | XRD 2$\theta$-$\omega$ linear scans measured from 35 nm thick SAO films.** All the films are deposited through laser ablation from a polycrystalline $Sr_3Al_2O_6$ target. For the growth of five sets of films, we varied the $P_{O2}$ and fixed the $F_L$ value at 2.5 (**a**), 2.0 (**b**), 1.6 (**c**), 1.3 (**d**), and 1.0 J/cm$^2$ (**e**). All samples are capped with a 3 nm STO protection layer.

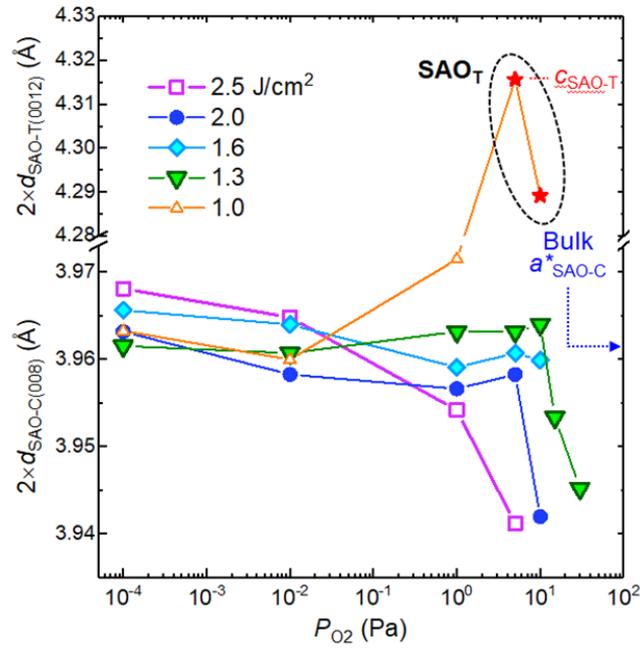

**Supplementary Fig. 2. | $P_{O2}$ and $F_L$-dependent $d$-spacing values of $SAO_C(008)$ and $SAO_T(0012)$ diffractions.** The $d$-spacing values [$d_{SAO-T(0012)}$ and $d_{SAO-C(008)}$] are derived from the XRD $2\theta$-$\omega$ scans shown in **Supplementary Fig. 1**. To directly compare with the lattice constant in pseudocubic perovskite unit-cell, both the $d_{SAO-T(0012)}$ and $d_{SAO-C(008)}$ values are multiplied by a factor of 2.

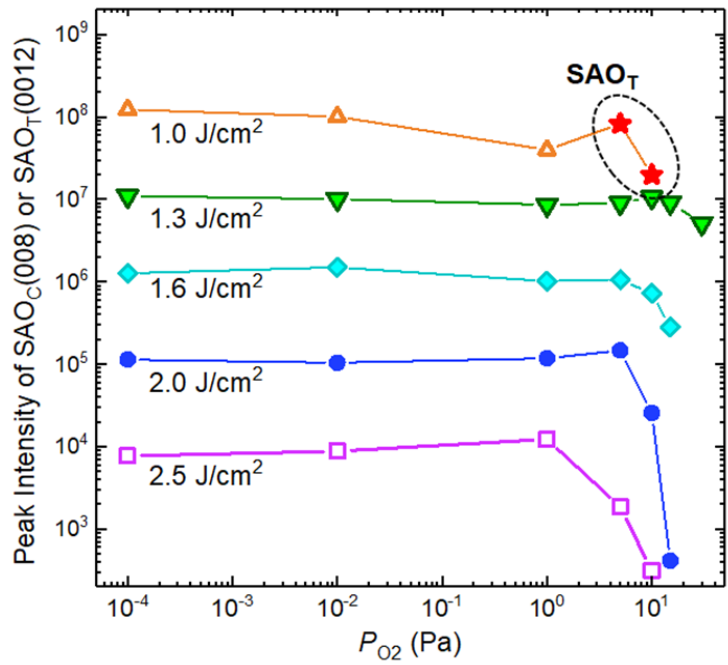

**Supplementary Fig. 3. | $P_{O2}$ and $F_L$-dependent peak intensities of the $SAO_C$(008) and $SAO_T$(0012) diffractions.** These values are also derived from the XRD $2\theta$-$\omega$ scans shown in **Supplementary Fig. 1**.

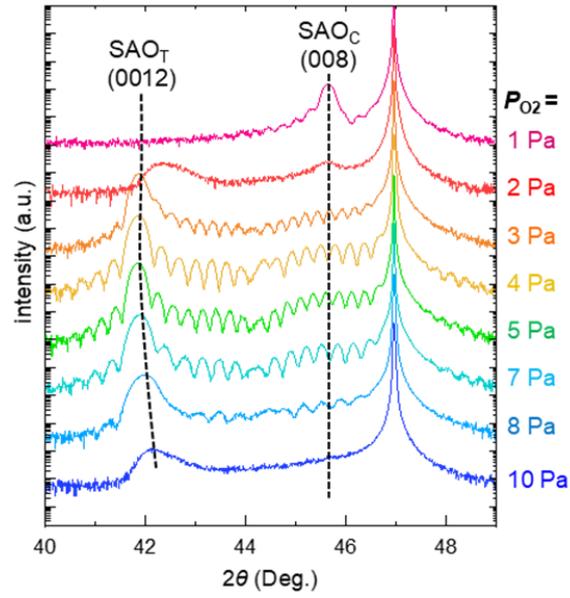

**Supplementary Fig. 4. | XRD 2θ-ω linear scans of STO (3 nm)/SAO$_T$ (30 nm)/LSAT(001) films grown at a fixed $F_L$ = 1.0 J/cm$^2$ at various $P_{O2}$.** All the films are deposited through laser ablation from a polycrystalline Sr$_3$Al$_2$O$_6$ target.

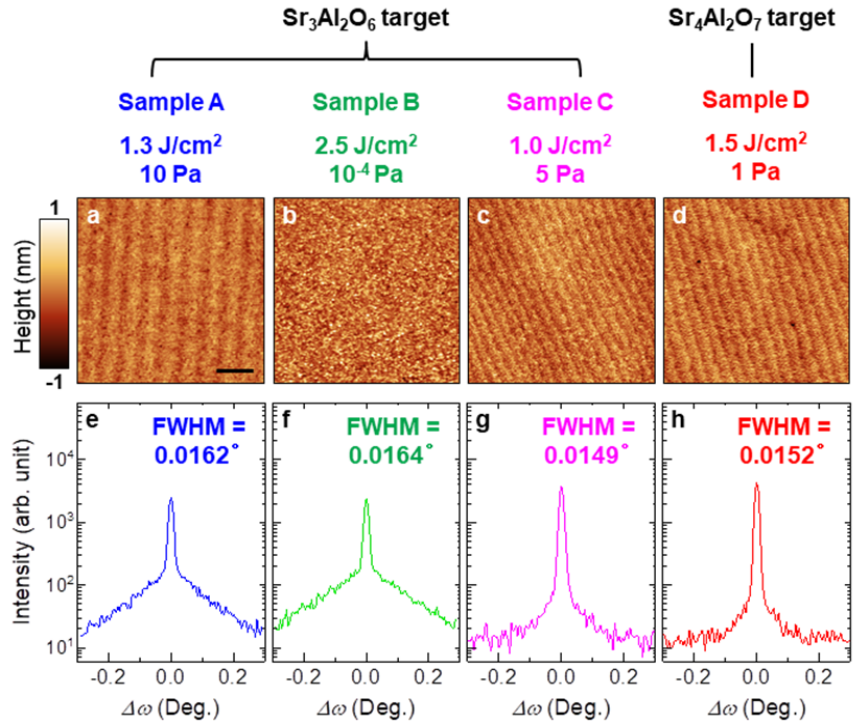

**Supplementary Fig. 5. | Characterizations on the epitaxial quality of four SAO/LSAT(001) film samples grown under different conditions.** Sample A (2.5 J·cm$^{-2}$, 10$^{-4}$ Pa), Sample B (1.0 J·cm$^{-2}$, 10 Pa), and Sample C (1.0 J·cm$^{-2}$, 5 Pa.) are deposited using a Sr$_3$Al$_2$O$_6$ target. Sample D (1.5 J·cm$^{-2}$, 1 Pa.) is deposited using a Sr$_4$Al$_2$O$_7$ target. Determined by the conditions, Sample A and B are SAO$_C$ films while Sample C and D are SAO$_T$ films. (**a-d**) Atomic force microscopy (AFM) topographic images of the four SAO samples. (**e,f**) XRD $\omega$-scan rocking curves of the SAO$_C$(008) diffractions. The curves display broad shoulders underneath the sharp SAO$_C$(008) peaks, corresponding to strain relaxation-induced diffused scattering. (**g,h**) XRD $\omega$-scan rocking curves of the SAO$_T$(0012) diffractions. The curves display sharp peaks only, which suggests that the strain relaxation in SAO$_T$/LSAT(001) film is negligible.

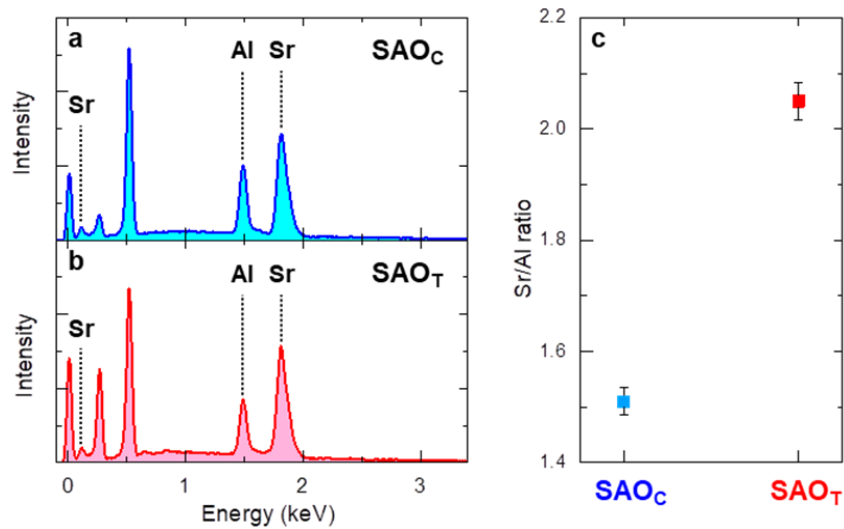

**Supplementary Fig. 6. | Cation stoichiometries of SAO$_C$ and SAO$_T$ films.** (**a,b**) Representative energy-dispersive X-ray spectroscopy EDS curves measured from the 100 nm thick SAO$_T$ (a) and SAO$_C$ (b) films grown on NdGaO$_3$(001) substrate. (**c**) The averaged Sr/Al molar ratios of SAO$_C$ and SAO$_T$ compounds, derived from **Supplementary Table 1**. The SAO$_C$ phase is close to the stoichiometric Sr$_3$Al$_2$O$_6$, while the SAO$_T$ phase turns out to be a Sr$_4$Al$_2$O$_7$ compound.

| SAO$_T$ | Sr | Al | Sr/Al |
| --- | --- | --- | --- |
| 1 | 32.3 | 15.8 | 2.04 |
| 2 | 32.3 | 15.6 | 2.07 |
| 3 | 32.1 | 15.5 | 2.07 |
| 4 | 32.0 | 15.4 | 2.08 |
| 5 | 31.8 | 15.9 | 2.00 |
| **Mean value (standard deviation)** | 32.1 (0.21) | 15.64 (0.21) | 2.05 (0.03) |

| SAO$_C$ | Sr | Al | Sr/Al |
| --- | --- | --- | --- |
| 1 | 29.2 | 19.2 | 1.52 |
| 2 | 29.1 | 19.3 | 1.51 |
| 3 | 28.8 | 19.2 | 1.50 |
| 4 | 29.0 | 19.5 | 1.48 |
| 5 | 29.4 | 19.3 | 1.52 |
| **Mean value (standard deviation)** | 29.1 (0.22) | 19.3 (0.22) | 1.51 (0.02) |

**Supplementary Table 1. | Molar ratios of Sr and Al in 100 nm thick SAO$_C$/NGO(001) and SAO$_T$/NGO(001) films, derived from energy-dispersive spectroscopy (EDS) results.** We choose to grow the thick films on NGO(001) substrate to exclude the influence of Sr and Al elements in the substrate [e.g. STO(001) or LSAT(001)] on the EDS results. To further ensure the reliability of EDS results, we repeated the measurements at five different regions in both samples.

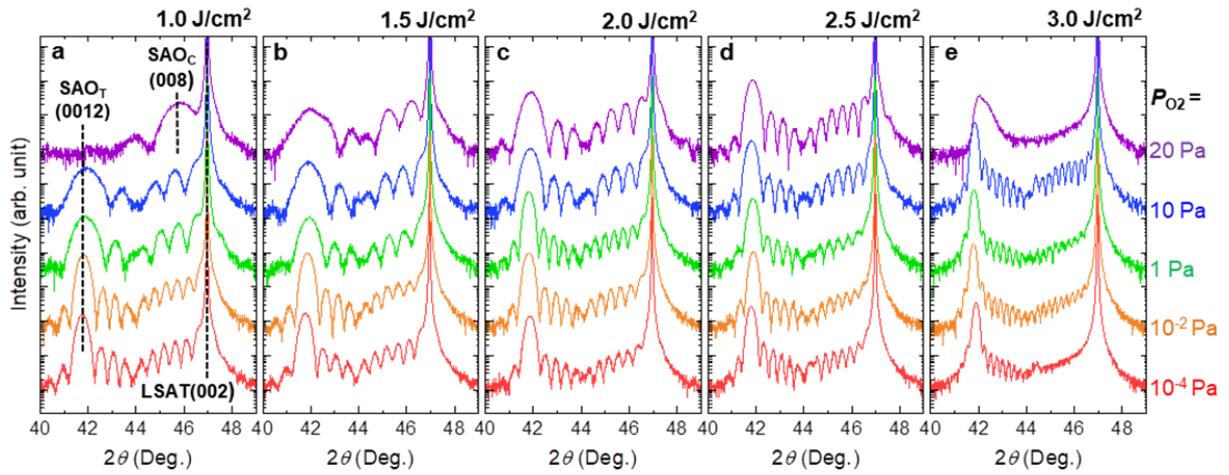

**Supplementary Fig. 7. | XRD 2θ-ω linear scans measured from a series of SAO films, deposited through laser ablation of a polycrystalline $Sr_4Al_2O_7$ target.** For the growth of the five sets of films, we varied the $P_{O2}$ and fixed the $F_L$ value at 1.0 (**a**), 1.5 (**b**), 2.0 (**c**), 2.5 (**d**), and 3.0 J/cm² (**e**). All samples are capped with a 3 nm STO protection layer.

# Section 2. Crystal structure of SAO$_T$ epitaxial films

In this section, we focus on probing the crystal structure of the SAO$_T$ film. We first determined the structural symmetry of the SAO$_T$ phase and then investigated its epitaxial strain states by structural characterizations. Next, we probe its atomic structure by detailed analyses of the scanning transmission electron microscopy (STEM) results. Lastly, we performed additional first-principles calculations to determine the crystal structure of the SAO$_T$ phase and further understand the capability of SAO$_T$ in sustaining epitaxial strain

## 2.1 Structure symmetry of the SAO$_T$ phase

We first measured the reciprocal space maps (RSMs) near the four LSAT{103} diffractions to investigate the structural symmetry of the SAO$_T$ phases (**Supplementary Fig. 8**). All of the SAO$_T${2218} diffractions show the same out-of-plane reciprocal vectors ($Q_z$) and share the same in-plane reciprocal vector ($Q_x$) with that of the LSAT{103} diffractions. These results signify a tetragonal symmetry for the SAO$_T$ lattice. We also performed second harmonic generation (SHG) characterizations on the structural symmetry. As shown in **Supplementary Fig. 9**, the experimental polar plots from SAO$_T$/LSAT(001) film can be well fitted using the tetragonal *4mm* point group, which is consistent with the RSMs results shown in **Supplementary Fig. 8**.

By taking a close look at the RSMs in **Supplementary Fig. 8**, we found the SAO$_T$(2218)/($\bar{2}\bar{2}$18) diffraction spots, compared to the ($\bar{2}$218)/(2$\bar{2}$18) diffraction spots, are slightly longer along the $Q_z$ axis and more diffused along the $Q_x$/$Q_y$ axis. This subtle difference implies that the structural symmetry of the SAO$_T$ phase could be even lower (i.e. in orthorhombic symmetry) in the atomic scale. This point can be further verified by the RSMs measured from SAO$_T$/NGO(001) film (**Supplementary Fig. 10**), in which the SAO$_T$ shows the same orthorhombic symmetry as the NGO(001) substrate.

## 2.2 Epitaxial strain states of SAO$_T$ films

The ability to sustain epitaxial strain is one of the most prominent features of the SAO$_T$ phase. For the SAO$_T$ films grown on LSAT(001), the strain coherency can be maintained up to 100 nm thick (**Supplementary Fig. 11**). The resultant structural coherency further ensures a sharp and defect-free SAO$_T$/LSAT(001) interface (**Supplementary Fig. 12**). To further attest this point, we grew the SAO$_T$ films under the optimized growth condition onto a variety of perovskite substrates, including LaAlO$_3$(001) [LAO(001)], SrLaGaO$_4$(001) [SLGO(001)], NdGaO$_3$(001) [NGO(001)], LSAT(001), STO(001), DyScO$_3$(001) [DSO(001)], and KTaO$_3$(001) [KTO(001)]. The structure characterizations of these films are shown in **Supplementary Fig. 13**. According to the strong SAO$_T$(2218) peak intensities in the XRD 2$\theta$-$\omega$ linear scans (**Supplementary Fig. 13a)**, the SAO$_T$ epitaxial films can grow well on most of the substrates except for KTO(001). For this substrate with a large lattice constant of up to 3.988 Å, the SAO$_T$(2218) peak disappears and the SAO$_C$(008) peak appears. Namely, the KTO(001)-imposed tensile strain can destabilize the SAO$_T$ phase and stabilize the SAO$_C$ phase. On this basis, we suggest that the substrate-imposed compressive strain also plays a critical role in stabilizing the SAO$_T$ phase.

The epitaxial strain states of SAO$_T$ films grown on various substrates are further characterized by reciprocal space mappings (RSMs). RSMs near SAO$_T$(2218) diffractions are shown in **Supplementary Fig. 13b-g**. As summarized in **Fig. 1f** in the main text, the in-plane lattice constants derived from RSMs (in pseudocubic notation, $a^*_{SAO-T}$) are plotted as a function of in-plane lattice constants of substrates ($a^*_{sub}$, in pseudocubic notation). As shown in **Supplementary Fig. 13c-g**, the SAO$_T$ films grown on SLGO(001), NGO(001), LSAT(001), STO(001) and DSO(001) substrates are fully strained, leading to $a^*_{SAO-T} \sim a^*_{sub}$. Nevertheless, the SAO$_T$ film grown on LAO(001) is strain relaxed (**Supplementary Fig. 13b**), leading to the unequal $a^*_{SAO-T} \sim a^*_{sub}$. As summarized in **Fig. 1f** in the main text, by coherently growing the high-quality SAO$_T$ film on a variety of (001)-oriented ABO$_3$ perovskite substrates, the in-plane $a^*_{SAO-T}$ can be continuously adjusted in a wide range (3.84~3.95 Å).

Inspired by the previous works on (Ba,Sr,Ca)$_3$Al$_2$O$_6$ compounds, we also tried the synthesis of other super-tetragonal 427 compounds by substituting or doping the Sr with Ca/Ba. Although synthesizing the Ca$_4$Al$_2$O$_7$ and Ba$_4$Al$_2$O$_7$ compounds is still challenging using the common solid-state reaction method (Ca$_4$Al$_2$O$_7$ structure could be thermodynamically unstable, and Ba$_4$Al$_2$O$_7$ decomposes rapidly in ambient), we do succeed in synthesizing the half-doped Ba$_2$Sr$_2$Al$_2$O$_7$ (BSAO$_T$) and Ca$_2$Sr$_2$Al$_2$O$_7$ (CSAO$_T$) films. As shown in **Supplementary Fig. 14**, the BSAO$_T$ and CSAO$_T$ films can be coherently strained to the KTO(001) and LAO(001) substrates, respectively. These doped films further extended the strain-tuned in-plane lattice constant range from 3.78 to 3.99 Å (**Fig. 1f** in the main text), which almost covers the entire lattice constant range of commercially available perovskite substrates.

The coexistence of strain coherency of SAO$_T$ films and the wide tuning range of $a^*_{SAO-T}$ in-plane lattice constant were rarely observed in the previously reported SAO$_C$ films. The SAO$_C$ films, even grown on the most lattice-matched substrates [e.g., STO(001)] are strain-relaxed, resulting in high-density dislocations at the heterointerface[9,10].

Interestingly, we also found that the SAO$_T$ film can be epitaxially grown on SrLaGaO$_4$(100) [SLGO(100)] substrate. The SLGO crystal has a tetragonal symmetry. The lattice constants are $a_{SLGO} = b_{SLGO} = 3.843$ Å and $c_{SLGO} = 12.712$ Å. According to the **supplementary Section 2.3**, the SAO$_T$ film has a tetragonal/orthorhombic structure, $a_{SAO-T} \sim b_{SAO-T} \sim 2\sqrt{2} a_{SLGO}$, and $c_{SAO-T} \sim 2c_{SLGO}$. Therefore, as shown in **Supplementary Fig. 15a,b**, the SAO$_T$ film grown on SLGO(100) is uniformly (110)-oriented. The RSM (**Supplementary Fig. 15c**) further revealed that the in-plane epitaxial relationship are SAO$_T$[110]//SLGO[100] and SAO$_T$[001]//SLGO[001] (schematically depicted in **Supplementary Fig. 15d**). The SAO$_T$(666) and SLGO(303) diffractions share the same in-plane $Q_x$ value. Hence, the SAO$_T$ film is fully strained to the SLGO(100) substrate, which further confirms the superior ability of SAO$_T$ to adapt epitaxial strain. More importantly, such a unique epitaxial growth enables an anisotropic structural template for the epitaxial growth of various perovskite-like oxides and corresponding freestanding membranes with elongated *c*-axis lies in-plane. Typical examples include YBa$_2$Cu$_3$O$_{7-\delta}$, T-phase BiFeO$_3$, and Ruddlesden-Popper compounds[11-13].

## 2.3 Additional analyses of the STEM results

The detailed atomic structure of the SAO$_T$ phase is examined by scanning transmission electron microscopy (STEM) in cross-sectional high-angle annular dark-field (HAADF) mode. As shown in **Supplementary Fig. 16**, the HAADF-STEM images were measured from the SAO$_T$ film along both LSAT[100] and [110] zone-axis. The HAADF-STEM images measured from the SAO$_T$ film along LSAT[100] zone-axis (**Supplementary Fig. 16a**) display an ABO$_3$ perovskite-like atomic contrast. More interestingly, the "A-site" atomic columns exhibit an intensity modulation along the out-of-plane [001] axis, which can be highlighted by the image with reduced brightness and enhanced contrast (**Supplementary Fig. 16b**). Specifically, as marked by the red arrows in **Supplementary Fig. 16**, the STEM intensity of the "A-site" atomic plane becomes higher for every other 3 perovskite-like unit-cells. HAADF-STEM images measured along the LSAT[110] zone axis (**Supplementary Fig. 16c**) display a much more complicated cation ordering. For clarity, we labelled the atomic layers along the out-of-plane direction from No. #1 to #13. The layers #13 to #8 are a mirror image of the layers #1 to #6. The perovskite "A-site" atomic layers are layers #2, #4, and #6. All of these three layers are fully occupied, and layer #4 is brighter than layers #2 and #6, consistent with the intensity modulation shown in **Supplementary Fig. 16a,b**. The perovskite "B-site" atomic layers are layers #1, #3, #5, and #7. All of these atomic layers are partially occupied. The layer #1 and #7 consist of alternatively occupied atomic columns, while layers #3 and #5 consist of atomic columns with distinct STEM intensities. According to these STEM results, we speculate the in-plane lattice constant of SAO$_T$ ($a_{\text{SAO-T}}$) should be $2\sqrt{2}$ or 4 times as large as that of the simple perovskite unit-cell, and the out-of-plane lattice constant ($c_{\text{SAO-T}}$) should be 3 or 6 times as large as that of the cubic perovskite unit-cell ($c_p$).

## 2.4 DFT simulation on the SAO$_T$ structure

We now try to determine the atomic structure of the SAO$_T$ phase by DFT calculation. To the best of our knowledge, none of the reported strontium aluminates can match the STEM and XRD results. We then searched numerous possible structure candidates, including Ba$_4$Al$_2$O$_7$[14], Sr$_4$Ga$_2$O$_7$[15], Sr$_3$Ga$_2$O$_6$[16,17], Sr$_{10}$Al$_6$O$_{19}$[18], etc., which have similar chemical formulas to the SAO$_T$ (Sr$_4$Al$_2$O$_7$) phase. Based on these parent compounds, we replaced the Ba (Ga) atoms with Sr (Al) and then performed density-functional-theory (DFT) structural relaxations to obtain convergent structures (as described in the **Methods and Materials section**). By checking these DFT-relaxed structure candidates, we found the Sr$_4$Al$_2$O$_7$ unit-cell, stimulated based on the orthorhombic

Ba$_4$Al$_2$O$_7$ compounds with space group *Cmca* (No.64), provide the best match with the STEM images measured from SAO$_T$. The DFT-level structure relaxation is convergent, providing a thermodynamically and energetically stable orthorhombic Sr$_4$Al$_2$O$_7$ unit-cell. Accordingly, we speculate that the SAO$_T$ could be a Sr$_4$Al$_2$O$_7$ compound with orthorhombic symmetry and atomic structure similar to Ba$_4$Al$_2$O$_7$.

As summarized in **Supplementary Table 2**, the simulated lattice constants of SAO$_T$ (Sr$_4$Al$_2$O$_7$) are $a_{SAO-T}$ = 10.798 Å, $b_{SAO-T}$ = 11.238 Å, and $c_{SAO-T}$ = 25.732 Å. As shown in **Supplementary Fig. 17a**, the atomic arrangements of simulated SAO$_T$ (Sr$_4$Al$_2$O$_7$) unit-cell viewed along SAO$_T$[100] and [010] axes match perfectly with the HAADF-STEM image viewed along LSAT[110] zone-axis.

As depicted in **Fig. 2g** in the main text, this consistency strongly suggests that the epitaxial relationship of coherently-grown SAO$_T$/LSAT(001) films should be SAO$_T$[100]//LSAT[110], SAO$_T$[010]//LSAT[1$\bar{1}$0], and SAO$_T$[001]//LSAT[001]. On this basis, we speculate that the SAO$_T$ unit-cell is $2\sqrt{2} \times 2\sqrt{2} \times 6$ times as large as the simple ABO$_3$ perovskite unit-cell. And we can obtain the reduced lattice constants as $a^*_{SAO-T} = b^*_{SAO-T}$ = 3.896 Å, and $c^*_{SAO-T}$ = 4.288 Å. As summarized in **Supplementary Table 2**, these simulated values are very close to those derived from the XRD results of the SAO$_T$/LSAT(001) film, further supporting the validity of our proposed SAO$_T$ (Sr$_4$Al$_2$O$_7$) structure.

The simulated SAO$_T$ structure and speculated epitaxial relationship between SAO$_T$ film and LSAT substrate have been further corroborated by 3-dimensional full reciprocal space mapping (3D-RSM) using synchrotron XRD. The 3D-RSMs were performed on the BL02U2 beamline at Shanghai Synchrotron Radiation Facility, with a Huber diffractometer, an EigerX 500K area detector, and incident monochromatic X-ray photons of 18.2 keV. As shown in **Supplementary Fig. 18a**, in the reciprocal range of $Q_x$: -2.6~2.6 Å$^{-1}$, $Q_y$: -2.6~2.6 Å$^{-1}$, $Q_z$: 1.2-5.2 Å$^{-1}$, all the diffraction information is collected and presented in the 3D-RSM. As shown in **Supplementary Fig. 18b**, a $Q_x$-$Q_y$ slice taken at $Q_z$ = 4.845 Å$^{-1}$ clearly shows the aforementioned epitaxial relationship and tetragonal structure of SAO$_T$.

Note that the simulated and fully-relaxed SAO$_T$ unit-cell should have orthorhombic symmetry, consistent with the two-fold symmetric in-plane atomic structure shown in **Supplementary Fig. 17b**. Therefore, we suggest that the observed tetragonal SAO$_T$ unit-cell should originate from the compressive biaxial strain imposed from LSAT(001) substrate. And the SAO$_T$ unit-cell should be restored to orthorhombic if grown on an orthorhombic substrate [e.g. NGO(001), as shown in **Supplementary Fig. 10**].

### 2.5 DFT simulations on the SAO unit-cells under strain

After determining the atomic structure of SAO$_T$, we can further understand the observed coherent strain state in the ABO$_3$/SAO$_T$ heterostructures by additional DFT calculations. The ABO$_3$/SAO$_T$ structure coherency is intuitively dominated by two factors: 1) the ability of SAO$_T$ to sustain strain and 2) the bonding strength at ABO$_3$/SAO$_T$ interface. We will evaluate both factors using DFT calculations.

First, to evaluate the ability of SAO$_T$ to sustain epitaxial strain, we calculated the energy changes of SAO$_T$ and SAO$_C$ unit-cells under strain. Since the commonly used ABO$_3$(001) perovskite substrates have either cubic or orthorhombic symmetry, we considered both biaxial and anisotropic strain cases. As mentioned in the **Materials and Methods** section, we manually altered the in-plane lattice constants $a_{SAO}$ and $b_{SAO}$ to impose biaxial/anisotropic strain, relaxed the lattice along the $c_{SAO}$ axis, and calculated the normalized energy change $\Delta E$. The results are summarized in **Supplementary Fig. 19** and **Fig. 3a,b** in the main text.

In the biaxial strain case, the relative change in lattice constant ($\varepsilon = \Delta a_{SAO} / a_{SAO}$) or pseudocubic lattice constant ($a^*_{SAO}$) serves as the control parameters. The $\Delta E - \varepsilon$ curves are shown in **Fig. 3a,b** in the main text, and the $\Delta E - a^*_{SAO}$ curves are shown in **Supplementary Fig. 19a**, respectively. Both figures reveal the parabolic line shapes of the $\Delta E - \varepsilon$ and $\Delta E - a^*_{SAO}$ curves calculated from SAO$_T$ and SAO$_C$ unit-cells. The local minima of $\Delta E - \varepsilon$ ($\Delta E - a^*_{SAO}$) curves are located at $\varepsilon = 0$ (the strain-free $a^*_{SAO}$ values). According to the $\Delta E - \varepsilon$ curves, the $\Delta E$ of the SAO$_T$ unit-cell is consistently smaller than that of the SAO$_C$ unit-cell for the entire $\varepsilon$ range. Especially for the compressive strain case ($\varepsilon < 0$), the $\Delta E/E$ value for the SAO$_T$ unit-cell is less than 60% of the value calculated for the SAO$_C$ unit-cell. We also labeled the $a^*_{sub}$ for various perovskite substrates in **Supplementary Fig. 19a**. The $\Delta E - a^*_{SAO}$ curves further confirm that the SAO$_T$ is structurally stable when epitaxially grown on substrates with relatively smaller $a^*_{sub}$.

In the anisotropic strain case, the control parameter is the orthorhombicity ($\gamma = b_{SAO}/a_{SAO}$). As shown in **Supplementary Fig. 19b**, both the $\Delta E - \gamma$ curves calculated from SAO$_T$ and SAO$_C$ unit-cells exhibit parabolic shapes. The curve calculated from the SAO$_C$ unit-cell is symmetric to the local minimum at $\gamma = 1$ due to its intrinsic cubic symmetry. By contrast, the $\Delta E - \gamma$ curve calculated from the SAO$_T$ unit-cell is highly asymmetric, showing a local minimum at $\gamma = 1.04$ due to its intrinsic orthorhombic symmetry. As $\gamma$ increases from 1 to 1.06, the $\Delta E$ of SAO$_T$ is limited below 0.03%, while the $\Delta E$ for SAO$_C$ surges up to 0.1%.

The above DFT calculations demonstrate that the low-symmetry SAO$_T$ unit-cell could have more structural degrees of freedom to deform and accommodate the ABO$_3$ perovskite substrate-induced bi-axial and anisotropic strain, whereas the high-symmetry cubic SAO$_C$ unit-cell could be more rigid under stress.

Second, to evaluate the ABO$_3$/SAO$_T$ interfacial bonding strength, we constructed SAO$_T$/STO(001) heterostructure supercells and calculated the interfacial bonding energy ($E_{bond}$). For comparison, we also calculated the $E_{bond}$ of an SAO$_C$/STO(001) supercell and an LAO/STO supercell. Given that the surface layers of SAO$_C$ and SAO$_T$ are Sr-rich, we speculate that the TiO$_2$-terminated SAO/STO interface should be energetically more favorable. Thus, we set the STO surface of these supercells to be TiO$_2$-terminated (**Supplementary Fig. 20**). We then calculated the interfacial bonding energy $E_{bond}$ (normalized by $n_{STO}$, the number of STO unit-cells bonded at the interface). As summarized in **Fig. 3c** in the main text, the calculated $E_{bond}$ at both SAO$_C$/STO(001) interface (-0.82 eV) and SAO$_T$/STO(001) interface (-1.97 eV) are negative, implying stable interfacial bonding states. More importantly, the $E_{bond}$ calculated from the SAO$_T$/STO(001) interface is more than twice as large as that of the SAO$_C$/STO(001) interface, and even comparable with the $E_{bond}$ calculated from the LAO/STO(001) interface (-2.34 eV). The strong interfacial bonding can help the epitaxial strain state propagate coherently in the ABO$_3$/SAO$_T$ heterostructures.

In brief, the DFT calculations revealed two unique features of SAO$_T$, which the SAO$_C$ phase does not process. First, the SAO$_T$ unit-cell can easily deform under both the biaxial and anisotropic strain. Second, the ABO$_3$/SAO$_T$ interface has strong bonding strength. Both effects work cooperatively and ensure the coherent strain states observed in various ABO$_3$/SAO$_T$ heterostructures shown in **Fig. 1** and **Supplementary Figs. 13-15**.

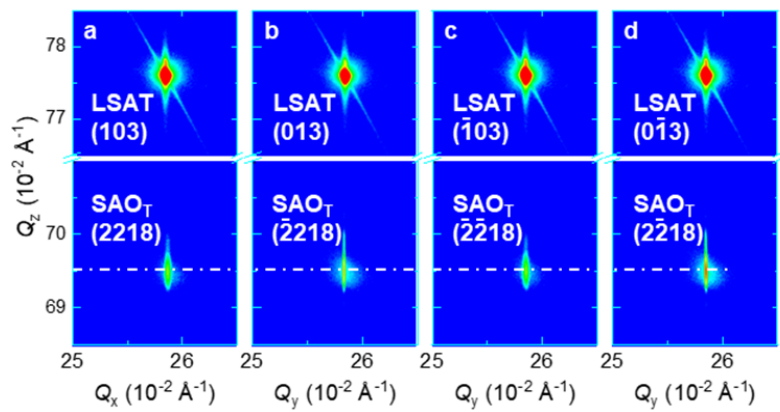

**Supplementary Fig. 8. | Structure symmetry of SAO$_T$/LSAT(001) films determined by reciprocal space mapping.** (**a-d**) Reciprocal space mappings (RSMs) measured from an STO(3 nm)/SAO$_T$(30 nm)/LSAT(001) film near the LSAT (103), (013), ($\bar{1}$03), and (0$\bar{1}$3) diffractions.

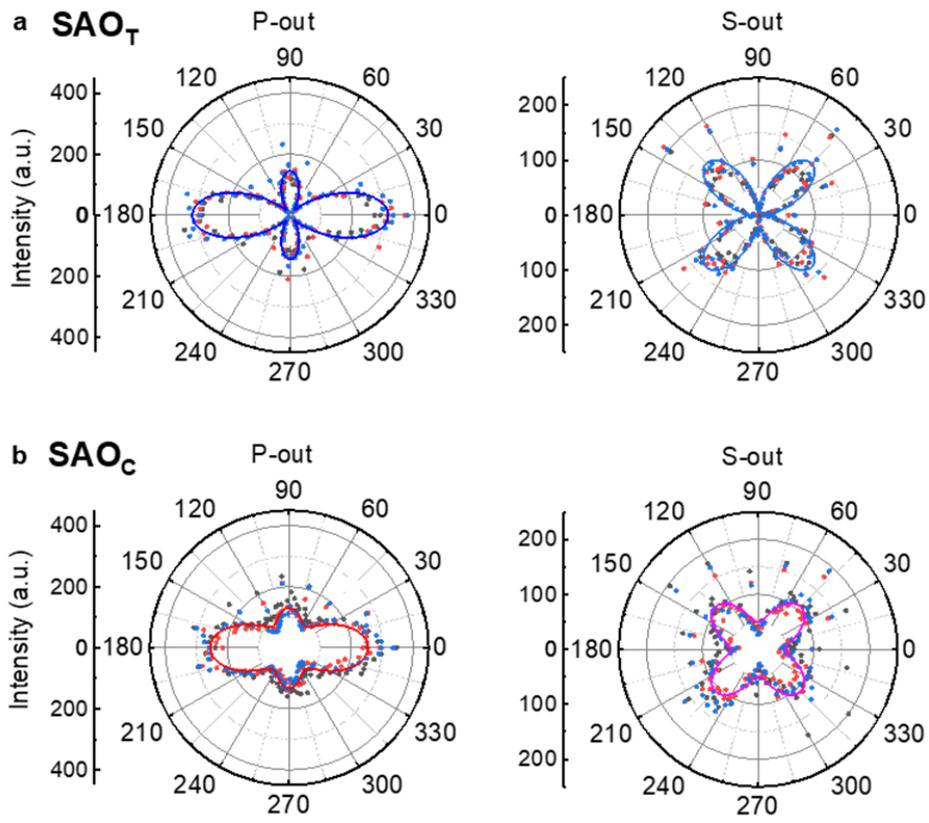

**Supplementary Fig. 9. | Structure symmetry of SAO$_T$/LSAT(001) films determined by optical second harmonic generation.** (**a,b**) Optical second harmonic generation (SHG) polar plots measured at room temperature as a function of incident beam polarization in (**a**) SAO$_T$ and (**b**) SAO$_C$ films. The solid dots are the experimental data, and the color lines indicate the fitting curves based on the tetragonal *4mm* point group.

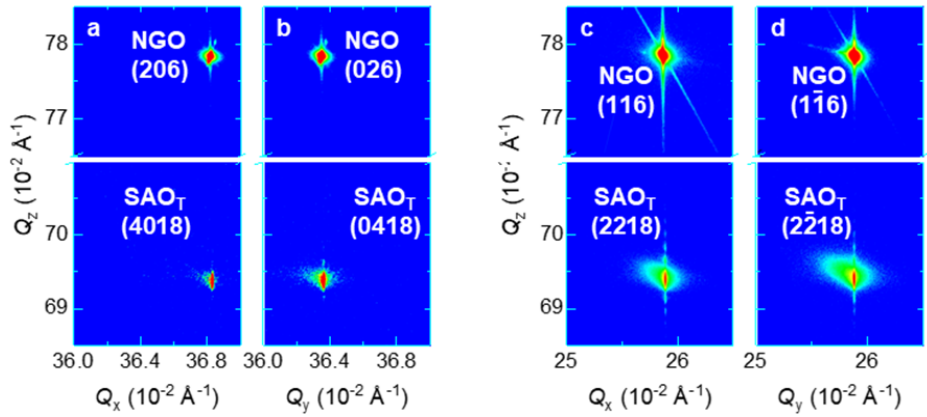

**Supplementary Fig. 10. | Structure symmetry of $SAO_T$/NGO(001) films determined by Reciprocal space mapping.** (**a,b**) Reciprocal space maps (RSMs) measured from an STO(3 nm)/$SAO_T$(50 nm)/NGO(001) film near NGO(206) and (026) diffractions. The notation of corresponding $SAO_T$ diffractions are (4018) and (0418). (**c,d**) RSMs measured near NGO(116) and (1$\bar{1}$6) diffractions. The notation of corresponding $SAO_T$ diffractions are (2218) and (2$\bar{2}$18).

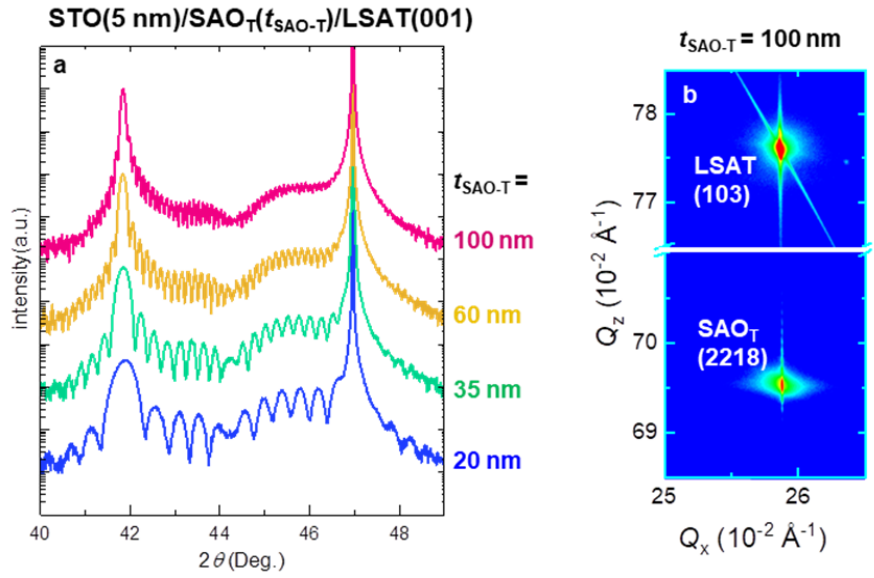

**Supplementary Fig. 11. | Structure characterizations of SAO$_T$/LSAT(001) films with various thicknesses.** XRD 2$\theta$-$\omega$ linear scans of STO (5 nm)/SAO$_T$/LSAT (001) films with various thicknesses ($t_{SAO-T}$) (**a**), and RSM around LSAT(103) diffraction measured from STO (5 nm)/SAO$_T$(100 nm)/LSAT film (**b**).

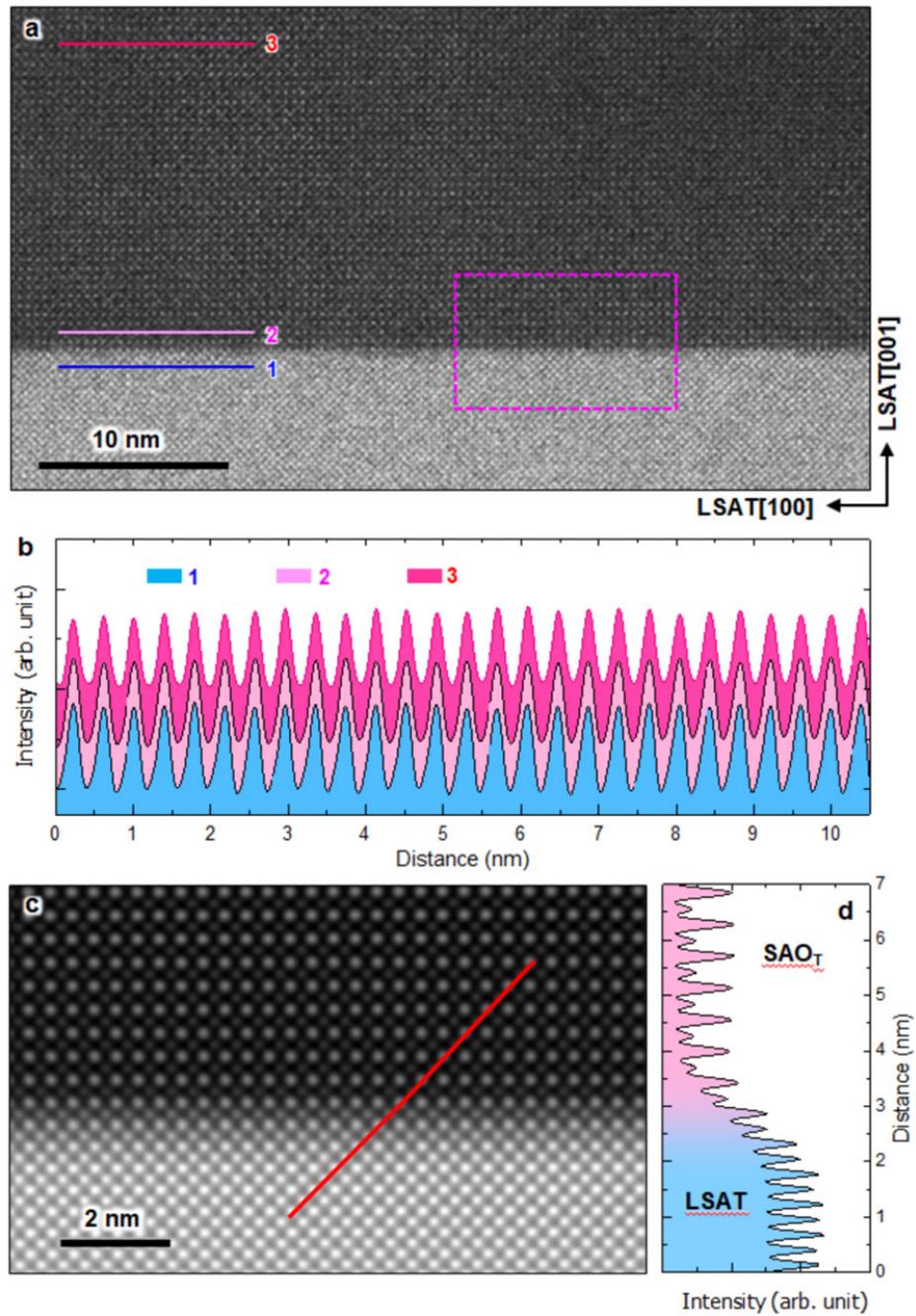

**Supplementary Fig. 12. | Microscopic characterizations on the SAO$_T$/LSAT(001) interface.** (**a**) HAADF-STEM image measured from a STO(3nm)/SAO$_T$(30 nm)/LSAT(001) film along LSAT[100] zone axis. (**b**) Lateral line profiles of STEM intensity extracted from the line marked in (**a**). Line #1 and #2 are located near the SAO$_T$/LSAT(001) interface, and line #3 is located near the top of the film. These atomic spacing shown in three line profiles are identical, further attesting a coherent strain state. (**c**) A zoom-in STEM image taken from the area marked in a dashed box shown in (**a**). (**d**) Line profile of STEM intensity extracted from the red line marked in (**b**). These STEM images clearly demonstrate that the SAO$_T$/LSAT(001) interface is defect-free, and the sharpness is limited within 1 to 2 unit-cells.

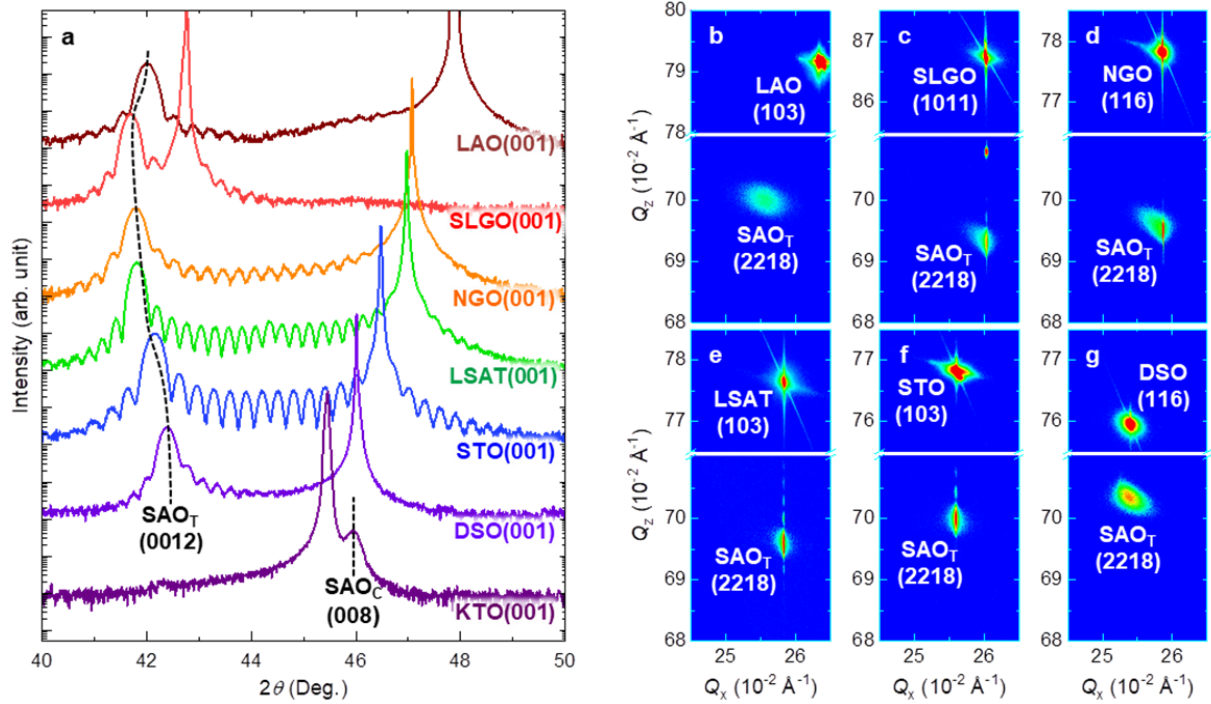

**Supplementary Fig. 13. | Structural characterizations of 30 nm thick SAO$_T$ films grown on various (001)-oriented perovskite substrates.** (**a**) XRD 2θ-ω linear scans of SAO$_T$ (30 nm) films grown on LAO(001), SLGO(001), NGO(001), LSAT(001), STO(001), DSO(001), and KTO(001) substrates. The SAO$_T$ phase can be stabilized in most of the substrates expect for KTO(001). (**b-g**) XRD RSMs around SAO$_T$(2218) diffractions measured from the SAO$_T$ films. For SAO$_T$/LAO(001) samples (**b**), the in-plane $Q_x$ values of SAO$_T$(2218) diffractions are different from the LAO(103), indicating fully relaxed strain states. For all the other SAO$_T$ films (**c-g**), the in-plane $Q_x$ of SAO$_T$(2218) diffractions and substrate diffractions are identical, signifying a coherent strain state.

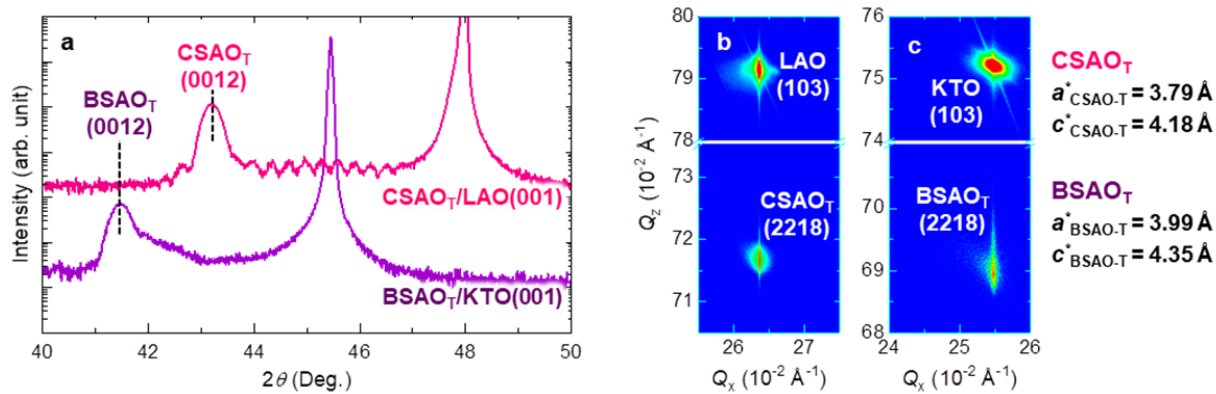

**Supplementary Fig. 14. | Structural characterizations of 30 nm thick Ba or Ca-doped SAO$_T$ films.** (a) XRD 2$\theta$-$\omega$ linear scans of 30 nm-thick Ca$_2$Sr$_2$Al$_2$O$_7$ (CSAO$_T$) and Ba$_2$Sr$_2$Al$_2$O$_7$ (CSAO$_T$) films grown on LAO(001) and KTO(001) substrates, respectively. (**b,c**) XRD RSMs around (**b**) CSAO$_T$(2218) and (**c**) BSAO$_T$(2218) diffractions measured from the CSAO$_T$ and BSAO$_T$ films. Both films are coherently strained. The lattice constants derived from RSMs are also labeled.

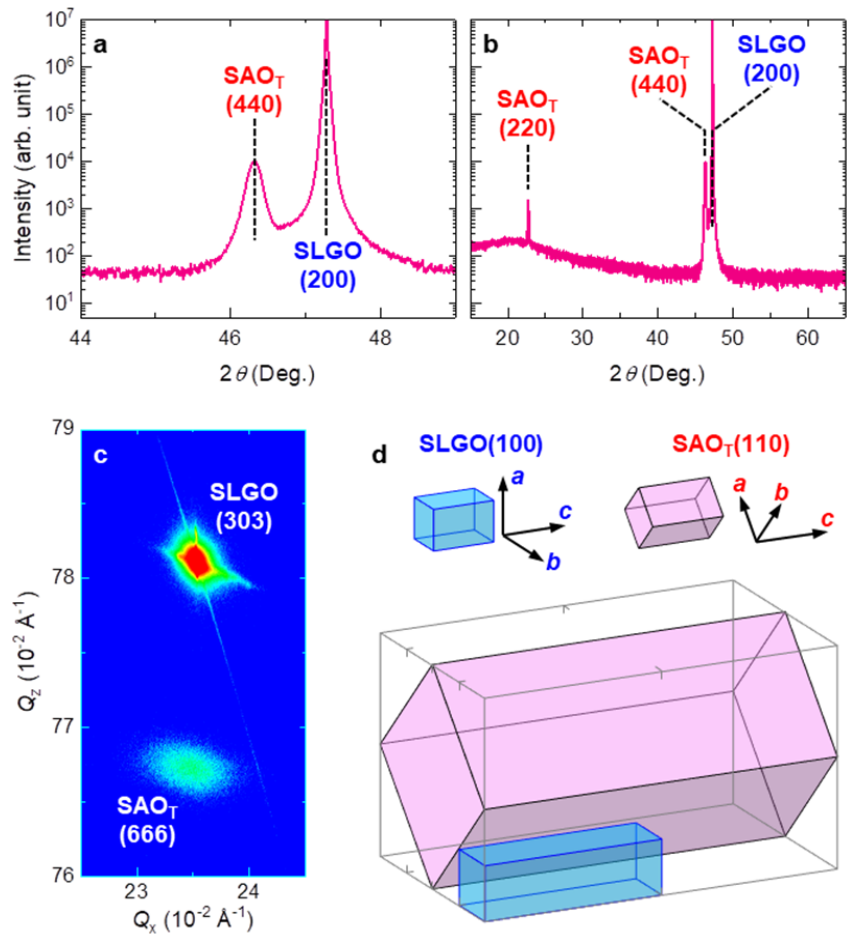

**Supplementary Fig. 15. | Structural characterizations of an SAO$_T$/SLGO(100) film.** (**a,b**) XRD $2\theta$-$\omega$ linear scans measured from a 30 nm thick SAO$_T$/SLGO(100) film near SLGO(200) diffraction (**a**) and in a wide Bragg angle range from 15 ~ 65° (**b**). In these two curves, only the SAO$_T$(220) and (440) peaks can be identified. (**c**) XRD RSM measured around SLGO(303) diffraction. The SAO$_T$(666) and SLGO(303) diffractions share the same in-plane $Q_x$ value. These results clearly signifies the epitaxial relationship of SAO$_T$/SLGO(100) film: SAO$_T$[110]//SLGO[100] and SAO$_T$[001]//SLGO[001]. And the SAO$_T$(110) film is coherently strained to the SLGO(100) substrate. (**d**) Schematic illustration of the epitaxial relationship in SAO$_T$/SLGO(100) film.

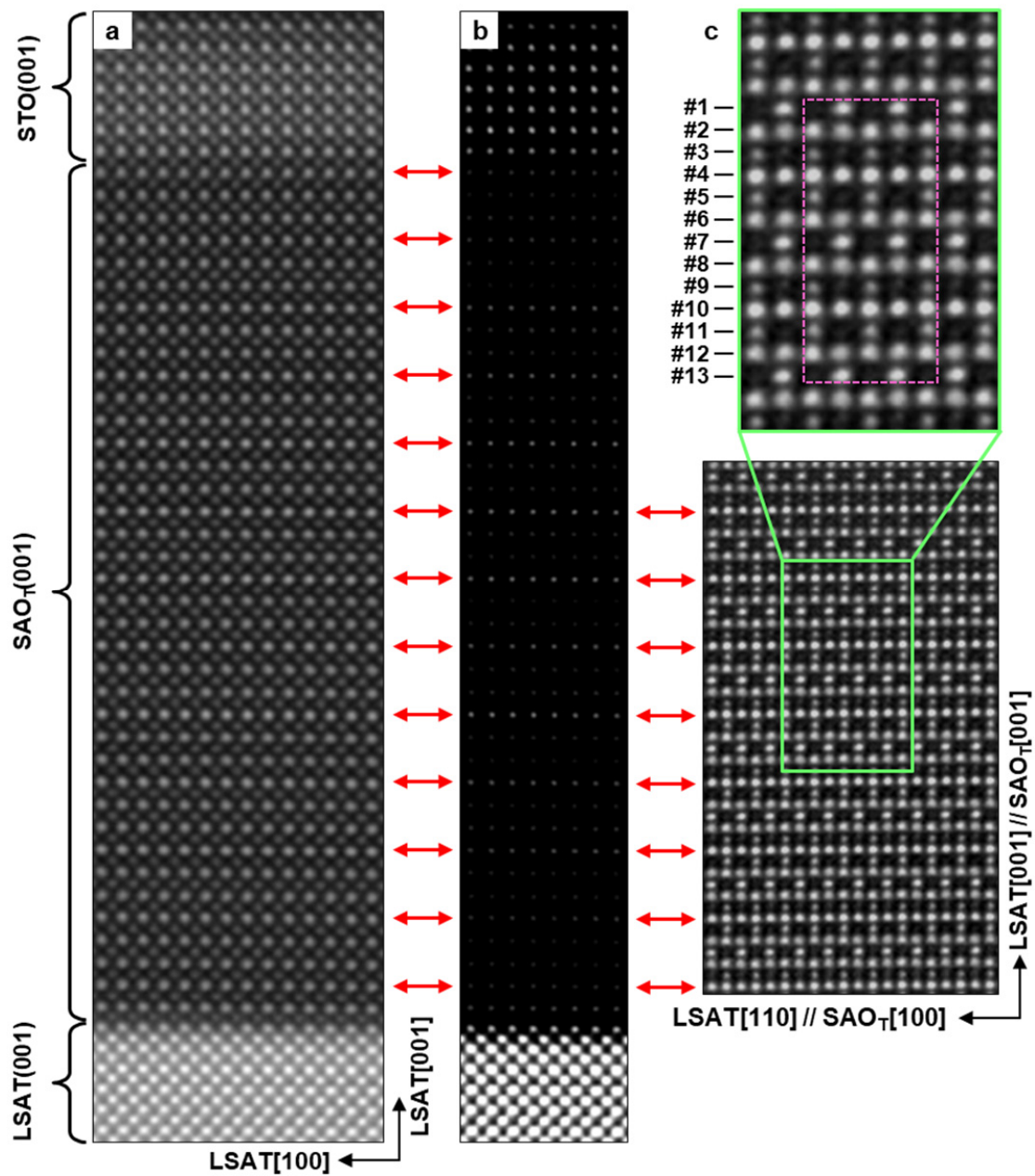

**Supplementary Fig. 16. | HAADF-STEM images of an SAO$_T$/LSAT(001) film.** (**a**) Large-scale HAADF-STEM image measured from an STO(3 nm)/SAO$_T$(15 nm)/LSAT(001) film along LSAT[100] zone axis. (**b**) A copy of STEM image shown in (**a**) with reduced brightness and enhanced contrast. The out-of-plane intensity modulations (marked by the red arrows) are highlighted by these image modifications. (**c**) HAADF-STEM image measured along LSAT[110] zone axis. The large-scale and zoomed-in images are plotted in the bottom and top panels, respectively. The different atomic layers are marked by numbers from #1 to #13. Layers #13 to #8 are the mirror images of layers #1 to #6.

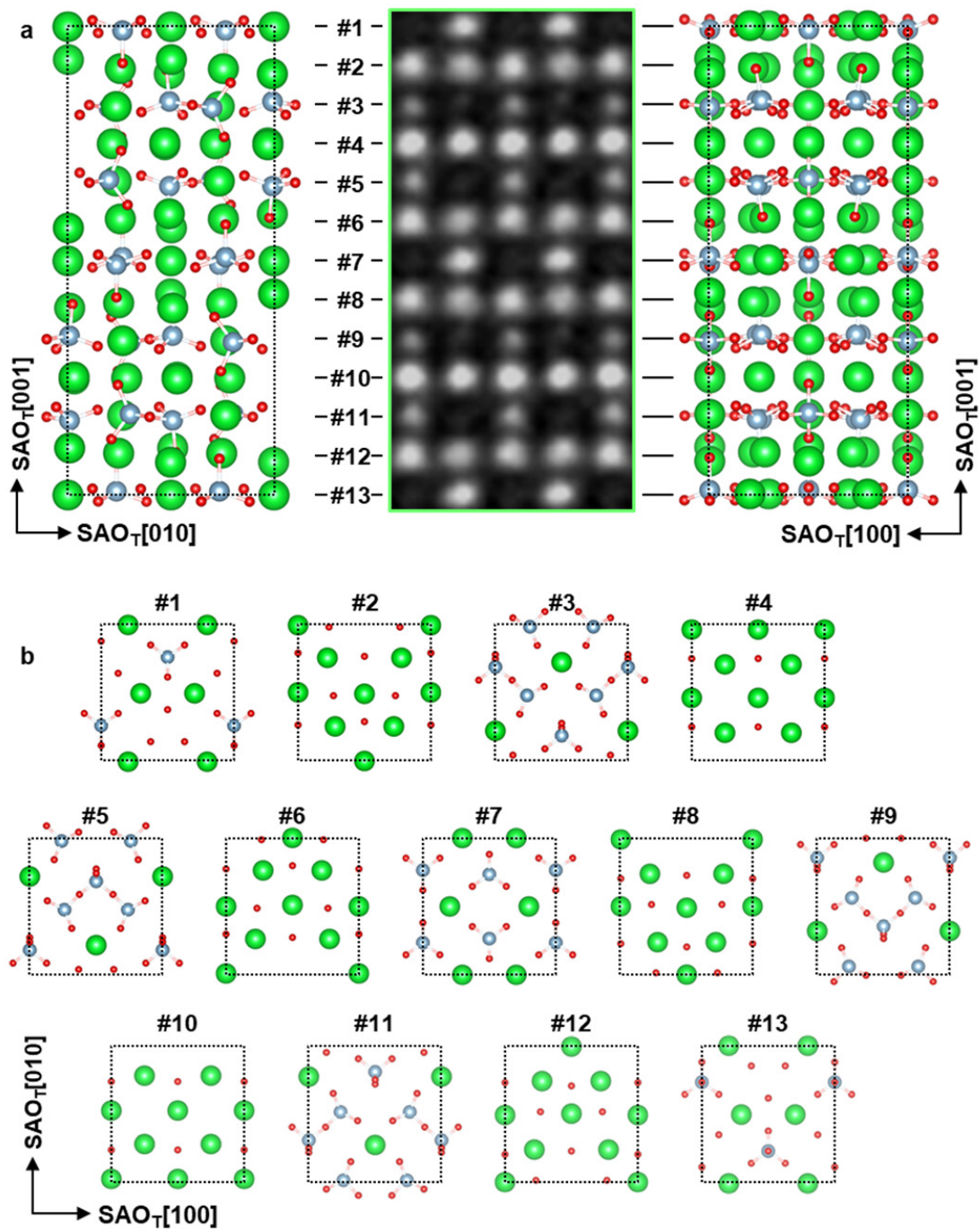

**Supplementary Fig. 17. | DFT-simulated crystal structure of SAO$_T$ unit-cell.** (**a**) Atomic structure of SAO$_T$ unit-cell viewed along [100] and [010] axes (left and right panels). The STEM image is inserted in the central panel for comparison. The STEM image shows nice correspondence with the simulated lattice viewed along both [100] and [010] axes. (**b**) Atomic structures of different in-plane atomic layers, labeled from #1 to #13. Layers #13 to #8 are a mirror image of layers #1 to #6.

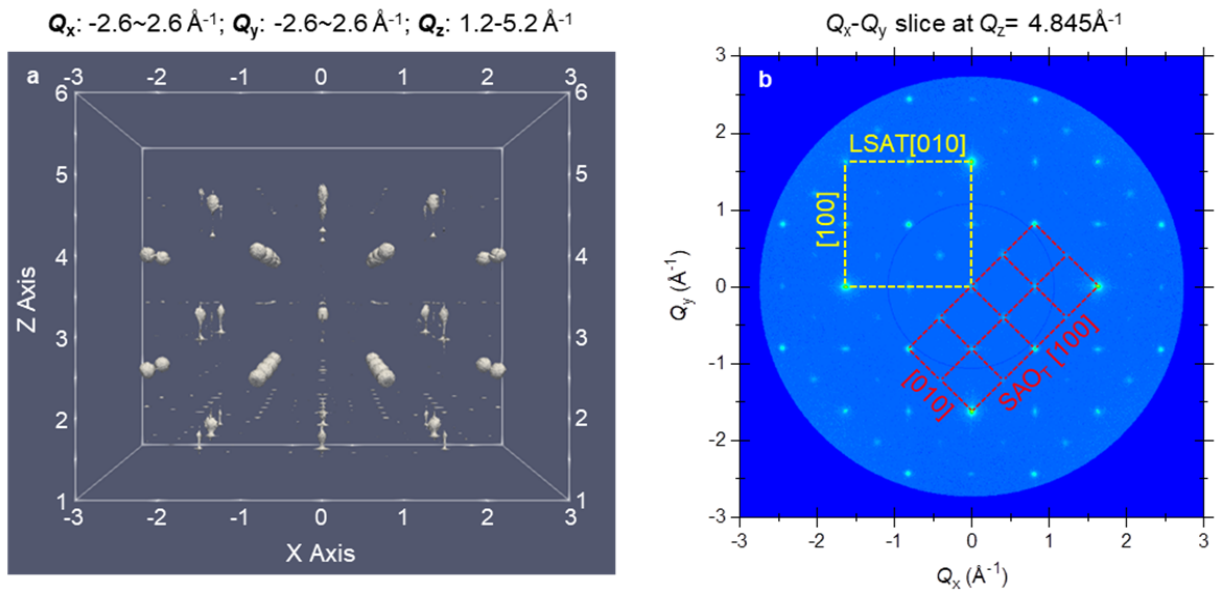

**Supplementary Fig. 18. | 3D-RSM on the epitaxial relationship between SAO$_T$ and LSAT(001).** (**a**) Full RSM covers the range of $Q_x$: -2.6~2.6 Å$^{-1}$, $Q_y$: -2.6~2.6 Å$^{-1}$, $Q_z$:1.2-5.2 Å$^{-1}$, view along [010]* direction in the reciprocal space. (**b**) The $Q_x$-$Q_y$ slice obtained at $Q_z$ = 4.845 Å$^{-1}$ from (**a**), showing the match relationship between the reciprocal lattices of SAO$_T$ film and LSAT substrate, i.e., SAO$_T$[110]// LSAT[100].

|  | $a$ (Å) | $a^*$ (Å) | $b$ (Å) | $b^*$ (Å) | $c$ (Å) | $c^*$ (Å) |
|---|---|---|---|---|---|---|
| $Sr_3Al_2O_6$ (Exp.) | 15.844 | 3.961 | 15.844 | 3.961 | 15.844 | 3.961 |
| $Sr_3Al_2O_6$ (DFT) | 15.815 | 3.954 | 15.815 | 3.954 | 15.815 | 3.954 |
| $Sr_4Al_2O_7$/LSAT(001) (Exp.) | 10.940 | 3.868 | 10.940 | 3.868 | 25.896 | 4.316 |
| $Sr_4Al_2O_7$/NGO(001) (Exp.) | 10.866 | 3.867 | 11.007 | 3.867 | 25.944 | 4.324 |
| $Sr_4Al_2O_7$/STO(001) (Exp.) | 11.045 | 3.905 | 11.045 | 3.905 | 25.758 | 4.293 |
| $Sr_4Al_2O_7$ (DFT) | 10.798 | 3.896 | 11.238 | 3.896 | 25.732 | 4.288 |

**Supplementary Table 2. | Lattice constants of $Sr_3Al_2O_6$ ($SAO_C$) and $Sr_4Al_2O_7$ ($SAO_T$) unit-cells determined by both experiments (Exp.) and DFT calculations.** The $a^*$, $b^*$, and $c^*$ are lattice constants normalized to the pseudocubic perovskite unit-cells. The cubic $SAO_C$ unit-cell is 4×4×4 times as large as the pseudocubic perovskite unit-cell. Thus we got $a^* = b^* = c^* = a/4 = b/4 = c/4$. The orthorhombic $SAO_T$ unit-cell is $2\sqrt{2} \times 2\sqrt{2} \times 6$ times as large as the pseudocubic perovskite unit-cell. Thus we got $a^* = b^* = (a^2 + b^2)^{1/2}/4$ and $c^* = c/6$.

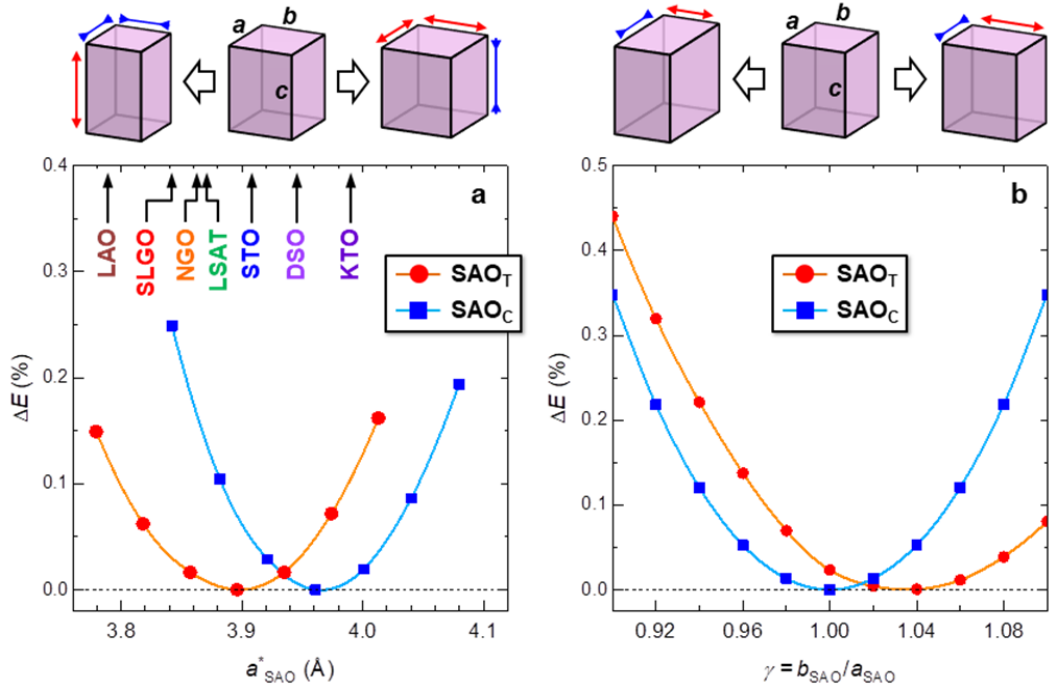

**Supplementary Fig. 19. | DFT calculated energy changes of $SAO_C$ and $SAO_T$ unit-cells under bi-axial and anisotropic strains.** We first calculate the total energy ($E_{total}$) of $SAO_C$ and $SAO_T$ unit-cells with fully relaxed lattices. We then also calculated the $E_{total}$ by imposing biaxial or anisotropic strain. The differences in $E_{total}$ between the strained and strain-free cases are defined as $\Delta E$. To directly compare the $SAO_C$ and $SAO_T$ unit-cells, the energy changes ($\Delta E$) are normalized by the $E_{total}$ of the strain-free unit-cells. For the bi-axial strain case, the control parameter is the normalized in-plane lattice constant change $\varepsilon = \Delta a_{SAO}/a_{SAO}$. The $\varepsilon$ versus $\Delta E$ curves obtained from $SAO_T$ and $SAO_C$ unit-cells are shown in Fig. 3A in the main text. **(a)** $\Delta E$ values plotted as a function of in-plane lattice constant in pseudo-cubic perovskite notation ($a^*_{SAO}$). Both reveal the parabolic line shapes and local minima at the fully relaxed $a^*_{SAO}$ values. The $a^*_{sub}$ for various perovskite substrates are also labeled. These results further confirm that the $SAO_T$ is structurally stable when epitaxially grown on substrates with relatively smaller $a^*_{sub}$. **(b)** $\Delta E - \gamma$ curves calculated from $SAO_C$ and $SAO_T$ unit-cells. The orthorhombicity ($\gamma = b_{SAO}/a_{SAO}$) is the control parameter for the anisotropic strain case. Both $\Delta E - \gamma$ curves exhibit parabolic shapes. The curve calculated from $SAO_C$ unit-cell is symmetric to the local minimum at $\gamma = 1$ due to its intrinsic cubic symmetry. By contrast, the $\Delta E - \gamma$ curve calculated from $SAO_T$ unit-cell is highly asymmetric, showing a local minimum at $\gamma = 1.04$ due to its intrinsic orthorhombic symmetry. As $\gamma$ increases from 1 to 1.06, the $\Delta E$ of $SAO_T$ is limited below 0.03%, while the $\Delta E$ for $SAO_C$ surges up to 0.1%. Clearly, the $SAO_T$ unit-cell shows substantial enhancement in structural flexibility under both biaxial and anisotropic strain.

Note that while the $\Delta E$ values calculated from $SAO_T$ at $\gamma < 1$ are higher than those calculated from $SAO_C$, they do not have practical significance. When epitaxially grown on orthorhombic substrates, the $SAO_T$ films always follow the most lattice-matched epitaxial relationship (e.g., $SAO_T[100]//NGO[100]$ and $SAO_T[010]//NGO[010]$ due to $b_{NGO} > a_{NGO}$). Accordingly, the experimental $\gamma$ values should be always equal or exceed 1. For clearer presentation, we omitted the $\Delta E$ data at $\gamma < 1$ range in **Fig. 3b** in the main text.

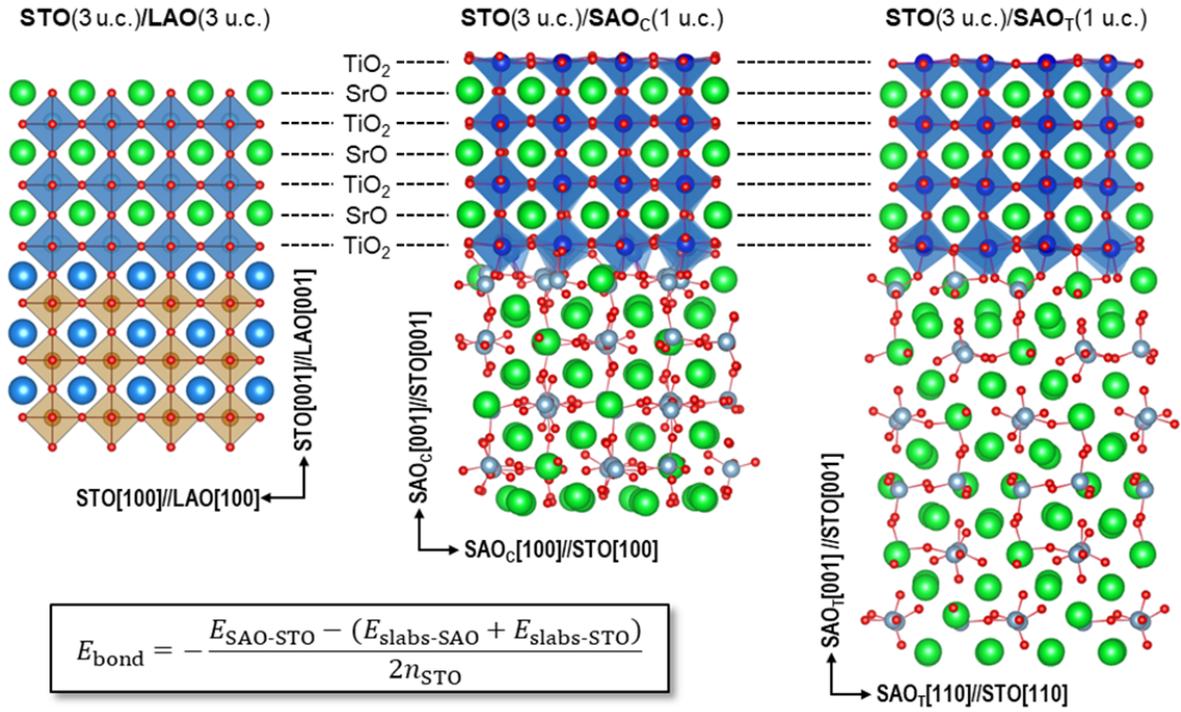

**Supplementary Fig. 20. | LAO/STO(001), $SAO_C$/STO(001), and $SAO_T$/STO(001) heterostructure supercells for interfacial bonding energy ($E_{bond}$) calculations.** All of the interfaces in these supercells are $TiO_2$-terminated. The formula for calculating the $E_{bond}$ is also inserted.

# Section 3: Characterizations of the freestanding oxide membranes

To further evaluate the feasibility of $SAO_T$ film as a versatile sacrificial layer, we have performed comparative studies on the perovskite-structured functional oxides grown on $SAO_T$ and $SAO_C$ films. As summarized in **Supplementary Table 3**, these oxides include NNO, LCMO, STO, SRO, BTO, and SSO, with a broad range of in-plane lattice constants ($a_p$, in pseudocubic notation) from 3.81 to 4.04 Å. To minimize the lattice mismatch between SAO and $ABO_3$ perovskite, we chose to grow the $ABO_3$/SAO bilayers on either LSAT(001) or STO(001) substrates. Characterizations of the structure and physical properties are listed and discussed as follows. We use standard PDMS-assisted release and transfer procedures of freestanding oxide membranes from both $SAO_C$ and $SAO_T$ (see Materials and Methods section for procedural details). To conduct a fair comparison, we fixed the thicknesses of both $SAO_C$ and $SAO_T$ ($t_{SAO-C}$ and $t_{SAO-T}$) layers at 30 nm. Notably, we found the quality of freestanding membrane also highly relies on the $t_{SAO-T}$. Therefore, we also included the characterizations of membranes exfoliated from 10 nm $SAO_T$, which shows the optimized membrane quality. This improvement could be attributed to both the enhanced structural coherency and slower release speed (see **Supplementary Section 4** for details) as $t_{SAO-T}$ decreases.

## 3.1 BaTiO₃ (BTO)

We first characterized the epitaxial quality and strain state of BTO (50 nm) films grown on $SAO_C$ and $SAO_T$. In this ferroelectric (FE) oxide, the ferroelastic domain transformation-induced superelasticity inherently accommodates stress and deformations generated during the lift-off process. For BTO film epitaxially grown on $SAO_C$, both the XRD $2\theta$–$\omega$ scans and RSMs (**Supplementary Fig. 21a-d**) signify the coexistence of $a$- and $c$-domains. Such an $ac$-domain structure persists in the released freestanding BTO membranes. Moreover, the RSM near LSAT(103) diffraction also signifies a relaxed strain state. The optical microscopic image (**Supplementary Fig. 21e**) displays a large-scale, crack-free, and wrinkled surface morphology. For BTO film grown on $SAO_T$, the XRD characterizations also demonstrate a satisfying epitaxial quality (**Supplementary Fig. 21f-i**) but a complete strain relaxation, similar to the BTO film grown on $SAO_C$. Moreover, as shown in **Supplementary Fig. 21j**, the optical microscopic image of the freestanding BTO membrane shows highly wrinkled morphology and even better integrity. The uncracked region is in millimeter-scale.

## 3.2 La₀.₇Ca₀.₃MnO₃ (LCMO)

The second oxide we tested is the rare-earth-doped manganite LCMO. As a typical strongly-correlated ferromagnetic metal, the physical properties of LCMO epitaxial films are sensitive to the strain states[19]. In addition, the LCMO is non-FE and thus does not process any ferroelastic domain-mediated superelasticity. The systematic structural characterizations are included in **Supplementary Figs. 22,23**. For the 35 nm thick LCMO film grown on $SAO_C$/LSAT(001), both the XRD characterizations (**Supplementary Fig. 22a-d** and **Supplementary Fig. 23a**) suggest a partially relaxed epitaxial strain at the LCMO/$SAO_C$ interface, which is consistent with the strain relaxation observed at $SAO_C$/LSAT(001) interface. As a result, the LCMO(004) diffractions of the exfoliated freestanding LCMO membrane are weak and broad, signifying poor crystallinity. Consistently, the optical microscopic image (**Supplementary Fig. 23d**) displays high-density cracks, which break the membrane into curvy stripes of ~10 μm wide. By contrast, the XRD characterizations from the LCMO/SAO$_T$(30 nm) bilayer (**Supplementary Fig. 22e-h** and **Supplementary Fig. 23b**) and LCMO/SAO$_T$(10 nm) bilayer (**Supplementary Fig. 23c**) demonstrate a coherent strain state and a superior epitaxial quality, which is comparable with the LCMO/LSAT(001) epitaxial films. The strong and sharp LCMO(004) Bragg diffraction of the freestanding LCMO membranes (**Supplementary Fig. 22g and Supplementary Fig. 23b,c**) confirms that the high crystallinity of LCMO prevails after water-assisted exfoliation. More surprisingly, the optical microscopic images of the LCMO membrane (**Supplementary Fig. 23e,f**) display a wrinkled but continuous morphology. The uncracked regions span from hundreds of micrometers to a few millimeters in scale, in sharp contrast with the cracked morphology observed in the membrane exfoliated from SAO$_C$. Note that reducing the $t_{SAO-T}$ can further enhance the membrane quality. As shown in **Supplementary Fig. 23g**, the membrane exfoliated from 10 nm SAO$_T$ appears highly wrinkled but has an almost crack-free morphology. Upon examining the high-resolution optical images over the entire 5×5 mm$^2$ membrane, we only observed a few crack regions near the top left corner (area #1) and the bottom edge (area #4). Such superior integrity should relate to the higher structural coherency of the LCMO/SAO$_T$ bilayer and slower release speed as $t_{SAO-T}$ decreases. And the release of compressive strain should be the main driving force of the wrinkled morphology.

We also characterized the representative physical properties of LCMO membranes, i.e., the ferromagnetic metallicity. For the LCMO/SAO$_T$/LSAT(001) epitaxial film, before exfoliation, the temperature-dependent magnetization (*M-T*) and resistivity (*ρ-T*) curves reveal a sharp paramagnetic-insulator (PMI) to ferromagnetic-metal (FMM) transition. The Curie temperature ($T_C$ ~ 267 K), saturated magnetization ($M_S$ ~ 3.5 $\mu_B$/Mn), and residual resistivity ($\rho_0$ ~ 2×10$^{-4}$ Ω·cm) are comparable to the bulk LCMO. After the water-assisted exfoliation, the sharpness and $T_C$ of the PMI-FMM transition in the corresponding LCMO membrane remain unchanged. The slight increase (decrease) of $\rho_0$ ($M_S$) is consistent with the continuous but wrinkled morphology. In contrast, for both the LCMO/SAO$_C$/LSAT(001) epitaxial film and corresponding LCMO membrane, the PMI-to-FMM transition becomes more slanted, and the key parameters ($T_C$, $M_S$, and $\rho_0$) also degrade significantly. These degradations can be attributed to the residual tensile strain at the LCMO/SAO$_C$ interface and the high-density cracks formed during exfoliation[19,20]. These results are shown in **Fig. 5** in the main text.

### 3.3 SrRuO$_3$ (SRO)

SRO is a representative itinerant ferromagnet. Due to its superior chemical stability and metallicity, it is widely used as the electrode layer in oxide-based heterostructures and electronic devices. The balance between electron-electron correlation and spin-orbital coupling also stimulates a variety of intriguing physical properties in SRO[21]. Therefore, achieving high-quality SRO freestanding membranes is essential in both application and fundamental research aspects. As shown in **Supplementary Fig. 24a-c**, the XRD 2θ–ω scans measured from SRO/SAO$_C$/LSAT(001) and SRO/SAO$_T$/STO(001) epitaxial films show strong SRO(220) diffraction peaks and sharp Laue fringes, signifying reasonable high epitaxial qualities. And the strong SRO(220) diffractions persist in the freestanding SRO membranes. The similar epitaxial quality for these three samples may relate to the small lattice mismatch among SRO, STO, SAO$_T$, and SAO$_C$. However, the optical microscopic images display drastic differences. As shown in **Supplementary Fig. 24d**, the SRO film released from SAO$_C$ exhibits high-density cracks. By contrast, the SRO film released by SAO$_T$ displays a wrinkled and continuous morphology (**Supplementary Fig. 24e,f**). The uncracked region spans from hundreds of micrometers to a few millimeters in scale. Notably, the membrane exfoliated from 10 nm SAO$_T$ shows highly wrinkled but almost crack-free morphology (**Supplementary Fig. 24g**). Upon examining the high-resolution optical images over the entire 5×5 mm$^2$ membrane, we only observed a few discrete cracks at the bottom left corner (area #4 in **Supplementary Fig. 24g**). This SRO membrane is even more wrinkled compared to the LCMO and STO counterparts (**Supplementary Fig. 23g** and **Fig. 25g**), probably originating from the release of larger compressive strain.

We then characterized the electrical transport and magnetic properties of SRO membranes, which shows a similar trend observed in freestanding LCMO membranes. For the SRO membrane exfoliated from SAO$_T$, the $M$-$T$ curves reveal a sharp FMM transition at ~150 K ($T_C$), accompanied by a strong perpendicular magnetic anisotropy (MA), which is dominated by the intrinsic magnetocrystalline anisotropy in SRO. The $\rho$-$T$ curve shows a clear Fermi-liquid to non-Fermi-liquid transition near $T_C$, and the residual resistance ratio (RRR) value is up to 4.83. Both the magnetic and electrical transport properties are comparable with high-quality SRO/STO(001) epitaxial films. In contrast, for the SRO membrane exfoliated from SAO$_C$, the $M$-$T$ curves show a demagnetization-dominated in-plane MA, which can be attributed to the degradation of crystallinity and the associated weak magnetocrystalline anisotropy. The residual resistance ratio (RRR) value reduces to 1.93, probably due to the high-density cracks. These results are shown in **Fig. 5** in the main text[22-24].

### 3.4 SrTiO$_3$ (STO)

STO is well-known as the "silicon in oxides", which is widely used as a substrate material and also consists of various intriguing functionalities[25]. We also characterized the structural integrity of freestanding STO membranes peeled from both SAO$_C$ and SAO$_T$. As shown in **Supplementary Fig. 25a-c**, the XRD $2\theta$–$\omega$ scans measured from STO/SAO$_C$/LSAT(001) and STO/SAO$_T$/STO(001) epitaxial films show strong STO(002) diffraction peaks and sharp Laue fringes, signifying reasonable high epitaxial qualities. And the strong STO(002) diffractions persist in the freestanding STO membranes released from SAO$_T$ (**Supplementary Fig. 25b,c**). Namely, the crystallinity of STO membranes released from SAO$_T$ is improved compared to the one released from SAO$_C$. The optical microscopic images measured from STO membranes also display drastic differences. As shown in **Supplementary Fig. 25d**, the STO film released from SAO$_C$ exhibits high-density cracks. By contrast, the STO film released by SAO$_T$ displays a wrinkled and continuous morphology (**Supplementary Fig. 25e,f**). The uncracked region spans from hundreds of micrometers to a few millimeters in scale. Notably, the membrane exfoliated from 10 nm SAO$_T$ appears highly wrinkled but crack-free morphology (**Supplementary Fig. 25g**). Upon examining the high-resolution optical images over the entire 5×5 mm$^2$ membrane, we only observed one discrete crack in area #3. Such superior integrity was rarely observed in the previously reported STO membranes.

### 3.5 LaNiO$_3$ (LNO) and NdNiO$_3$ (NNO)

LNO and NNO are typical perovskite-structured nickelates. LNO is metallic, while NNO is a typical charge-transfer insulator, which exhibits a metal-insulator transition (MIT) upon cooling[26]. Moreover, LNO and NNO have relatively small $a_p$ of 3.853 and 3.808 Å, resulting in considerate lattice mismatches with both SAO$_T$ and SAO$_C$. To test the feasibility of SAO$_T$ for oxides with relatively small $a_p$, we characterized the structural integrity of freestanding LNO and NNO membranes peeled from both SAO$_C$ and SAO$_T$. The XRD 2$\theta$–$\omega$ scans measured from LNO/SAO$_C$/LSAT(001) and LNO/SAO$_T$/LSAT(001) epitaxial films are shown in **Supplementary Fig. S26a-c**, respectively. The XRD 2$\theta$–$\omega$ linear scans measured from LNO/SAO$_C$/LSAT(001) and LNO/SAO$_T$/LSAT(001) epitaxial films show strong LNO(004) diffraction peaks, signifying high epitaxial qualities. And the strong LNO(004) diffractions persist in the freestanding membranes. However, the optical microscopic images measured from LNO membranes display drastic differences. As shown in **Supplementary Fig. 26d**, the LNO film released from SAO$_C$ exhibits high-density cracks, breaking the membranes into stripes less than 10 μm wide. By contrast, the LNO film released by SAO$_T$ displays significantly improved integrity. For the membrane released from 30 nm SAO$_T$, the uncracked region spans up to hundreds of micrometers in scale (**Supplementary Fig. 26e**). By further reducing the $t_{SAO-T}$ to 10 nm, the uncracked region can reach a millimeter scale (**Supplementary Fig. 26f,g**). Note that the LNO membranes do not show obvious wrinkling features as the LCMO, STO, and SRO films, which are consistent with the tensile strain state of the LNO/SAO$_T$ interfaces.

The XRD 2$\theta$–$\omega$ scans and optical microscopic images measured from NNO/SAO$_C$/LSAT(001) and NNO/SAO$_T$/LSAT(001) epitaxial films are shown in **Supplementary Fig. 27**. The crystallinities of NNO/SAO$_T$/LSAT(001) epitaxial films (**Supplementary Fig. 27b,c**) are better compared to the one grown on SAO$_C$/LSAT(001) (**Supplementary Fig. 27a**). The improvement in crystallinity becomes more obvious in the released freestanding NNO membranes. The optical microscopic image measured from the NNO membrane released from SAO$_C$ also exhibits high-density cracks (**Supplementary Fig. 27d**), while the crack density reduces significantly in the membrane released from 30 nm SAO$_T$ (**Supplementary Fig. 27e**). As aforementioned, NNO has rather large lattice mismatches with both SAO$_C$ and SAO$_T$. Therefore the integrity of freestanding NNO membranes released from 30 nm SAO$_T$ is not as good as the aforementioned LNO, LCMO, SRO, and STO membranes. As shown in **Supplementary Fig. 27g**, for the membrane released from 10 nm SAO$_T$, the uncracked regions can reach a few millimeters in scale. This improvement could be attributed to both the enhanced structural coherency and slower release speed as $t_{SAO-T}$ decreases.

We also characterized the electrical transport of the freestanding LNO and NNO membranes. For the LNO/SAO$_C$/LSAT(001) epitaxial film, the temperature-dependent resistivity ($\rho$-$T$) curve shows an upturn at ~90 K (**Supplementary Fig. 29a**). This unexpected MIT can be explained by the weak-localization effect, which is related to the high-density defects generated during misfit strain relaxation at the LNO/SAO$_C$ interface. After being released from SAO$_C$, the overall $\rho$ of the freestanding membrane increases drastically, consistent with the observed degradations in crystallinity and integrity. In contrast, for LNO/SAO$_T$/LSAT(001) epitaxial film, the $\rho$-$T$ curve signifies robust metallicity down to 10 K (**Supplementary Fig. 29b**). And the strong metallicity persists in the released freestanding membranes. The overall $\rho$ only increases slightly. These results strongly suggest that SAO$_T$ can effectively improve the metallicity of LNO membranes, consistent with the improvements in crystallinity and integrity.

For the NNO/SAO$_C$/LSAT(001) epitaxial film, the $\rho$-$T$ curve reveals a rather slanted MIT at ~180 K, accompanied by a small thermal hysteresis (**Supplementary Fig. 30a**). The degradation in MIT can be attributed to the partially relaxed misfit strain at the NNO/SAO$_C$ interface and associated degradation in crystallinity. For the NNO/SAO$_T$/LSAT(001) epitaxial film, the $\rho$-$T$ curve shows a sharper MIT and a larger thermal hysteresis (**Supplementary Fig. 30b**), which is consistent with the previously reported NNO/LSAT(001) epitaxial films under a slight tensile strain[27]. After being released from SAO$_C$, the high-density cracks in freestanding NNO membranes make the electrical transport measurements extremely challenging. We cannot measure a reliable $\rho$-$T$ curve from the SAO$_C$-released membrane. For the NNO membrane released from SAO$_T$, the improved integrity enables a reliable electrical transport measurement. The $\rho$-$T$ curve reveals a sharp MIT, accompanied by a significant increment in $\rho$ by over three orders of magnitudes. And the thermal hysteresis also becomes smaller. This $\rho$-$T$ behavior is similar to the strain-free NNO thick films.

### 3.6 SrSnO$_3$ (SSO)

SSO is a typical perovskite-structured wide-band-gap semiconductors. By introducing La doping, it can become perovskite-structured transparent conductive oxide layers[28]. Moreover, SSO has a large $a_p$ of 4.035 Å, which can result in large lattice mismatches with both SAO$_T$ and SAO$_C$ (**Supplementary Table 3**). To test the feasibility of the SAO$_T$ sacrificial layer for oxides with relatively large $a_p$, we also characterized the SSO membranes exfoliated from SAO$_C$ and SAO$_T$. As shown in **Supplementary Fig. 28a-c**, the XRD $2\theta$–$\omega$ scans measured from SSO/SAO$_C$/LSAT(001) and SSO/SAO$_T$/STO(001) epitaxial films show strong SSO(004) diffraction peaks and sharp Laue fringes, signifying reasonable high epitaxial qualities. The strong SSO(004) diffractions persist in the freestanding membranes released from SAO$_T$. The optical microscopic images display drastic differences. As shown in **Supplementary Fig. 28d**, the SSO membrane released from 30 nm SAO$_C$ exhibits high-density cracks, which break the membrane into rectangular pieces of several tens of micrometers in dimension. By contrast, the SSO membrane released from 30 nm SAO$_T$ displays significant improvement in integrity (**Supplementary Fig. 28e**). The uncracked region spans up to hundreds of micrometers in scale. By further reducing the $t_{SAO-T}$ to 10 nm, the uncracked region can reach a few millimeters in scale (**Supplementary Fig. 28f,g**).

|  | Lattice Constant (Å) | Lattice mismatch (%) | | |
| --- | --- | --- | --- | --- |
|  |  | $SAO_C$ | $SAO_T$/LSAT | $SAO_T$/STO |
| **$NdNiO_3$** | 3.808 | 4.02 | 1.63 | - |
| **$LaNiO_3$** | 3.853 | 2.80 | 0.44 | - |
| **$La_{0.7}Ca_{0.3}MnO_3$** | 3.874 | 2.51 | -0.16 | - |
| **$SrTiO_3$** | 3.905 | 1.43 | - | 0 |
| **$SrRuO_3$** | 3.932 | 0.79 | - | -0.69 |
| **$BaTiO_3$** | 3.998 | -0.93 | - | -2.33 |
| **$SrSnO_3$** | 4.035 | -1.83 | - | -3.22 |

**Supplementary Table 3. | Lattice constants (bulk values, in pseudocubic perovskite notation) of various functional oxides for preparing freestanding oxide membranes, and the lattice mismatches between these oxides and $SAO_T$/$SAO_C$.** For $SAO_C$, since the films are strain-relaxed, the lattice mismatch is calculated using the bulk $a_{SAO-C}$ = 3.961 Å. For $SAO_T$, however, all the films are coherently strained to the substrate. Thus, the lattice mismatches between $SAO_T$ and target oxides are calculated using either $a_{SAO-T}$ = $a_{LSAT}$ = 3.868 Å or $a_{SAO-T}$ = $a_{STO}$ = 3.905 Å, depending on the substrate used.

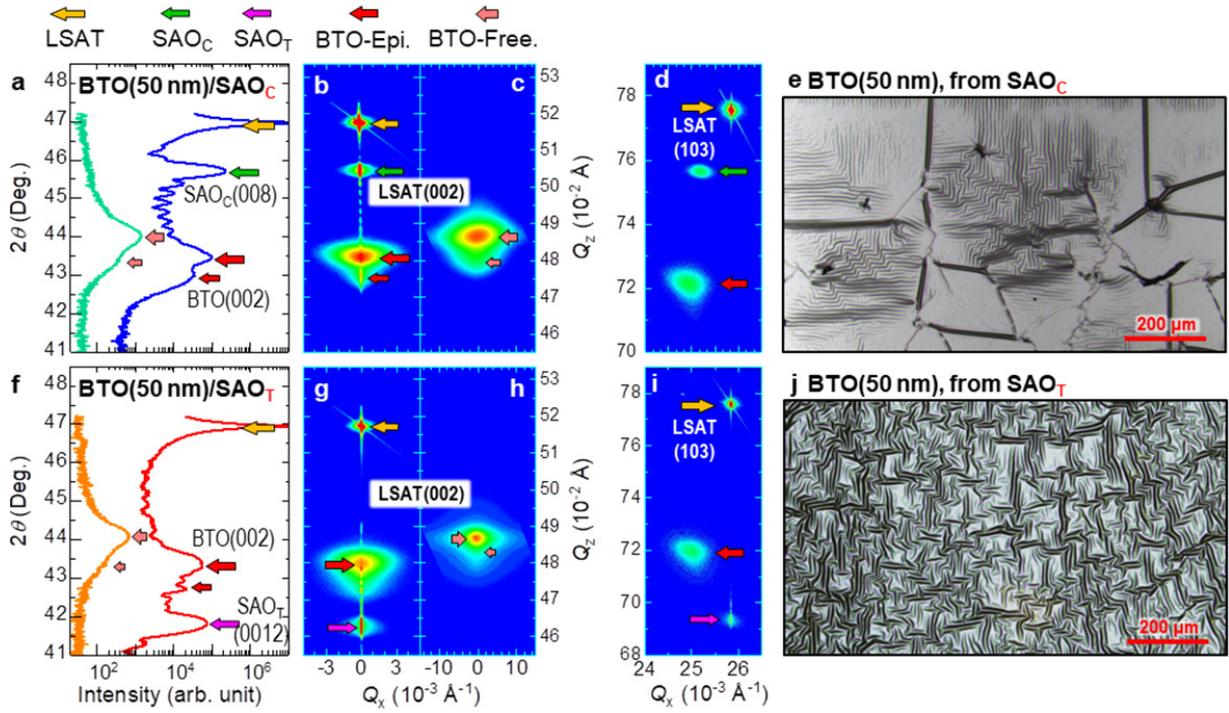

**Supplementary Fig. 21. | Structural characterizations of BaTiO$_3$ (BTO) epitaxial films and freestanding membranes.** (**a-d**) XRD 2$\theta$-$\omega$ linear scans (**a**), RSMs around BTO(002) diffractions (**b,c**), and RSM around BTO(103) diffraction (**d**) measured from BTO epitaxial films (BTO-Epi.) and freestanding membranes (BTO-Free.) grown on/exfoliated from SAO$_C$/LSAT(001). On top of (**a-d**), Bragg diffractions assigned to LSAT, SAO$_C$, SAO$_T$, BTO-Epi., and BTO-Free are marked by open arrows with different sizes and colors. (**f-i**) XRD 2$\theta$-$\omega$ linear scans (f), RSMs around BTO(002) Bragg diffractions (**g,h**), and RSM around BTO(103) diffraction (**i**) measured from BTO-Epi and BTO-Free. grown on/exfoliated from SAO$_T$/LSAT(001). (**e,j**) Optical microscopic images of BTO freestanding membranes exfoliated from SAO$_C$ (e) and SAO$_T$ (j).

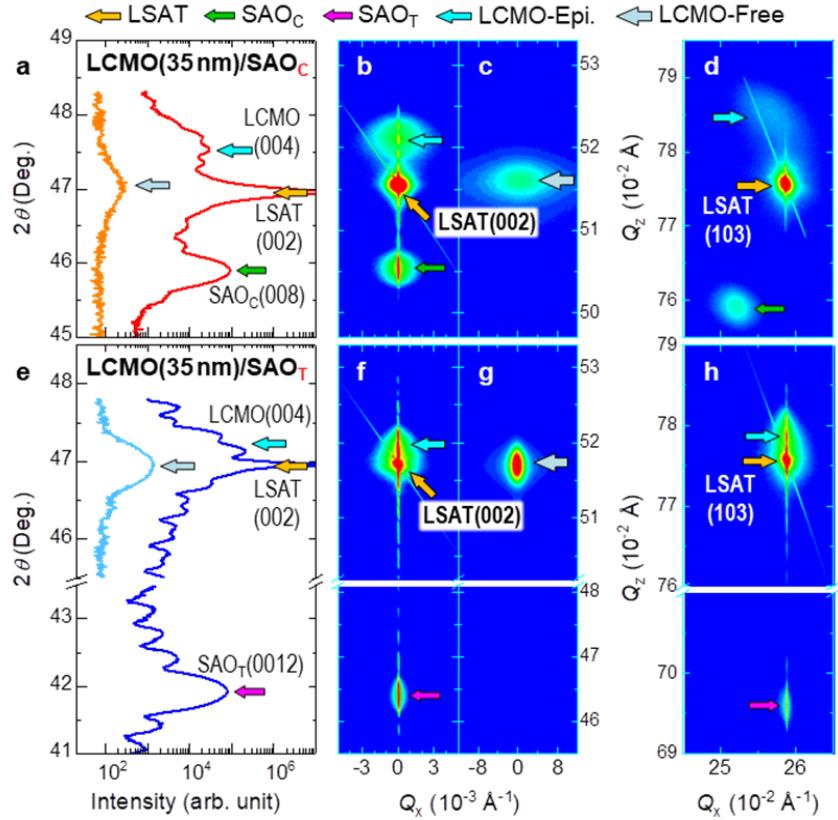

**Supplementary Fig. 22. | Strain states and epitaxial qualities of 35 nm thick LCMO epitaxial films and freestanding membranes.** (**a-d**) XRD $2\theta\text{-}\omega$ linear scans (**a**), RSMs around LCMO(004) diffractions (**b,c**), and RSM around LCMO(116) diffraction (**d**) measured from LCMO epitaxial film (LCMO-Epi.) and freestanding membranes (LCMO-Free.) grown on/exfoliated from $SAO_C$/LSAT(001). On top of (**a-d**), Bragg diffractions assigned to LSAT, $SAO_C$, $SAO_T$, LCMO-Epi., and LCMO-Free are marked by open arrows with different sizes and colors. The LCMO(004) Bragg diffraction spots shown in (**a**) and (**b**) are weak and diffused, implying a degradation in crystallinity for both LCMO-Epi. and LCMO-Free. In (**c**), the LCMO(116) diffraction is also diffused. The in-plane reciprocal space vector ($Q_x$) values span between the corresponding ones of LSAT(103) and $SAO_C$(4012), which strongly suggests the LCMO film grown on $SAO_C$ is partially strain-relaxed. (**e-h**) XRD $2\theta\text{-}\omega$ linear scans (**e**), RSMs around LCMO(004) Bragg diffractions (**f,g**), and RSM around LCMO(116) diffraction (**h**) measured from LCMO-Epi. and LCMO-Free. grown on/exfoliated from $SAO_T$/LSAT(001). The LCMO Bragg diffraction spots in (**f-h**) are sharp and strong, suggesting a superior crystallinity for both LCMO-Epi. and LCMO-Free. In (**h**), the $Q_x$ values of LCMO(116) and $SAO_T$(2218) are the same as the one of LSAT(103) substrate, which confirms that the LCMO-Epi./$SAO_T$ bilayer is coherently strained with the LSAT(001) substrate.

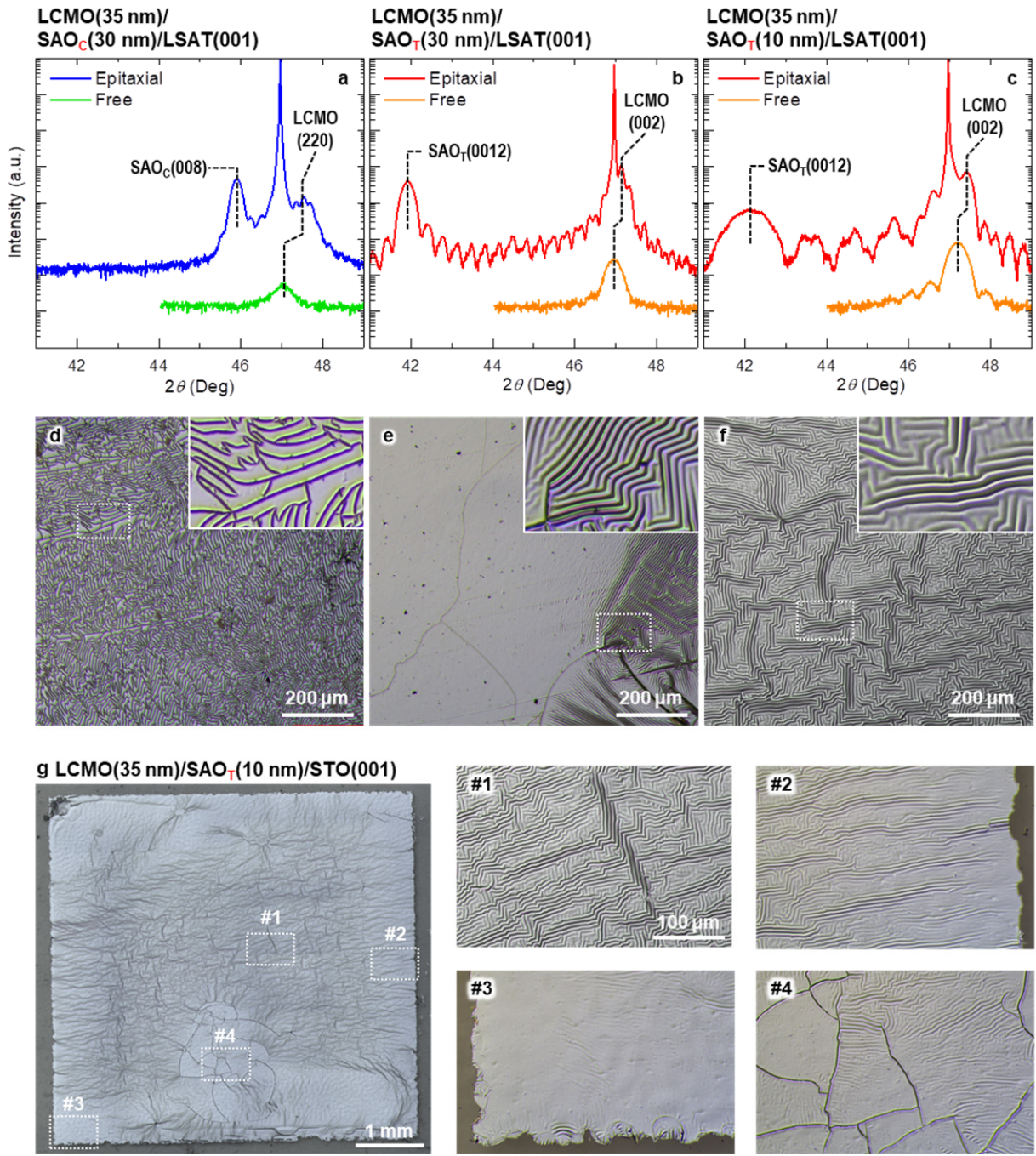

**Supplementary Fig. 23. | Structure characterizations of 35 nm thick LCMO epitaxial films and freestanding membranes.** (**a-c**) XRD $2\theta$-$\omega$ linear scans of 35 nm thick $La_{0.7}Ca_{0.3}MnO_3$ (LCMO) epitaxial films and freestanding membranes exfoliated from 30 nm thick $SAO_C$ (**a**), 30 nm thick $SAO_T$ (**b**), and 10 nm thick $SAO_T$ (**c**). (**d-f**) Corresponding optical microscopic images of these 35 nm thick LCMO freestanding membranes. The insets are zoomed-in images taken from the regions marked in dashed boxes. (**g**) Large-scale optical microscopic image of the 35 nm thick LCMO freestanding membranes released from 10 nm $SAO_T$. The 4 zoom-in images taken from the areas marked as #1 ~ #4 are also inserted.

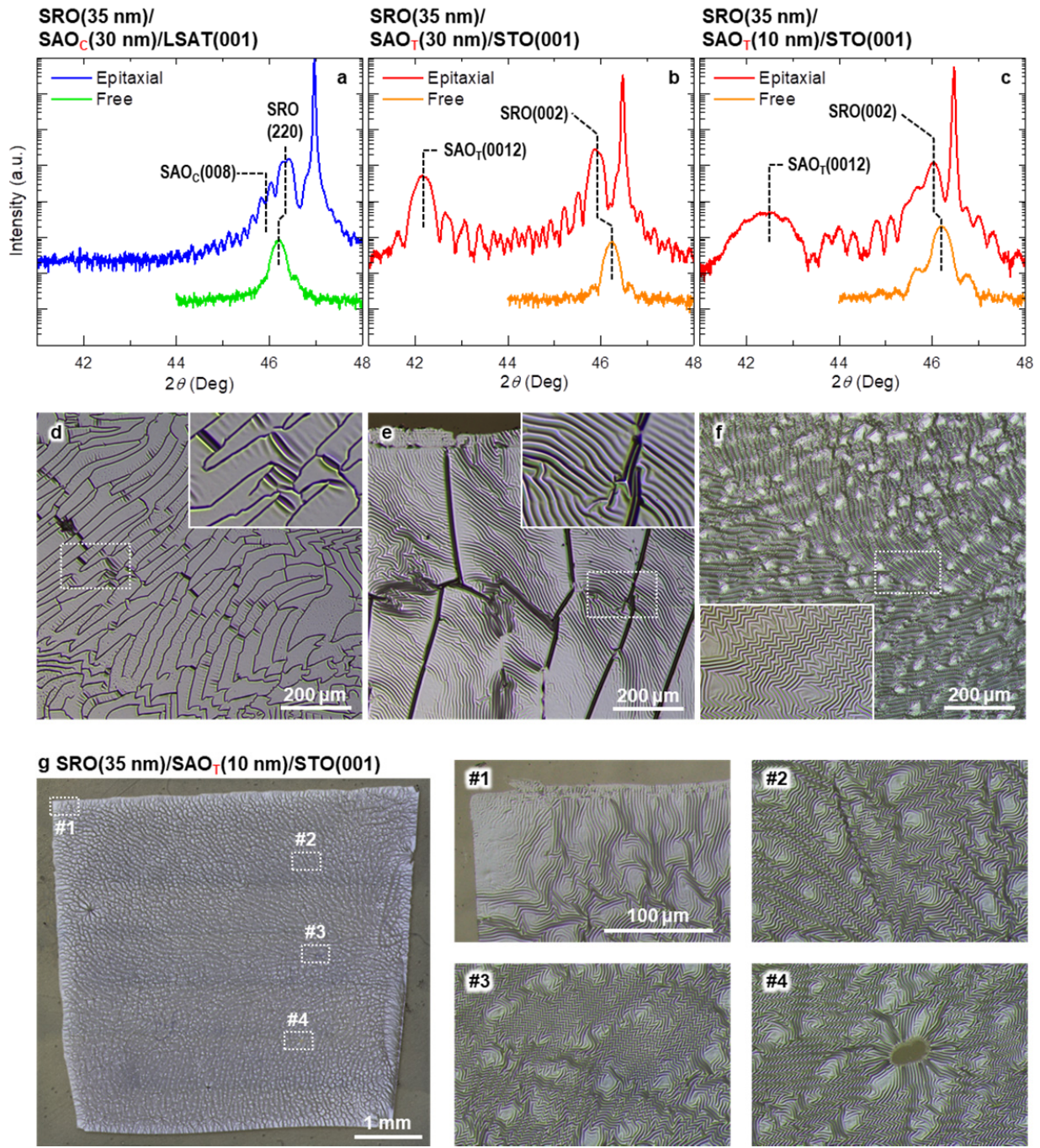

**Supplementary Fig. 24. | Structure characterizations of 35 nm thick SrRuO₃ epitaxial films and freestanding membranes.** (a-c) XRD 2θ-ω linear scans of 35 nm thick SrRuO₃ (SRO) epitaxial films and freestanding membranes exfoliated from 30 nm thick SAO$_C$ (a), 30 nm thick SAO$_T$ (b), and 10 nm thick SAO$_T$ (c). (d-f) Corresponding optical microscopic images of these 35 nm thick SRO freestanding membranes. The insets are zoomed-in images taken from the regions marked in dashed boxes. (g) Large-scale optical microscopic image of the 35 nm thick SRO freestanding membranes released from 10 nm SAO$_T$. The 4 zoom-in images taken from the areas marked as #1 ~ #4 are also inserted.

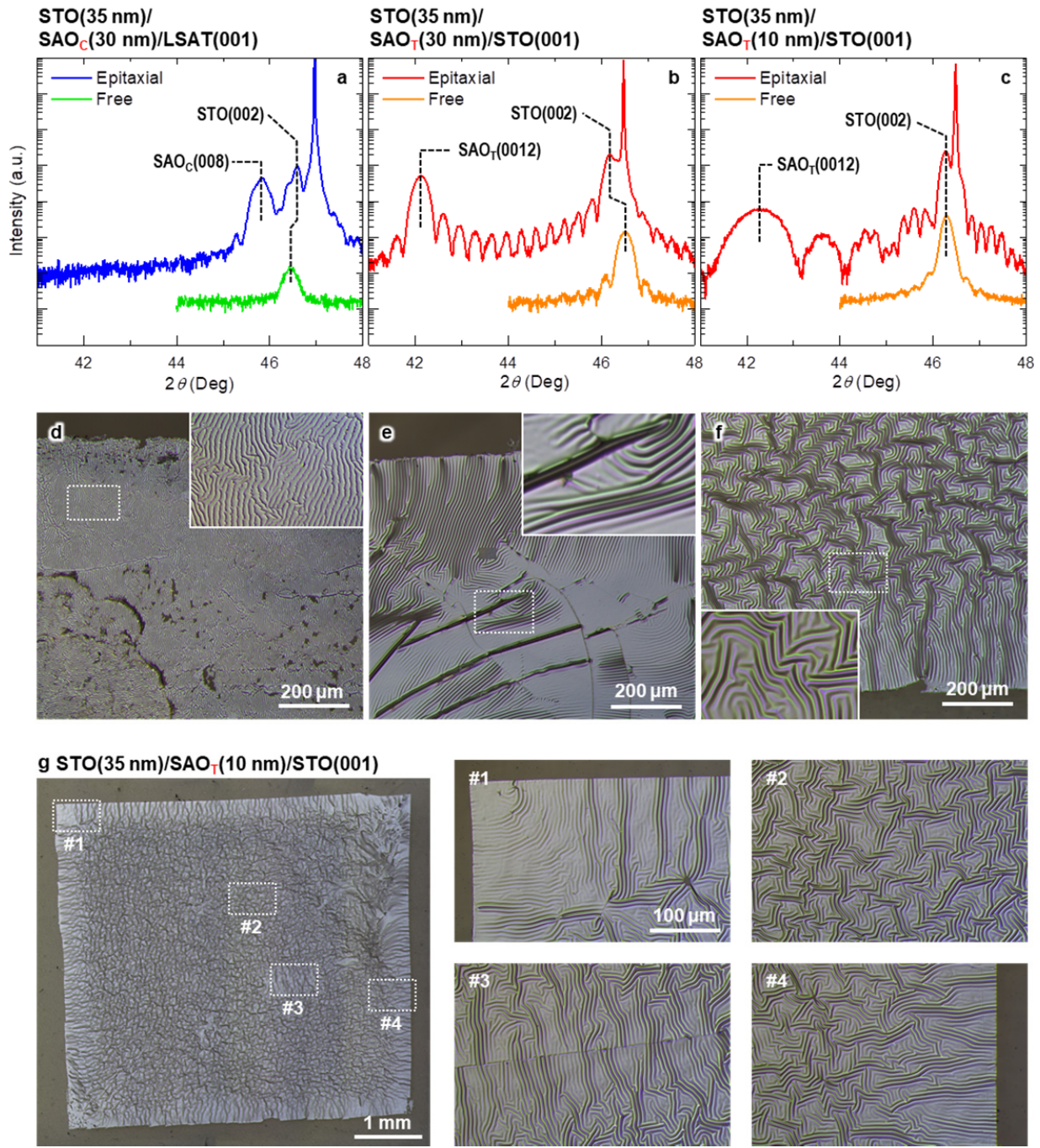

**Supplementary Fig. 25. | Structure characterizations of 35 nm thick SrTiO₃ epitaxial films and freestanding membranes.** (**a-c**) XRD $2\theta$-$\omega$ linear scans of 35 nm thick SrTiO₃ (STO) epitaxial films and freestanding membranes exfoliated from 30 nm thick $SAO_C$ (**a**), 30 nm thick $SAO_T$ (**b**), and 10 nm thick $SAO_T$ (**c**). (**d-f**) Corresponding optical microscopic images of these 35 nm thick STO freestanding membranes. The insets are zoomed-in images taken from the regions marked in dashed boxes. (**g**) Large-scale optical microscopic image of the 35 nm thick STO freestanding membranes released from 10 nm $SAO_T$. The 4 zoom-in images taken from the areas marked as #1 ~ #4 are also inserted.

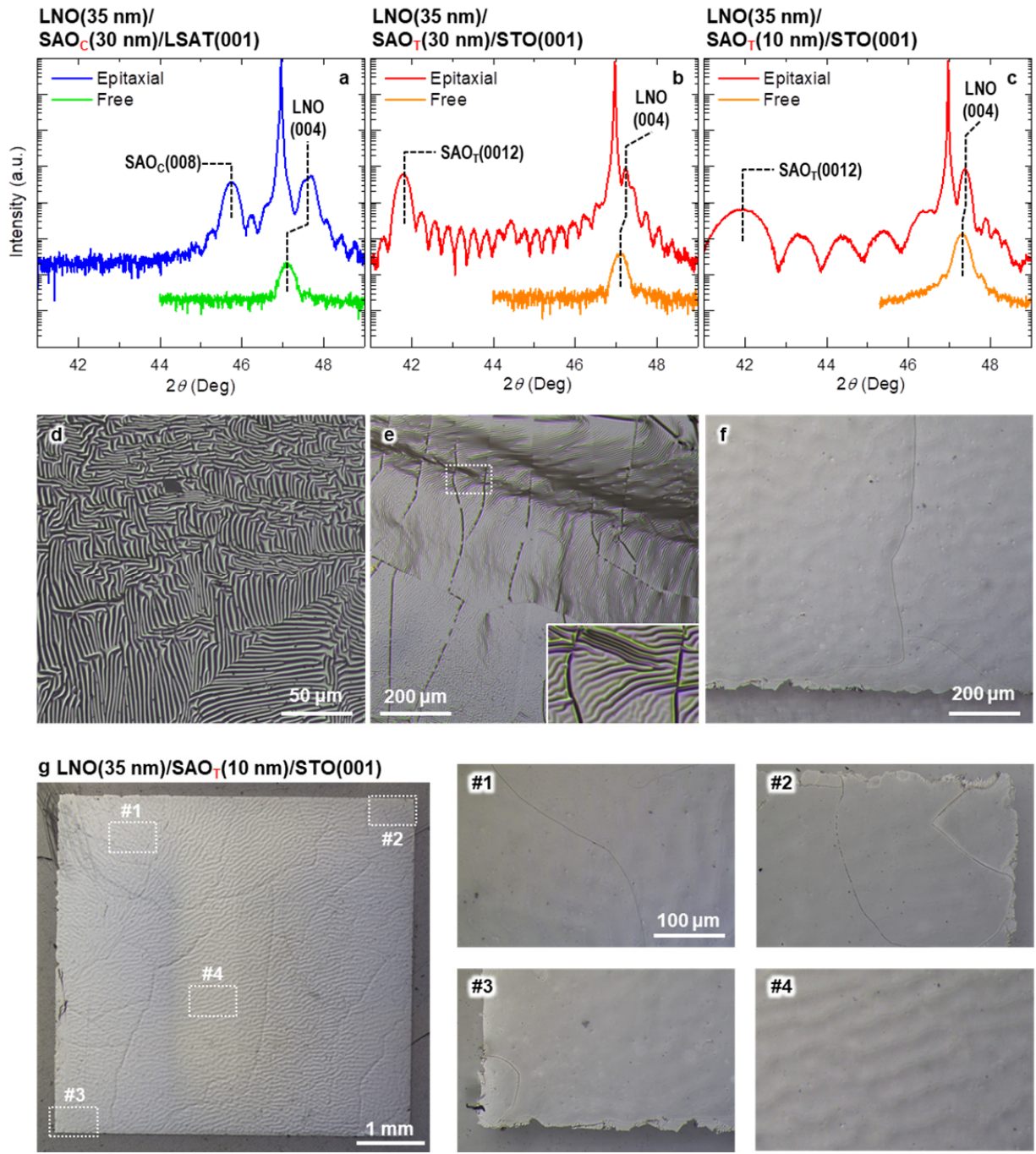

**Supplementary Fig. 26. | Structure characterizations of 35 nm thick LaNdO$_3$ epitaxial films and freestanding membranes.** (**a-c**) XRD 2$\theta$-$\omega$ linear scans of 35 nm thick LaNdO$_3$ (LNO) epitaxial films and freestanding membranes exfoliated from 30 nm thick SAO$_C$ (**a**), 30 nm thick SAO$_T$ (**b**), and 10 nm thick SAO$_T$ (**c**). (**d-f**) Corresponding optical microscopic images of these 35 nm thick LNO freestanding membranes. The insets are zoomed-in images taken from the regions marked in dashed boxes. (**g**) Large-scale optical microscopic image of the 35 nm thick LNO freestanding membranes released from 10 nm SAO$_T$. The 4 zoom-in images taken from the areas marked as #1 ~ #4 are also inserted.

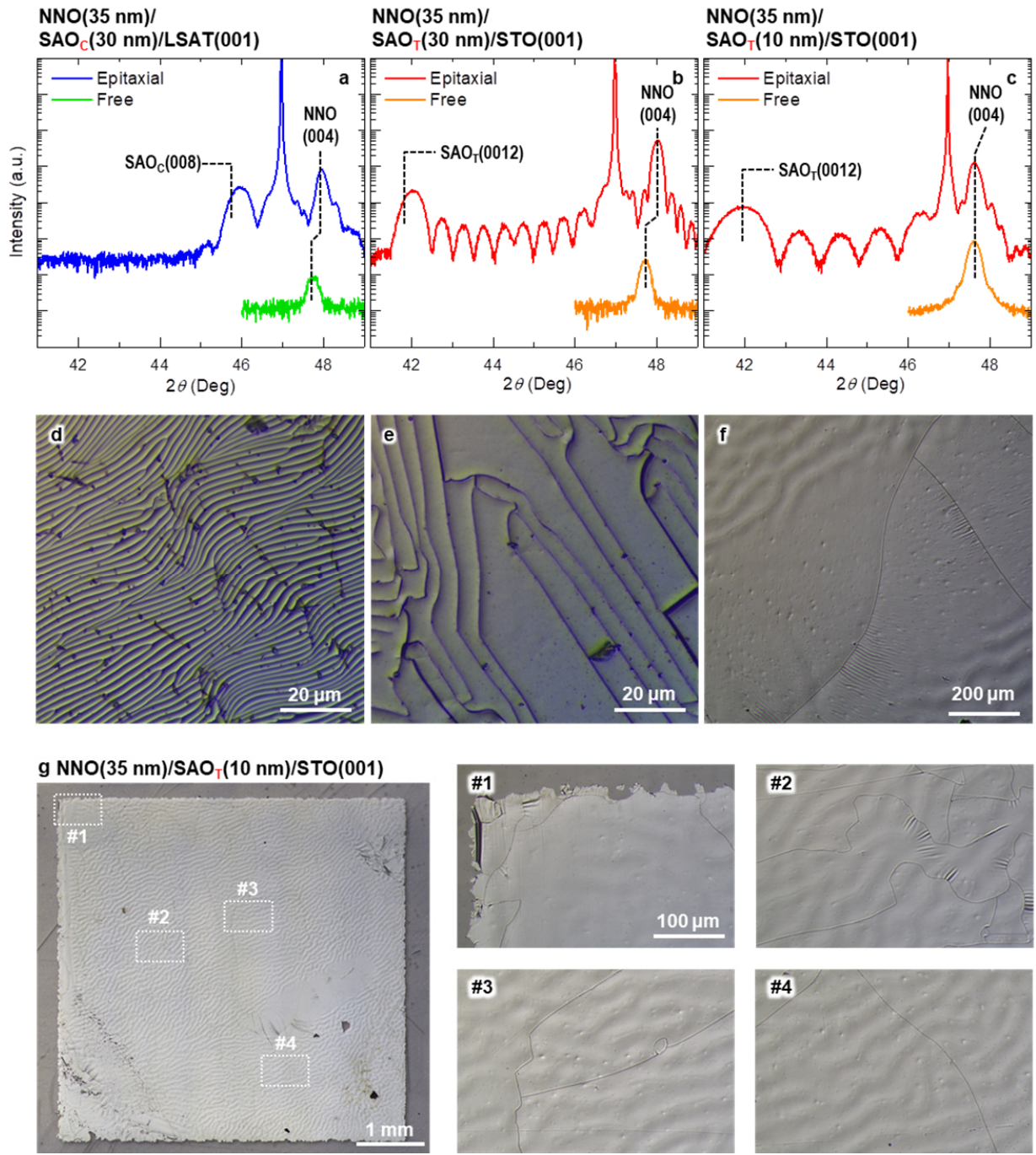

**Supplementary Fig. 27. | Structure characterizations of 35 nm thick NdNdO$_3$ epitaxial films and freestanding membranes.** (**a-c**) XRD 2$\theta$-$\omega$ linear scans of 35 nm thick NdNdO$_3$ (NNO) epitaxial films and freestanding membranes exfoliated from 30 nm thick SAO$_C$ (**a**), 30 nm thick SAO$_T$ (**b**), and 10 nm thick SAO$_T$ (**c**). (**d-f**) Corresponding optical microscopic images of these 35 nm thick NNO freestanding membranes. The insets are zoomed-in images taken from the regions marked in dashed boxes. (**g**) Large-scale optical microscopic image of the 35 nm thick NNO freestanding membranes released from 10 nm SAO$_T$. The 4 zoom-in images taken from the areas marked as #1 ~ #4 are also inserted.

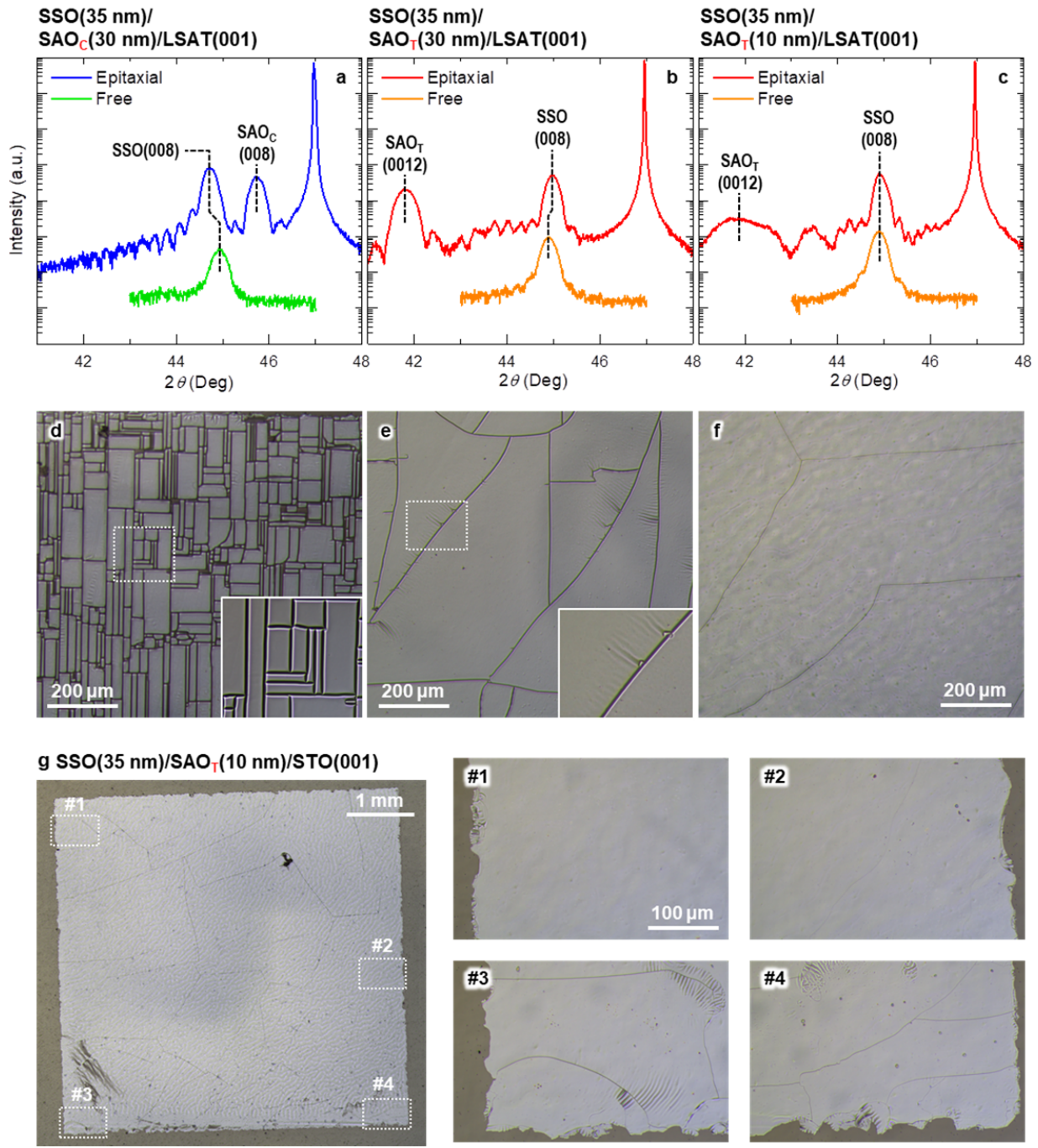

**Supplementary Fig. 28. | Structure characterizations of 35 nm thick SrSnO₃ epitaxial films and freestanding membranes.** (**a-c**) XRD 2θ-ω linear scans of 35 nm thick SrSnO$_3$ (SSO) epitaxial films and freestanding membranes exfoliated from 30 nm thick SAO$_C$ (**a**), 30 nm thick SAO$_T$ (**b**), and 10 nm thick SAO$_T$ (**c**). (**d-f**) Corresponding optical microscopic images of these 35 nm thick SSO freestanding membranes. The insets are zoomed-in images taken from the regions marked in dashed boxes. (**g**) Large-scale optical microscopic image of the 35 nm thick SSO freestanding membranes released from 10 nm SAO$_T$. The 4 zoom-in images taken from the areas marked as #1 ~ #4 are also inserted.

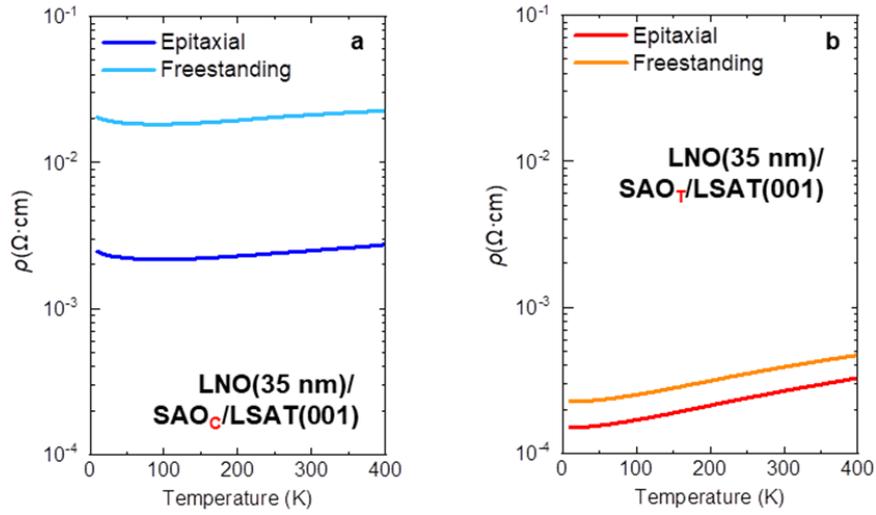

**Supplementary Fig. 29. | Characterizations on the electrical transport properties of 35 nm thick LNO epitaxial films and freestanding membranes.** (**a**) $\rho$-$T$ curve of LNO epitaxial film and freestanding membranes grown on/exfoliated from $SAO_C$/LSAT(001). (**b**) $\rho$-$T$ curves of NNO epitaxial film and freestanding membranes grown on/exfoliated from $SAO_T$/LSAT(001).

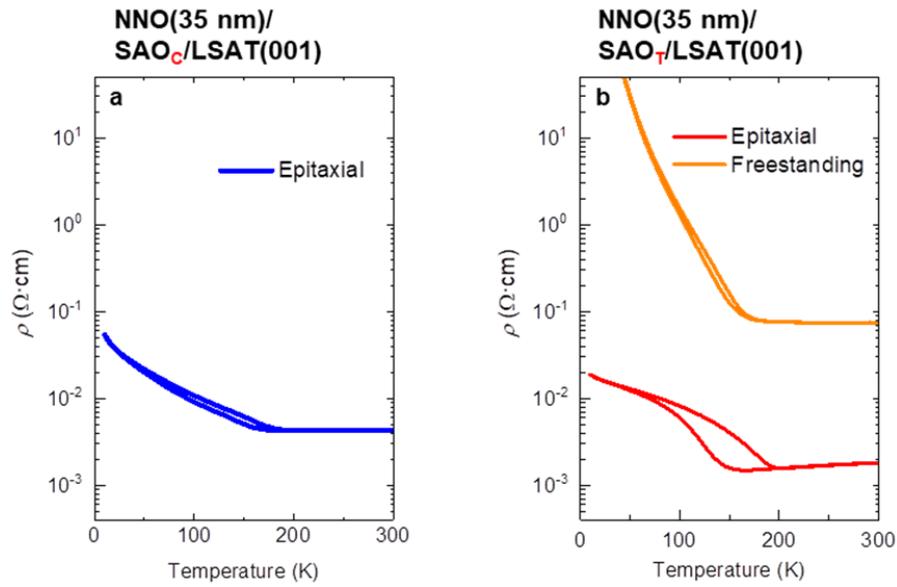

**Supplementary Fig. 30. | Characterizations on the electrical transport properties of 35 nm thick NNO epitaxial films and freestanding membranes.** (**a**) $\rho$-$T$ curve of NNO epitaxial film grown on $SAO_C$/LSAT(001). We cannot measure a reliable $\rho$-$T$ curve from the corresponding freestanding membrane due to the formation of high-density cracks. (**b**) $\rho$-$T$ curves of NNO epitaxial film and freestanding membranes grown on/exfoliated from $SAO_T$/LSAT(001).

# Section 4: Water-solubility and stability test of the SAO$_T$ epitaxial films

From the perspective of water-soluble sacrificial layers, water-solubility is one of the most important properties. In this section, we performed a comparison study on the SAO$_T$ and SAO$_C$ films about the water-solubility and chemical stability against moisture.

We first choose to use BTO films to compare the water-solubility of SAO$_T$ and SAO$_C$ films. The BTO films are FE and ferroelastic. The resultant superelasticity can ensure the crystallinity and integrity of the freestanding BTO membranes exfoliated from both SAO$_T$ and SAO$_C$ (See **Supplementary Fig. 21** in **Supplementary Section 3**). This feature makes it an ideal material to compare the water-solubility of SAO$_T$ and SAO$_C$ films.

As shown in **Supplementary Fig. 31**, we measured a series of optical microscopic images from the BTO(50 nm)/SAO$_T$(30 nm)/LSAT(001) [**Supplementary Fig. 31a**] and BTO(50 nm)/SAO$_C$(30 nm)/LSAT(001) [**Supplementary Fig. 31b**] bilayer heterostructures in-situ during the peeling-off processes in deionized water. The corresponding videos for the peeling-off processes of SAO$_T$ and SAO$_C$ are included in **Supplemental Movie 1** and **Movie 2**. As soon as dipping the bilayers in water, the SAO layers start to dissolve. The dissolution releases the BTO film on top, leading to a wrinkled surface morphology. As a result, the released region turns dark in the optical microscopic image, while the bonded region remains bright. For the SAO$_T$ case, the dissolved region almost expands over the entire image in 3 minutes. For the SAO$_C$ case, by contrast, more than 20% region remains undissolved even after 20 minutes. Namely, the membrane release using SAO$_T$ is nearly one order of magnitude faster than that using SAO$_C$ (**Supplementary Fig. 31c**).

We also monitored the peeling-off processes of other oxides from SAO$_T$ and SAO$_C$. The time durations for fully releasing various oxide membranes are summarized in **Supplementary Fig. 31d**. For all of the oxides we examined, the exfoliations from a 30 nm thick SAO$_T$ are much faster than those from SAO$_C$. Given a 5×5 mm$^2$ sized film sample, complete exfoliation from SAO$_T$ takes a few to ten minutes only, while the exfoliation from a 30 nm thick SAO$_C$ takes tens of minutes to hours. For other publications, the typical time scale can be as long as tens of hours to days (**Supplementary Table 4**). Compared to the (Sr,Ca)$_3$Al$_2$O$_6$ compounds with smaller lattice constants and poorer water-solubility (*13*), the advantage of exfoliation speed in SAO$_T$ could be even more prominent.

In addition, as shown in **Supplementary Fig. 32a**, the release duration of oxide membranes also strongly depends on the SAO film thickness ($t_{SAO}$). And the release speed using SAO$_T$ is much faster than that using SAO$_C$ for the entire $t_{SAO}$ range. As shown in **Supplementary Fig. 32b**, we also found the release duration using BSAO$_T$ (CSAO$_T$) is faster (slower) than that using SAO$_T$ film. The trend is similar to the (Ca,Sr,Ba)$_3$Al$_2$O$_6$ compounds. These results further attest the higher water solubility of $SAO_T$, and it also provides another independent parameter for controlling the dissolution speed and further optimizing the freestanding membrane quality.

The high water-solubility of $SAO_T$ can be understood based on its unique atomic structure. As illustrated in **Supplementary Fig. 33a-c**, the key aspect of the water-solubility of $SAO_C$ is the Al-O network that consists of discrete 12-membered $Al_6O_{18}^{18-}$ rings. These rings can hydrolyze in water. As shown in **Supplementary Fig. 33d-f**, the Al-O network of $SAO_T$ consists of more isolated $AlO_4^{5-}$ and $Al_3O_{10}^{11-}$ groups. Compared to the $Al_6O_{18}^{18-}$ rings in $SAO_C$, these discrete chemical groups are expected to be hydrolyzed in water more easily[14,15], thus leading to higher water-solubility.

We then turn to check the stability of $SAO_T$ and $SAO_C$ films against moisture in the ambient. As shown in **Supplementary Fig. 34a,b**, we first measured a series of the XRD $2\theta$-$\omega$ scans from the $SAO_T$/LSAT(001) and $SAO_C$/LSAT(001) single layer films stored in ambient (24 °C, relative humidity of 65%) for different durations. The peak intensity of $SAO_T$(0012) or $SAO_C$(008) diffraction is a reliable indicator of the structure and chemical stabilities. Without any capping layer, the $SAO_T$ film structure degrades dramatically within 1 hour when stored in the ambient, while the $SAO_C$ film remains intact after 12 hours of storage in the ambient. This trend is consistent with the high water-solubility of the $SAO_T$ phase as depicted in **Supplementary Figs. 31 and 32**. We also checked the stability of SAO films with a 2 nm thick STO capping layer (**Supplementary Fig. 34c,d**). In this case, the XRD $2\theta$-$\omega$ scan of the $SAO_T$ film surprisingly remains intact even after 40 days of storage in ambient, while the structure degradation of $SAO_C$ films occurs shortly after 5 days. Such a surprisingly reversed trend should relate to the extraordinarily high epitaxial quality of $SAO_T$ film and STO capping layer grown on top, which prevents the penetration of humidity. On the contrary, we speculate that the large lattice mismatches between STO, $SAO_C$, and LSAT(001) substrate will create high-density defects (i.e. dislocations) at the heterointerfaces, which helps the humidity in the ambient to gradually dissolve the $SAO_C$ layer.

In brief, compared to the $SAO_C$ phase, the new $SAO_T$ phase shows higher water-solubility, which enables a more efficient exfoliation of freestanding oxide membranes in water. Moreover, the high epitaxial qualities of $ABO_3$/$SAO_T$ heterostructures also promote the structure and chemical stabilities against moisture in the ambient. These appealing features further suggest that the $SAO_T$ film could be a more suitable water-soluble sacrificial layer for perovskite oxide membranes.

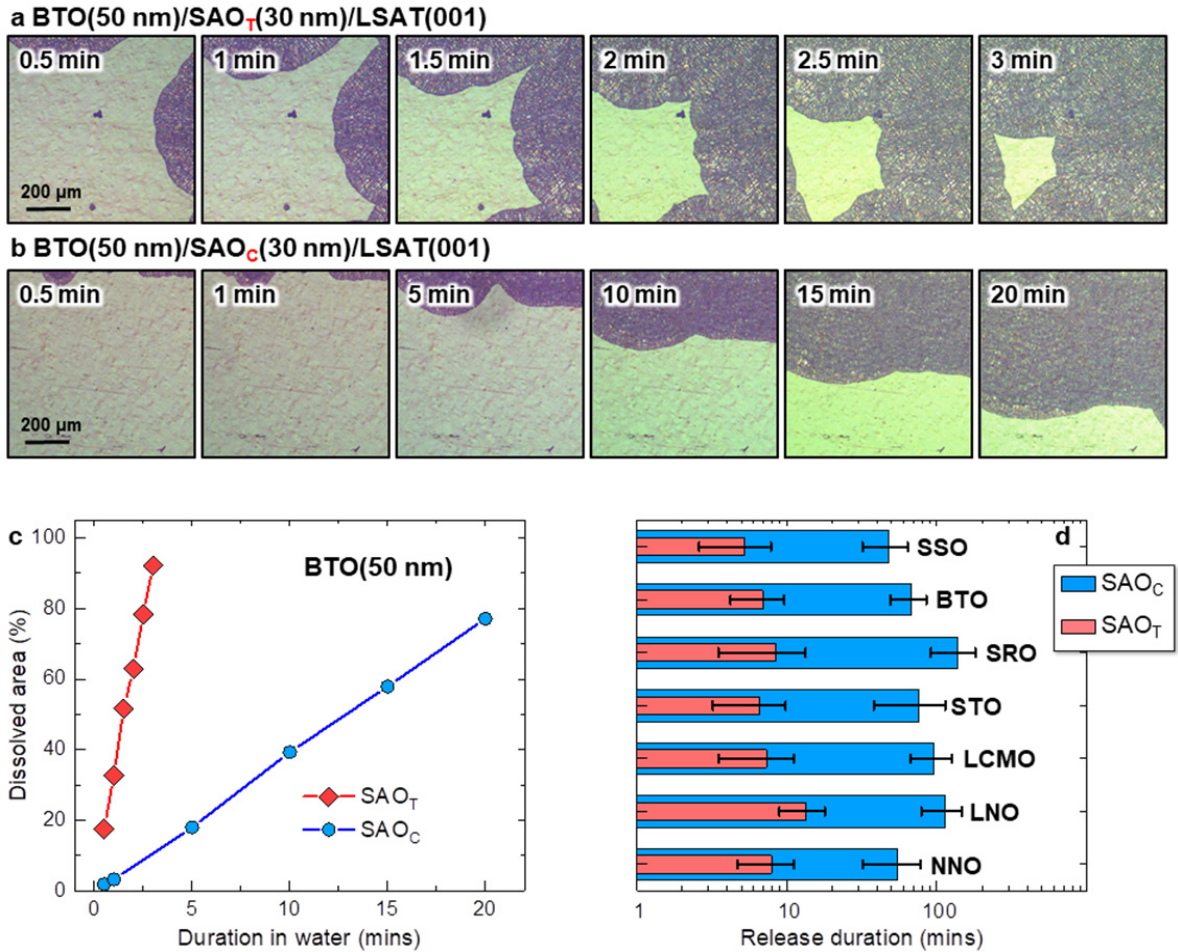

**Supplementary Fig. 31. | Release speed of oxide membranes from SAO$_C$ and SAO$_T$.** (**a,b**) In-situ optical microscopic images measured from (**a**) BTO(50 nm)/SAO$_T$(30 nm)/LSAT(001) and (**b**) BTO(50 nm)/SAO$_C$(30 nm)/LSAT(001) bilayer heterostructures during the dissolution processes in deionized water. The bright and dark regions correspond to the undissolved and dissolved regions, respectively. The monitored area for all the optical images is 1 mm$^2$. (**c**) Relative dissolved areas of BTO membranes plotted as a function of time duration in water. (**d**) Time duration for complete release of various oxide membranes (35 nm) from SAO$_C$ and SAO$_T$. The sizes for all the samples are 5×5 mm$^2$. The error bars represent the standard deviation of release time durations of 5 samples.

| Transferred films | Thickness (nm) | Exfoliation Duration | Reference |
|---|---|---|---|
| LSMO | 20~40 | ~1 day | *Nat. Mater.* **15**, 1255 (2016) |
| BTO (2.8 nm)/LSMO (10 nm)/BTO (2.8 nm) | 15.6 | 24 h | *Nano Lett.* **19**, 3999 (2019) |
| BTO | 8~70 | several hours | *Cryst.* **10**, 733 (2020) |
| $BiFeO_3$ | 110 | few hours | *Sci. Adv.* **6**, eaba5847 (2020) |
| SRO, $CeO_2$, $CeO_2$/STO | 100 | 24 h | *Nano-Micro Lett.* **13**, 39 (2021) |
| $Sr_2IrO_4$ | 24 | few minutes ~ hours | *ACS Appl. Nano Mater.* **3**, 6310 (2020) |
| SRO (BTO) | 100~300 | 1 day | *Adv. Funct. Mater.* **30**, 2001236 (2020) |
| SRO | 15~30 | 12 h | *npj Flex. Electron.* **6**, 24 (2022) |
| $Fe_3O_4$ | 160 | 5 min | *Adv. Funct. Mater.* **30**, 2003495 (2020) |
| STO ($Sr_2CaAl_2O_6$) | 14 | Not mentioned | *Nat. Commun.* **11**, 3141 (2020) |
| LCMO ($SrCa_2Al_2O_6$) | 4~8 | Not mentioned | *Science* **368**, 71 (2020) |
| La:$BaSnO_3$ ($Ba_3Al_2O_6$) | 60 | 5~10 min | *ACS Appl. Electron. Mater.* **1**, 1269 (2019) |

**Supplementary Table 4. | Summary of the etching time of some crystal functional oxides with $SAO_C$ and its derivatives serve as sacrificial layer reported.** We quote the same terminologies about exfoliation time durations mentioned in the publications.

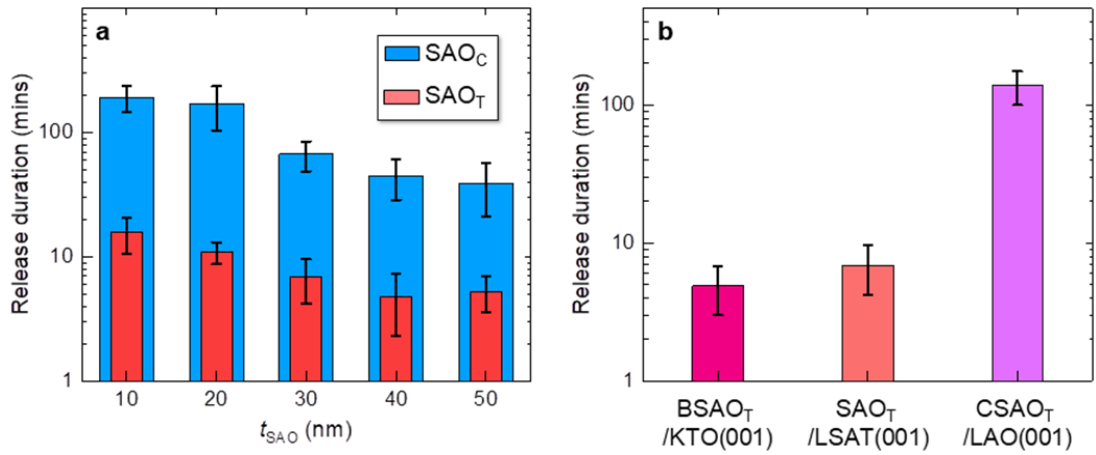

**Supplementary Fig. 32.** | (**a**) Time duration for complete release of 50 nm thick BTO films grown on $SAO_C$ and $SAO_T$ layers with various thicknesses ($t_{SAO}$). (**b**) Time duration for complete release of 50 nm thick BTO films grown on 30 nm thick $BSAO_T$/KTO(001), $SAO_T$/LSAT(001) and $CSAO_T$/LAO(001) films.

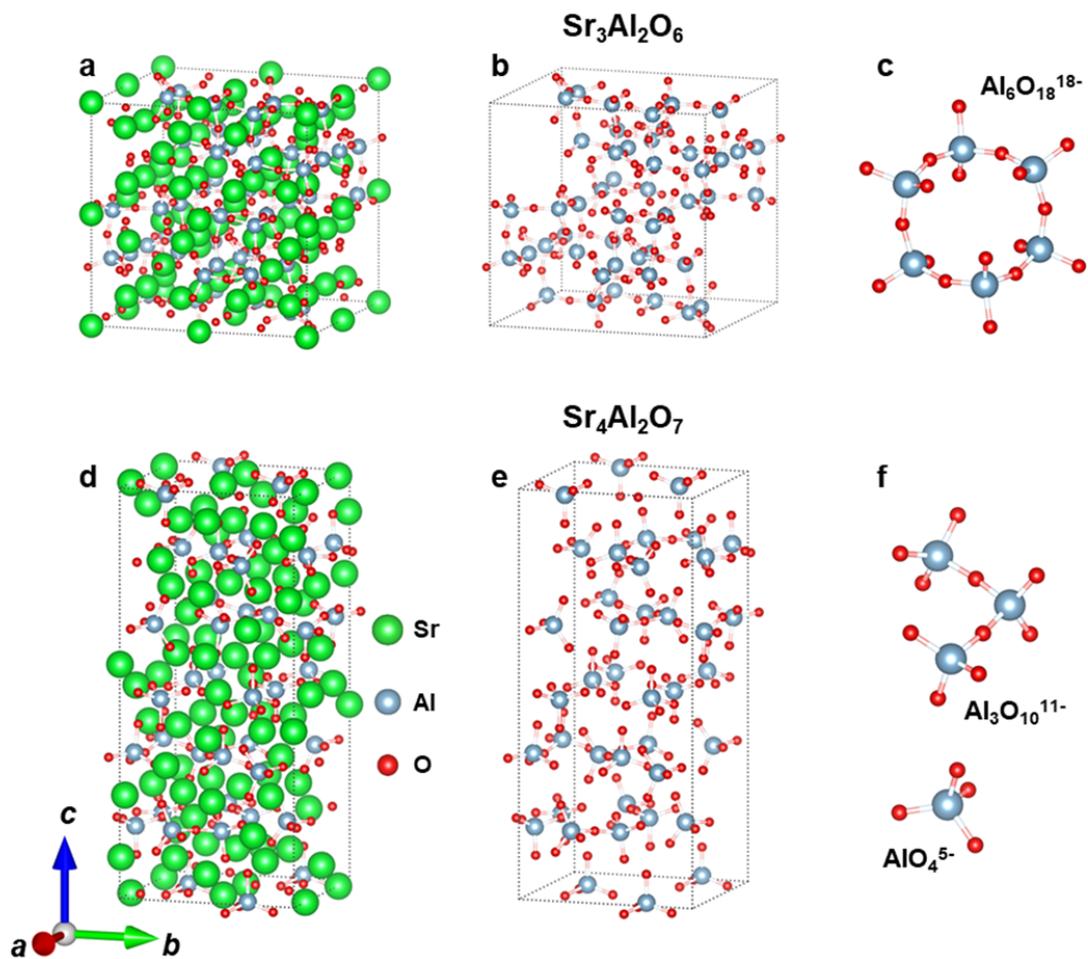

**Supplementary Fig. 33. | Schematics on the structural origin of the high water-solubility of $SAO_T$.**
(**a**) Schematic illustration of the $SAO_C$ unit-cell. (**b**) Schematic illustration of the Al-O networks in the $SAO_C$ unit-cell. (**c**) Schematic illustration of the discrete 12-membered $Al_6O_{18}^{18-}$ ring in the $SAO_C$ unit-cell. (**d**) Schematic illustration of the $SAO_T$ unit-cell. (**e**) Schematic illustration of the Al-O networks in the $SAO_T$ unit-cell. (**f**) Schematic illustration of the isolated $AlO_4^{5-}$ and $Al_3O_{10}^{11-}$ groups in the $SAO_T$ unit-cell.

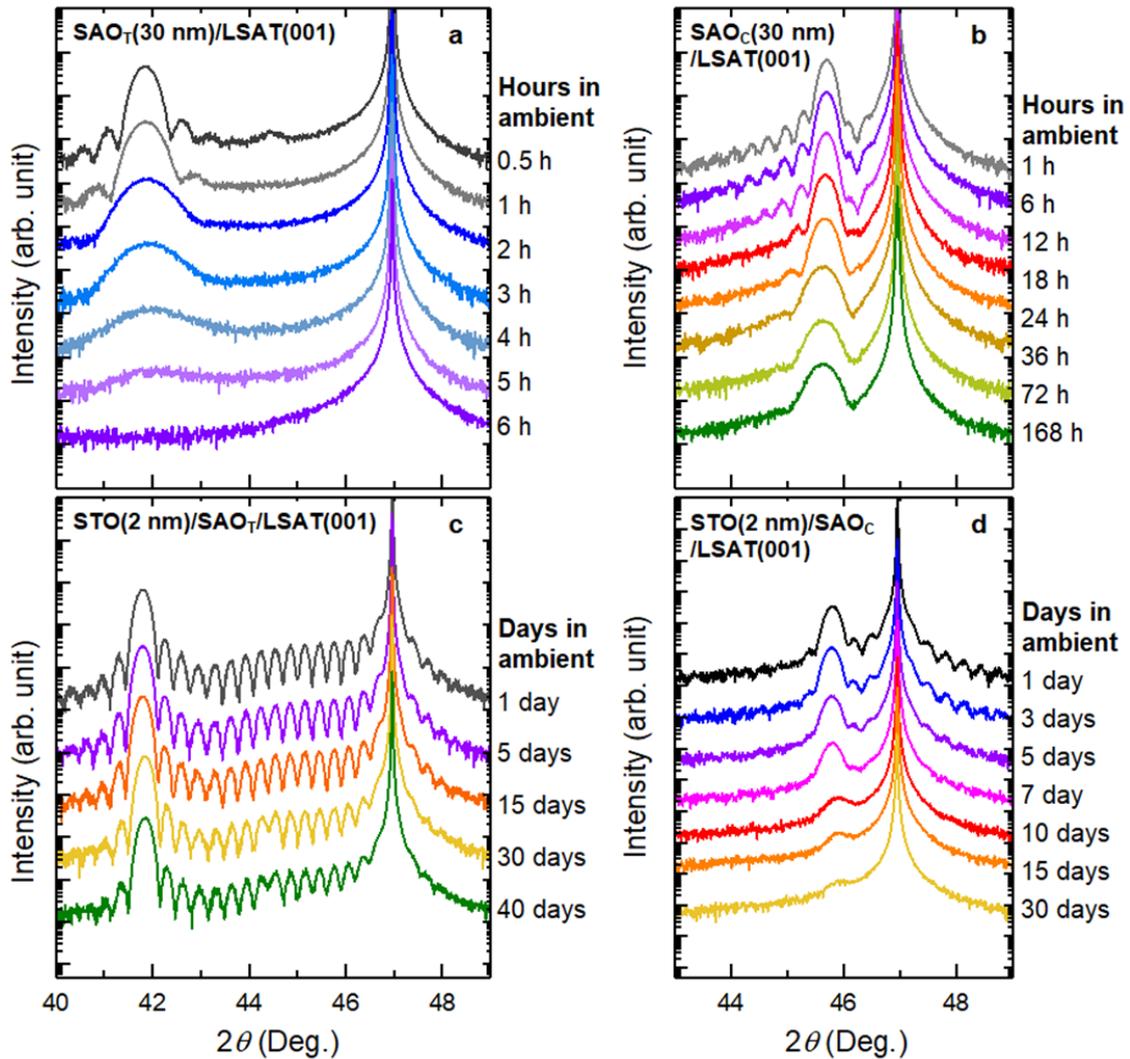

**Supplementary Fig. 34. | Stability test for SAO$_T$ and SAO$_C$ thin films against humidity in ambient.** XRD 2θ-ω scans measured from (**a**) SAO$_T$(30 nm)/LSAT(001) single layer film, (**b**) SAO$_C$(30 nm)/LSAT(001) single layer film, (**c**) STO(2 nm)/SAO$_T$(30 nm)/LSAT(001) bilayer, and (**d**) STO(2 nm)/SAO$_C$(30 nm)/LSAT(001) bilayer after stored in ambient for various durations. The samples are stored in a controlled ambient environment: a temperature of 24 °C and a relative humidity of 65%.

**Supplementary Movie 1.**

In-situ monitoring movie of the exfoliation processes of BTO(50 nm)/SAO$_T$(30 nm)/LSAT(001)

**Supplementary Movie 2.**

In-situ monitoring movie of the exfoliation processes of BTO(50 nm)/SAO$_C$(30 nm)/LSAT(001)

**Reference**

1. Chrisey, D. B. Editor & Hubler, G. K. Editor, *Pulsed Laser Deposition of Thin Films*.(John Wiley & Sons, New York, 1994).

2. Guo, H. et al. Growth diagram of $La_{0.7}Sr_{0.3}MnO_3$ thin films using pulsed laser deposition. *J. Appl. Phys.* **113**, 234301 (2013).

3. Gonzalo, J., Gómez San Román, R., Perrière, J., Afonso, C. N. & Pérez Casero R. Pressure effects during pulsed-laser deposition of barium titanate thin films. *Appl. Phys. A Mater.* **66**, 487–491 (1998).

4. Xu, C. et al. Impact of the interplay between nonstoichiometry and kinetic energy of the plume species on the growth mode of $SrTiO_3$ thin films. *J. Phys. D: Appl. Phys.* **47**, 034009 (2014).

5. Dushman, S., Lafferty, J. M. & Pasternak, R. A. Scientific Foundations of Vacuum Technique. *Phys. Today*. **15**, 53-54 (1962).

6. Hau, S. K., Wong, K. H., Chan, P. W. & Choy, C. L. Intrinsic resputtering in pulsed-laser deposition of lead-zirconate-titanate thin films. *Appl. Phys. Lett.* **66**, 245 (1995).

7. Liu, G. Z., Lei Q. Y. & Xi, X. X. Stoichiometry of $SrTiO_3$ films grown by pulsed laser deposition. *Appl. Phys. Lett.* **100**, 202902 (2012).

8. Dam, B., Rector, J. H., Johansson, J., Huijbregtse, J. & De Groot, D. G. Mechanism of incongruent ablation of $SrTiO_3$. *J. Appl. Phys.* **83**, 3386–3389 (1998).

9. Baek, D. J., Lu, D., Hikita, Y., Hwang, H. Y. & Kourkoutis, L. F. Mapping cation diffusion through lattice defects in epitaxial oxide thin films on the water-soluble buffer layer $Sr_3Al_2O_6$ using atomic resolution electron microscopy. *APL Mater.* **5**, 096108 (2017).

10. Lu, D. et al. Strain Tuning in Complex Oxide Epitaxial Films Using an Ultrathin Strontium Aluminate Buffer Layer. Phys. *Status Solidi - Rapid Res. Lett.* **12**, 1700339 (2018).

11. Foltyn, S. R. et al. Materials science challenges for high-temperature superconducting wire. *Nat. Mater.* **6**, 631 (2007).

12. Rossell, M. D. et al. Atomic structure of highly strained $BiFeO_3$ thin films. *Phys. Rev. Lett.* **108**, 047601 (2012).

13. Liu, P. et al. High-Quality Ruddlesden–Popper Perovskite Film Formation for High-Performance Perovskite Solar Cells. *Adv. Mater.* **33**, 2002582 (2021).

14. Kahlenberg, V. Crystal structure of $Ba_8[Al_3O_{10}][AlO_4]$, a novel mixed-anion Ba aluminate related to kilchoanite. *Mineral. Mag.* **65**, 533–541 (2001).

15. Kahlenberg, V., Lazić, B. & Krivovichev, S. V. Tetrastrontium-digalliumoxide ($Sr_4Ga_2O_7$)-Synthesis and crystal structure of a mixed anion strontium gallate related to perovskite. *J. Solid State Chem.* **178**, 1429–1439 (2005).

16. Ishihara, T., Matsuda, H., Bin Bustam, M. A. & Takita Y. Oxide ion conductivity in doped Ga based perovskite type oxide. *Solid State Ionics*. **86–88**, 197–201 (1996).

17. Kahlenberg, V. The crystal structures of the strontium gallates $Sr_{10}Ga_6O_{19}$ and $Sr_3Ga_2O_6$. *J. Solid State Chem.* **160**, 421–429 (2001).

18. Kahlenberg, V. Synthesis and crystal structure of $Sr_{10}Al_6O_{19}$: A derivative of the perovskite structure type in the system $SrO-Al_2O_3$. *Mater. Res. Bull.* **37**, 715–726 (2002).

19. Gao, G., Jin, S. & Wu, W. Lattice-mismatch-strain induced inhomogeneities in epitaxial $La_{0.7}Ca_{0.3}MnO_3$ films. *Appl. Phys. Lett.* **90**, 012509 (2007).

20. Hong, S. S. et al. Extreme tensile strain states in La$_{0.7}$Ca$_{0.3}$MnO$_3$ membranes. *Science.* **368**, 71–76 (2020).

21. Koster, G. et al. Structure, physical properties, and applications of SrRuO$_3$ thin films. *Rev. Mod. Phys.* **84**, 253–298 (2012).

22. Wang, L. et al. Ferroelectrically tunable magnetic skyrmions in ultrathin oxide heterostructures. *Nat. Mater.* **17**, 1087–1094 (2018).

23. Lee, H. G. et al. Atomic-Scale Metal-Insulator Transition in SrRuO$_3$ Ultrathin Films Triggered by Surface Termination Conversion. *Adv. Mater.* **32**, 1905815 (2020).

24. Lu, J. et al. Defect-Engineered Dzyaloshinskii-Moriya Interaction and Electric-Field-Switchable Topological Spin Texture in SrRuO$_3$. *Adv. Mater.* **33**, 2102525 (2021).

25. Kawasaki, M. et al. Atomic control of the SrTiO$_3$ crystal surface. *Science.* **266**, 1540–1542 (1994).

26. Middey, S. et al. Physics of Ultrathin Films and Heterostructures of Rare-Earth Nickelates. *Annu. Rev. Mater. Res.* **46**, 305–334 (2016).

27. Chandra, M., Das, S., Aziz, F., Prajapat, M. & Mavani, K. R. Induced metal-insulator transition and temperature independent charge transport in NdNiO$_{3-\delta}$ thin films. *J. Alloys Compd.* **696**, 423–427 (2017).

28. Wang, H. et al. Transparent and conductive oxide films of the perovskite La$_x$Sr$_{1-x}$SnO$_3$ (x ≤ 0.15): Epitaxial growth and application for transparent heterostructures. *J. Phys. D. Appl. Phys.* **43**, 035403 (2010).